\documentclass[reqno,centertags]{amsart}
\usepackage{amssymb}
\usepackage{hyperref}
\usepackage{upref}

\newtheorem{theorem}{Theorem}[section]
\newtheorem{lemma}[theorem]{Lemma}

\newtheorem{hypothesis}[theorem]{Hypothesis}
\newtheorem{example}[theorem]{Example}
\theoremstyle{remark}
\newtheorem{remark}[theorem]{Remark}

\allowdisplaybreaks

\newcommand{\bbN}{{\mathbb{N}}}
\newcommand{\bbR}{{\mathbb{R}}}
\newcommand{\bbZ}{{\mathbb{Z}}}
\newcommand{\bbC}{{\mathbb{C}}}

\newcommand{\calD}{{\mathcal D}}
\newcommand{\calE}{{\mathcal E}}
\newcommand{\calF}{{\mathcal F}}
\newcommand{\calI}{{\mathcal I}}
\newcommand{\calK}{{\mathcal K}}

\newcommand{\calM}{{\mathcal M}}

\newcommand{\f}{\frac}
\newcommand{\lb}{\label}
\newcommand{\no}{\nonumber}
\newcommand{\be}{\begin{equation}}
\newcommand{\ee}{\end{equation}}
\newcommand{\bea}{\begin{eqnarray}}
\newcommand{\eea}{\end{eqnarray}}
\newcommand{\ul}{\underline}
\newcommand{\ol}{\overline}
\newcommand{\ti}{\widetilde}

\newcommand{\frh}{\mathfrak{h}}
\newcommand{\frf}{\mathfrak{f}}
\newcommand{\frg}{\mathfrak{g}}

\newcommand{\tr}{\mathrm{tr}}

\DeclareMathOperator{\sAL}{s-AL}
\DeclareMathOperator{\shAL}{s-\hatt {AL}}
\DeclareMathOperator{\AL}{AL}
\DeclareMathOperator{\hAL}{\hatt{AL}}

\newcommand{\humu}{{ \hat{\underline{\mu} }}}
\newcommand{\hunu}{{\underline{\hat{\nu}}}}
\newcommand{\hmu}{{\hat{\mu} }}
\newcommand{\hnu}{{\hat{\nu}}}

\newcommand{\dpm}{\delta_\pm}

\newcommand{\Div}{\operatorname{Div}}

\newcommand{\Oh}{O}

\newcommand{\dott}{\,\cdot\,}
\newcommand{\hatt}{\widehat}  
\newcommand{\Pinfp}{{P_{\infty_+}}}
\newcommand{\Pinfm}{{P_{\infty_-}}}
\newcommand{\Pzp}{{P_{0,+}}}
\newcommand{\Pzm}{{P_{0,-}}}
\newcommand{\Pinfpm}{{P_{\infty_\pm}}}
\newcommand{\Pzpm}{{P_{0,\pm}}}

\DeclareMathOperator{\sym}{Sym}

\newcommand{\symq}{{\sym^{p} (\calK_p)}}

\newcommand{\pgam}{\Gamma}

\newcommand{\gam}{\gamma}

\numberwithin{equation}{section}

\begin{document}

\title[The Ablowitz--Ladik Hierarchy Revisited]{The Ablowitz--Ladik Hierarchy Revisited}
\author[F.\ Gesztesy]{Fritz Gesztesy}
\address{Department of Mathematics,
University of Missouri,
Columbia, MO 65211, USA}
\email{\href{mailto:fritz@math.missouri.edu}{fritz@math.missouri.edu}}
\urladdr{\href{http://www.math.missouri.edu/personnel/faculty/gesztesyf.html}{http://www.math.missouri.edu/personnel/faculty/gesztesyf.html}}

\author[H.\ Holden]{Helge Holden}
\address{Department of Mathematical Sciences,
Norwegian University of
Science and Technology, NO--7491 Trondheim, Norway}
\email{\href{mailto:holden@math.ntnu.no}{holden@math.ntnu.no}}
\urladdr{\href{http://www.math.ntnu.no/\~{}holden/}{http://www.math.ntnu.no/\~{}holden/}}

\author[J. Michor]{Johanna Michor}
\address{Faculty of Mathematics\\
University of Vienna\\
Nordbergstrasse 15\\ 1090 Wien\\ Austria\\ and International Erwin Schr\"odinger
Institute for Mathematical Physics, Boltzmanngasse 9\\ 1090 Wien\\ Austria}
\email{\href{mailto:Johanna.Michor@esi.ac.at}{Johanna.Michor@esi.ac.at}}
\urladdr{\href{http://www.mat.univie.ac.at/~jmichor/}{http://www.mat.univie.ac.at/\~{}jmichor/}}

\author[G. Teschl]{Gerald Teschl}
\address{Faculty of Mathematics\\
University of Vienna\\
Nordbergstrasse 15\\ 1090 Wien\\ Austria\\ and International Erwin Schr\"odinger
Institute for Mathematical Physics, Boltzmanngasse 9\\ 1090 Wien\\ Austria}
\email{\href{mailto:Gerald.Teschl@univie.ac.at}{Gerald.Teschl@univie.ac.at}}
\urladdr{\href{http://www.mat.univie.ac.at/~gerald/}{http://www.mat.univie.ac.at/\~{}gerald/}}

\thanks{Research supported in part by the Research Council of Norway,  
the US National Science Foundation under Grant No.\ DMS-0405526, and
the Austrian Science Fund (FWF) under Grant No.\ Y330.}

\thanks{To appear in {\it  Proceedings of the conference on 
Operator Theory, Analysis in Mathematical Physics - OTAMP 2006}, J.\ Janas, P.\ Kurasov, A.\ Laptev, S.\ Naboko, and G.\ Stolz (eds.), Operator Theory: Advances and Applications, Birkh\"auser, Basel.}

\date{\today}

\begin{abstract}
We provide a detailed recursive construction of the Ablowitz--Ladik (AL) hierarchy and its zero-curvature formalism. The two-coefficient AL hierarchy under investigation can be considered a complexified version of the discrete nonlinear 
Schr\"odinger equation and its hierarchy of nonlinear evolution equations.

Specifically, we discuss in detail the stationary Ablowitz--Ladik formalism in connection with the underlying hyperelliptic curve and the stationary Baker--Akhiezer function and separately the corresponding time-dependent Ablowitz--Ladik formalism. 
\end{abstract}

\maketitle

\section{Introduction}\lb{sAL1}

The prime example of an integrable nonlinear differential-difference system to be discussed in this paper is the Ablowitz--Ladik system, 
\begin{align}
\begin{split}
-i\alpha_t - (1 - \alpha \beta) (\alpha^- +\alpha^+) + 2\alpha &=0, \lb{AL1.12} \\
-i\beta_t + (1 - \alpha\beta)(\beta^- + \beta^+) - 2\beta &=0   
\end{split}
\end{align}
with $\alpha=\alpha(n,t)$, $\beta=\beta(n,t)$, $(n,t)\in\bbZ\times\bbR$. 
Here we used the notation $f^{\pm}(n) = f(n\pm 1)$, $n\in\bbZ$, for complex-valued 
sequences $f=\{f(n)\}_{n\in\bbZ}$. The system \eqref{AL1.12} arose in the mid-seventies when Ablowitz and Ladik, in a series of papers \cite{AblowitzLadik:1975}--\cite{AblowitzLadik2:1976} (see also
\cite{Ablowitz:1977}, \cite[Sect.\ 3.2.2]{AblowitzClarkson:1991},
\cite[Ch.\ 3]{AblowitzPrinariTrubatch:2004},
\cite{ChiuLadik:1977}), used inverse scattering methods to analyze certain
integrable differential-difference systems. In particular, Ablowitz and Ladik 
\cite{AblowitzLadik:1976} (see also \cite[Ch.\ 3]{AblowitzPrinariTrubatch:2004}) showed that in the focusing case, where $\beta = -\ol\alpha$, and in the defocusing case, where 
$\beta = \ol\alpha$, \eqref{AL1.12} yields the discrete analog of the nonlinear 
Schr\"odinger equation
\begin{equation}
-i\alpha_t - (1 \pm |\alpha|^2)(\alpha^- + \alpha^+) +2\alpha =0. \lb{AL1.13}
\end{equation}
We will refer to \eqref{AL1.12} as the Ablowitz--Ladik system. The principal theme of this paper will be to derive a detailed recursive construction of the Ablowitz--Ladik hierarchy, a completely integrable sequence of systems of nonlinear evolution equations on the lattice $\bbZ$ whose first nonlinear member is the Ablowitz--Ladik system 
\eqref{AL1.12}. In addition, we discuss the zero-curvature formalism for the Ablowitz--Ladik (AL) hierarchy in detail.

Since the original discovery of Ablowitz and Ladik in the mid-seventies, there has been great interest in the area of integrable differential-difference equations. Two principal directions of research are responsible for this development: Originally, the development  was driven by the theory of completely integrable systems and its applications to fields such as nonlinear optics, and more recently, it gained additional momentum due to its intimate connections with the theory of orthogonal polynomials. In this paper we will not discuss the connection with orthogonal polynomials (see, however, the introduction of \cite{GesztesyHoldenMichorTeschl:2007}) and instead refer to the recent references 
\cite{BertolaGekhtman:2005}, \cite{Deift:2007}, \cite{{KillipNenciu:2006}}, \cite{Li:2005}, \cite{Nenciu:2005}, \cite{Nenciu:2005a}, \cite{Nenciu:2006}, \cite{Simon:2005}, 
\cite{Simon:2005a}, \cite{Simon:2006}, and the literature cited therein.

The first systematic discussion of the Ablowitz--Ladik (AL) hierarchy appears to be due to Schilling \cite{Schilling:1989} (cf.\ also \cite{TamizhmaniMa:2000}, 
\cite{Vekslerchik:2002}, \cite{ZengRauchWojciechowski:1995}); infinitely many conservation laws are derived, for instance, by Ding, Sun, and Xu \cite{DingSunXu:2006}; the bi-Hamiltonian structure of the AL hierarchy is considered by Ercolani and Lozano \cite{ErcolaniLozano:2006}; connections
between the AL hierarchy and the motion of a piecewise linear curve have
been established by Doliwa and Santini \cite{DoliwaSantini:1995}; B\"acklund and Darboux transformations were studied by Geng \cite{Geng:1989} and 
Vekslerchik \cite{Vekslerchik:2006}; the Hirota bilinear formalism, AL $\tau$-functions, etc., were considered by Vekslerchik \cite{Vekslerchik:2002}. The initial value problem for half-infinite AL systems was discussed by Common \cite{Common:1992}, for an application of the inverse scattering method to \eqref{AL1.13} we refer to Vekslerchik and Konotop \cite{VekslerchikKonotop:1992}. This just scratches the surface of these developments and the interested reader will find much more material in the references cited in these papers and the ones discussed below. Algebro-geometric (and periodic) solutions of the AL system \eqref{AL1.12} have briefly been studied by Ahmad and Chowdhury \cite{AhmadChowdhury:1987}, 
\cite{AhmadChowdhury:1987a}, Bogolyubov, Prikarpatskii, and Samoilenko
\cite{BogolyubovPrikarpatskiiSamoilenko:1981},  Bogolyubov and
Prikarpatskii \cite{BogolyubovPrikarpatskii:1982}, 
Chow, Conte, and Xu \cite{ChowConteXu:2006}, Geng, Dai, and Cao 
\cite{GengDaiCao:2003}, and Vaninsky \cite{Vaninsky:2001}. 

In an effort to analyze models describing oscillations in nonlinear dispersive wave systems, Miller, Ercolani, Krichever, and Levermore \cite{MillerErcolaniKricheverLevermore:1995} (see also \cite{Miller:1995}) gave a detailed analysis of algebro-geometric solutions of the AL system 
\eqref{AL1.12}. Introducing 
\begin{equation}
U(z)=\begin{pmatrix} z & \alpha \\ \beta z & 1 \end{pmatrix},
\quad
V(z)=i\begin{pmatrix}
z-1-\alpha\beta^- & \alpha - \alpha^- z^{-1}\\ \beta^- z -\beta &
1+ \alpha^- \beta -z^{-1} \end{pmatrix}    \lb{AL1.20}
\end{equation}
with $z\in\bbC\setminus\{0\}$ a spectral parameter,  the authors in
\cite{MillerErcolaniKricheverLevermore:1995} relied on the fact that the 
Ablowitz--Ladik system \eqref{AL1.12} is equivalent to the zero-curvature equations
\begin{equation}
U_t+UV -V^+ U=0.  \lb{AL1.21}
\end{equation} 
Miller, Ercolani, Krichever, and Levermore
\cite{MillerErcolaniKricheverLevermore:1995} then derived the theta function representations of $\alpha,\beta$ satisfying the AL system \eqref{AL1.12}. Vekslerchik \cite{Vekslerchik:1999} also studied finite-genus solutions for the AL hierarchy by establishing connections with Fay's identity for theta functions. Recently, a detailed study of algebro-geometric solutions for the entire AL hierarchy has been provided in \cite{GesztesyHoldenMichorTeschl:2007}. The latter reference also contains an extensive discussion of the connection between the Ablowitz--Ladik system \eqref{AL1.12} and orthogonal polynomials on the unit circle. The algebro-geometric initial value problem for the Ablowitz--Ladik hierarchy with complex-valued initial data, that is, the construction 
of $\alpha$ and $\beta$ by starting from a set of initial data (nonspecial divisors) of full measure, will be presented in \cite{GesztesyHoldenMichorTeschl:2007a}. 
The Hamiltonian and Lax formalisms for the AL hierarchy will be revisited in 
\cite{GesztesyHoldenMichorTeschl:2007b}. 

In addition to these recent  developments on the AL system and the AL hierarchy, we offer a variety of results in this paper apparently not covered before. These include: 

$\bullet$ An effective recursive construction of the AL hierarchy using Laurent polynomials.

$\bullet$ The detailed connection between the AL hierarchy and a ``complexified'' version of transfer matrices first introduced by Baxter \cite{Baxter:1960}, \cite{Baxter:1961}.

$\bullet$ A detailed treatment of the stationary and time-dependent Ablowitz--Ladik formalism.

The structure of this paper is as follows: In Section \ref{sAL2} we
describe our zero-curvature formalism for the Ablowitz--Ladik (AL) hierarchy.
Extending a recursive polynomial approach discussed in great detail in 
\cite{GesztesyHolden:2003} in the continuous case and in 
\cite{BullaGesztesyHoldenTeschl:1997}, \cite[Ch.\ 4]{GesztesyHolden:2005}, 
\cite[Chs.\ 6, 12]{Teschl:2000} in the discrete context to the case of Laurent polynomials with respect to the spectral parameter, we derive the AL hierarchy of systems of nonlinear evolution equations whose first nonlinear member is the Ablowitz--Ladik system \eqref{AL1.12}. Section \ref{sAL3} is devoted to a detailed study of the stationary AL hierarchy. We employ the recursive Laurent polynomial formalism of Section \ref{sAL2} to describe nonnegative divisors of degree $p$ on a hyperelliptic curve $\calK_p$ of genus $p$ associated with the $\ul p$th system in the stationary AL hierarchy. The corresponding time-dependent results for the AL hierarchy are presented in detail in Section \ref{sAL4}. Finally, Appendix \ref{ALApp.high} is of a technical nature and summarizes expansions of various key quantities related to the Laurent polynomial recursion formalism as the spectral parameter tends to zero or to infinity.

\section{The Ablowitz--Ladik Hierarchy, Recursion Relations, 
Zero-Curvature Pairs, and Hyperelliptic Curves} \label{sAL2}

In this section we provide the construction of the Ablowitz--Ladik hierarchy employing a polynomial recursion formalism and derive the associated sequence of Ablowitz--Ladik zero-curvature pairs. Moreover, we discuss the hyperelliptic curve underlying the stationary Ablowitz--Ladik hierarchy.

We denote by $\bbC^{\bbZ}$ the set of complex-valued sequences indexed by $\bbZ$. 

Throughout this section we suppose the following hypothesis. 

\begin{hypothesis} \lb{hAL2.1} 
In the stationary case we assume that $\alpha, \beta$ satisfy
\begin{equation}
\alpha, \beta\in \bbC^{\bbZ}, \quad  \alpha(n)\beta(n)\notin \{0,1\}, \; n\in\bbZ.   
\lb{AL2.01}
\end{equation}
In the time-dependent case we assume that $\alpha, \beta$ satisfy
\begin{align}
\begin{split}
& \alpha(\dott,t), \beta(\dott,t) \in \bbC^{\bbZ}, \; t\in\bbR, \quad 
\alpha(n,\dott), \beta(n,\dott)\in C^1(\bbR), \; n\in\bbZ,   \lb{AL2.01a}  \\
& \alpha(n,t)\beta(n,t)\notin \{0,1\}, \; (n,t)\in\bbZ\times \bbR.
\end{split}
\end{align}
\end{hypothesis}

Actually, up to Remark \ref{rAL2.12} our analysis will be time-independent and hence only the lattice variations of $\alpha$ and $\beta$ will matter. 

We denote by $S^\pm$ the shift operators acting on complex-valued sequences 
$f=\{f(n)\}_{n\in\bbZ} \in\bbC^{\bbZ}$ according to
\begin{equation}
(S^\pm f)(n)=f(n\pm1), \quad n\in\bbZ. \lb{AL2.02}
\end{equation}
Moreover, we will frequently use the notation
\begin{equation}
f^\pm = S^{\pm} f, \quad f\in\bbC^{\bbZ}. 
\end{equation}

To construct the  Ablowitz--Ladik hierarchy we will try to generalize
\eqref{AL1.20} by considering the $2\times2$ matrix
\begin{equation}
U(z) = \begin{pmatrix} z & \alpha  \\ z \beta & 1\\ \end{pmatrix},  
\quad z\in\bbC,    \lb{AL2.03}
\end{equation}
and making the ansatz
\begin{equation} \lb{AL_v}
V_{\ul p}(z) = i  \begin{pmatrix}
G_{\ul p}^-(z) & - F_{\ul p}^-(z)     \\[1.5mm]
H_{\ul p}^-(z) & - K_{\ul p}^-(z)  \\
\end{pmatrix},  \quad \ul p =(p_-,p_+)\in \bbN_0^2,
\end{equation}
where $G_{\ul p}$, $K_{\ul p}$, $F_{\ul p}$, and $H_{\ul p}$ are chosen as Laurent polynomials\footnote{In this paper, a sum is interpreted as zero whenever the upper limit in the sum is strictly less than its lower limit.} (suggested by the appearance of $z^{-1}$ in the matrix $V$ in \eqref{AL1.20})
\begin{align}  \label{ALansatz}
\begin{split}
G_{\ul p}(z) &= \sum_{\ell=1}^{p_-} z^{-\ell} g_{p_- -\ell,-} 
+ \sum_{\ell=0}^{p_+} z^\ell g_{p_+ -\ell,+}, \\
F_{\ul p}(z) &= \sum_{\ell=1}^{p_-} z^{-\ell} f_{p_- -\ell,-} 
+ \sum_{\ell=0}^{p_+} z^\ell f_{p_+ -\ell,+}, \\
H_{\ul p}(z) &= \sum_{\ell=1}^{p_-} z^{-\ell} h_{p_- -\ell,-} 
+ \sum_{\ell=0}^{p_+} z^\ell h_{p_+ -\ell,+}, \\
K_{\ul p}(z) &= \sum_{\ell=1}^{p_-} z^{-\ell} k_{p_- -\ell,-} 
+ \sum_{\ell=0}^{p_+} z^\ell k_{p_+ -\ell,+}.
\end{split}
\end{align}
Without loss of generality we will only look at the time-independent case and add time
later on. Then the stationary zero-curvature equation, 
\begin{equation}   \lb{ALstatzc}
0=U V_{\ul p} - V_{\ul p}^+ U,
\end{equation}
is equivalent to the following relationships between the Laurent polynomials
\begin{equation} \label{ALzc0}
U V_{\ul p} - V_{\ul p}^+ U=i\begin{pmatrix} z (G_{\ul p}^- - G_{\ul p}) + z \beta F_{\ul p} + \alpha H_{\ul p}^- & 
 F_{\ul p} - z F_{\ul p}^- - \alpha (G_{\ul p} + K_{\ul p}^-) \\[1mm]
  z \beta (G_{\ul p}^- + K_{\ul p}) - z H_{\ul p} + H_{\ul p}^- &
  -z \beta F_{\ul p}^- - \alpha H_{\ul p} + K_{\ul p} - K_{\ul p}^-  \end{pmatrix},
\end{equation}
respectively, to  
\begin{align} \label{AL1,1}
z (G_{\ul p}^- - G_{\ul p}) + z \beta F_{\ul p} + \alpha H_{\ul p}^- &= 0,\\ \label{AL2,2}
z \beta F_{\ul p}^- + \alpha H_{\ul p} - K_{\ul p} + K_{\ul p}^- &= 0,\\
\label{AL1,2}
 - F_{\ul p} + z F_{\ul p}^- + \alpha (G_{\ul p} + K_{\ul p}^-) &= 0,\\ \label{AL2,1}
 z \beta (G_{\ul p}^- + K_{\ul p}) - z H_{\ul p} + H_{\ul p}^- &= 0.
\end{align}

\begin{lemma}
Suppose the Laurent polynomials defined in \eqref{ALansatz} satisfy the zero-curvature equation \eqref{ALstatzc}, then
\begin{align}
& f_{0,+} = 0, \quad h_{0,-} = 0, \quad g_{0,\pm} = g_{0,\pm}^-,   
\quad k_{0,\pm} = k_{0,\pm}^-,   \lb{ALfhg}  \\
& k_{\ell,\pm} - k_{\ell,\pm}^- = g_{\ell,\pm} - g_{\ell,\pm}^-, \; \ell=0, \dots, p_\pm-1, 
\quad g_{p_+,+}-g^-_{p_+,+} = k_{p_+,+}-k^-_{p_+,+}.   \lb{ALkg}
\end{align}
\end{lemma}
\begin{proof}
Comparing coefficients at the highest order of $z$ in \eqref{AL2,2} and 
the lowest in \eqref{AL1,1} immediately yields $f_{0,+} = 0$, $h_{0,-} = 0$. 
Then $g_{0,+}=g^-_{0,+}$, $k_{0,-}=k^-_{0,-}$ are necessarily lattice constants by 
\eqref{AL1,1}, \eqref{AL2,2}. Since $\det(U(z))\neq 0$ for $z\in\bbC \setminus\{0\}$ by 
\eqref{AL2.01}, \eqref{ALstatzc} yields $\tr(V_{\ul p}^+)=\tr(UV_{\ul p}U^{-1})=\tr(V_{\ul p})$ and hence
\begin{equation}
G_{\ul p} - G^-_{\ul p} = K_{\ul p} - K^-_{\ul p}, 
\end{equation}
implying \eqref{ALkg}. Taking $\ell=0$ in \eqref{ALkg} then yields $g_{0,-}=g^-_{0,-}$ 
and $k_{0,+}=k^-_{0,+}$. 
\end{proof}

In particular, this lemma shows that we can choose 
\begin{equation}
k_{\ell,\pm} = g_{\ell,\pm}, \; 0\le \ell \le p_\pm-1, \quad k_{p_+,+}=g_{p_+,+}  \lb{ALg=k}
\end{equation}
without loss of generality (since this can always be achieved by adding a Laurent
polynomial times the identity to $V_{\ul p}$, which does not affect the zero-curvature
equation). Hence the ansatz \eqref{ALansatz} can be refined as follows (it is more convenient in the following to re-label 
$h_{p_+,+}=h_{p_- -1,-}$ and $k_{p_+,+}=g_{p_-,-}$, and hence, 
$g_{p_-,-}=g_{p_+,+}$),   
\begin{align}
F_{\ul p}(z) &= \sum_{\ell=1}^{p_-} f_{p_- -\ell,-} z^{-\ell} 
+ \sum_{\ell=0}^{p_+ -1} f_{p_+ -1-\ell,+}z^\ell,  
\label{ALF_p} \\ 
G_{\ul p}(z) &= \sum_{\ell=1}^{p_-} g_{p_- -\ell,-}z^{-\ell}  
+ \sum_{\ell=0}^{p_+} g_{p_+ -\ell,+}z^\ell,  
 \label{ALG_p}  \\ 
H_{\ul p}(z) &= \sum_{\ell=0}^{p_- -1} h_{p_- -1-\ell,-}z^{-\ell}  
+ \sum_{\ell=1}^{p_+} h_{p_+ -\ell,+}z^\ell, 
 \label{ALH_p}  \\
K_{\ul p}(z) & = G_{\ul p}(z) \, \text{ since } \, g_{p_-,-}=g_{p_+,+}.   \label{ALK_p}
\end{align}
In particular, \eqref{ALK_p} renders $V_{\ul p}$ in \eqref{AL_v} traceless in the stationary context. We emphasize, however, that equation \eqref{ALK_p} ceases to be valid in the 
time-dependent context: In the latter case \eqref{ALK_p} needs to be replaced by
\begin{equation}
K_{\ul p}(z) = \sum_{\ell=0}^{p_-} g_{p_- -\ell,-}z^{-\ell}  
+  \sum_{\ell=1}^{p_+} g_{p_+ -\ell,+}z^\ell 
= G_{\ul p}(z)+g_{p_-,-}-g_{p_+,+}.   \label{ALK_pt}
\end{equation}

Plugging the refined ansatz \eqref{ALF_p}--\eqref{ALK_p} into the zero-curvature equation \eqref{ALstatzc} and comparing coefficients then yields the following result.

\begin{lemma} \lb{ALlreczc}
Suppose that $U$ and $V_{\ul p}$ satisfy the zero-curvature equation \eqref{ALstatzc}. Then the coefficients $\{f_{\ell,\pm}\}_{\ell=0,\dots,p_{\pm}-1}$, 
$\{g_{\ell,\pm}\}_{\ell=0,\dots,p_{\pm}}$, and $\{h_{\ell,\pm}\}_{\ell=0,\dots,p_{\pm}-1}$ of $F_{\ul p}$, $G_{\ul p}$, $H_{\ul p}$, and $K_{\ul p}$ in 
\eqref{ALF_p}--\eqref{ALK_p} satisfy the following relations
\begin{align} \label{AL0+zc}
g_{0,+} &= \tfrac12 c_{0,+}, \quad f_{0,+} = - c_{0,+}\alpha^+, 
\quad h_{0,+} = c_{0,+}\beta, \\ \label{ALg_l+zc}
g_{\ell+1,+} - g_{\ell+1,+}^- &= \alpha h_{\ell,+}^- + \beta f_{\ell,+}, 
\quad 0 \le \ell \le p_+ -1,\\ \label{ALf_l+zc}
f_{\ell+1,+}^- &= f_{\ell,+} - \alpha (g_{\ell+1,+} + g_{\ell+1,+}^-), 
\quad 0 \le \ell \le p_+ -2, \\  \label{ALh_l+zc}
h_{\ell+1,+} &= h_{\ell,+}^- + \beta (g_{\ell+1,+} 
+ g_{\ell+1,+}^-), \quad 0 \le \ell \le p_+ -2,  
\end{align}
and
\begin{align} \label{AL0-zc}
g_{0,-} &= \tfrac12 c_{0,-}, \quad f_{0,-} = c_{0,-}\alpha, 
\quad h_{0,-} = - c_{0,-}\beta^+, \\ \label{ALg_l-zc}
g_{\ell+1,-} - g_{\ell+1,-}^- &= \alpha h_{\ell,-} + \beta f_{\ell,-}^-, 
\quad 0 \le \ell \le p_- -1,\\ \label{ALf_l-zc}
f_{\ell+1,-} &= f_{\ell,-}^- + \alpha (g_{\ell+1,-} + g_{\ell+1,-}^-), 
\quad 0 \le \ell \le p_- -2, \\ \label{ALh_l-zc}
h_{\ell+1,-}^- &= h_{\ell,-} - \beta (g_{\ell+1,-} + g_{\ell+1,-}^-), 
\quad 0 \le \ell \le p_- -2.
\end{align}
Here $c_{0,\pm}\in\bbC$ are given constants. In addition, \eqref{ALstatzc} reads
\begin{align} \label{ALzc}
 0&=U V_{\ul p} - V_{\ul p}^+ U   \no \\
  &=i \begin{pmatrix}0& \begin{matrix} -\alpha(g_{p_+,+} + g_{p_-,-}^-) \\
+ f_{p_+ -1,+} - f_{p_- -1,-}^- \end{matrix} \\[3mm]  
\begin{matrix} z(\beta(g_{p_+,+}^- + g_{p_-,-}) \\
- h_{p_- -1,-} + h_{p_+ -1,+}^-) \end{matrix} &0 \end{pmatrix}.  
\end{align}
\end{lemma}

Given Lemma \ref{ALlreczc}, we now introduce the sequences $\{f_{\ell,\pm}\}_{\ell\in \bbN_0}$, $\{g_{\ell,\pm}\}_{\ell\in \bbN_0}$, and $\{h_{\ell,\pm}\}_{\ell\in \bbN_0}$ 
recursively by
\begin{align} \label{AL0+}
g_{0,+} &= \tfrac12 c_{0,+}, \quad f_{0,+} = - c_{0,+}\alpha^+, 
\quad h_{0,+} = c_{0,+}\beta, \\ \label{ALg_l+}
g_{\ell+1,+} - g_{\ell+1,+}^- &= \alpha h_{\ell,+}^- + \beta f_{\ell,+}, \quad \ell \in\bbN_0,   
\\ \label{ALf_l+}
f_{\ell+1,+}^- &= f_{\ell,+} - \alpha (g_{\ell+1,+} + g_{\ell+1,+}^-), \quad \ell \in\bbN_0, 
\\  \label{ALh_l+}
h_{\ell+1,+} &= h_{\ell,+}^- + \beta (g_{\ell+1,+} + g_{\ell+1,+}^-), \quad \ell \in\bbN_0,  
\end{align}
and
\begin{align} \label{AL0-}
g_{0,-} &= \tfrac12 c_{0,-}, \quad f_{0,-} = c_{0,-}\alpha, 
\quad h_{0,-} = - c_{0,-}\beta^+, \\ \label{ALg_l-}
g_{\ell+1,-} - g_{\ell+1,-}^- &= \alpha h_{\ell,-} + \beta f_{\ell,-}^-, \quad  \ell \in\bbN_0,  
\\ \label{ALf_l-}
f_{\ell+1,-} &= f_{\ell,-}^- + \alpha (g_{\ell+1,-} + g_{\ell+1,-}^-), \quad  \ell \in\bbN_0, 
\\ \label{ALh_l-}
h_{\ell+1,-}^- &= h_{\ell,-} - \beta (g_{\ell+1,-} + g_{\ell+1,-}^-), \quad  \ell \in\bbN_0.
\end{align}
For later use we also introduce
\begin{equation}\lb{ALminus}
f_{-1,\pm}= h_{-1,\pm}=0.         
\end{equation}

\begin{remark}\lb{rAL2.2}
The sequences $\{f_{\ell,+}\}_{\ell\in \bbN_0}$, 
$\{g_{\ell,+}\}_{\ell\in \bbN_0}$, and
$\{h_{\ell,+}\}_{\ell\in \bbN_0}$ can be computed recursively as follows: 
Assume that $f_{\ell,+}$,
$g_{\ell,+}$, and $h_{\ell,+}$ are known.  Equation \eqref{ALg_l+} is a 
first-order difference equation in $g_{\ell+1,+}$ that can be solved directly
and yields a local lattice function that is determined up to a new constant denoted
by $c_{\ell+1,+}\in\bbC$. Relations \eqref{ALf_l+} and \eqref{ALh_l+}
then determine $f_{\ell+1,+}$ and $h_{\ell+1,+}$, etc.  The sequences 
$\{f_{\ell,-}\}_{\ell\in \bbN_0}$, $\{g_{\ell,-}\}_{\ell\in \bbN_0}$, and 
$\{h_{\ell,-}\}_{\ell\in \bbN_0}$ are determined similarly.
\end{remark}

Upon setting 
\begin{equation}
\gamma = 1 - \alpha \beta, \lb{ALgamma}
\end{equation}
one explicitly obtains 
\begin{align}
\begin{split}
f_{0,+} &= c_{0,+}(-\alpha^+), \\
f_{1,+} &= c_{0,+}\big(- \gamma^+ \alpha^{++} + (\alpha^+)^2 \beta\big) 
+ c_{1,+} (-\alpha^+), \\
g_{0,+} &= \tfrac{1}{2}c_{0,+},  \\
 g_{1,+} &= c_{0,+}(-\alpha^+ \beta) + \tfrac{1}{2}c_{1,+}, \\
g_{2,+}  &= c_{0,+}\big((\alpha^+ \beta)^2- \gamma^+ \alpha^{++} \beta - \gamma \alpha^+ \beta^- \big) + c_{1,+} (- \alpha^+ \beta) + \tfrac{1}{2}c_{2,+}, \\
h_{0,+} &= c_{0,+}\beta, \\
 h_{1,+} &= c_{0,+}\big(\gamma \beta^- - \alpha^+ \beta^2\big) 
+ c_{1,+} \beta, \\
f_{0,-} &= c_{0,-}\alpha, \\ 
f_{1,-} &= c_{0,-}\big(\gamma \alpha^- - \alpha^2 \beta^+\big) + c_{1,-} \alpha, \\
g_{0,-} &= \tfrac{1}{2}c_{0,-}, \\ 
g_{1,-} &= c_{0,-}(-\alpha \beta^+) + \tfrac{1}{2}c_{1,-}, \\
g_{2,-} &= c_{0,-}\big((\alpha \beta^+)^2- \gamma^+ \alpha \beta^{++} - \gamma \alpha^- \beta^+ \big) + c_{1,-} (- \alpha \beta^+) + \tfrac{1}{2}c_{2,-}, \\
h_{0,-} &= c_{0,-}(-\beta^+), \\ 
h_{1,-} &= c_{0,-}\big(- \gamma^+ \beta^{++} 
+ \alpha (\beta^+)^2 \big) + c_{1,-} (- \beta^+), \, \text{ etc.}
\end{split}
\end{align}
Here $\{c_{\ell,\pm}\}_{\ell \in \bbN}$ denote summation constants
which naturally arise when solving the difference equations for 
$g_{\ell, \pm}$ in \eqref{ALg_l+}, \eqref{ALg_l-}.  

In particular, by \eqref{ALzc}, the stationary zero-curvature relation \eqref{ALstatzc}, $0=U V_{\ul p} - V_{\ul p}^+ U$, is equivalent to 
\begin{align}
 -\alpha(g_{p_+,+} + g_{p_-,-}^-) + f_{p_+ -1,+} - f_{p_- -1,-}^-&=0,  \lb{AL2.50}\\ 
 \beta(g_{p_+,+}^- + g_{p_-,-}) + h_{p_+ -1,+}^- - h_{p_- -1,-} &=0.  \lb{AL2.51}
\end{align}
Thus, varying $p_\pm \in \bbN_0$,  equations \eqref{AL2.50} and \eqref{AL2.51} give rise to the stationary Ablowitz--Ladik (AL) hierarchy which we introduce as follows
\begin{align}\lb{ALstat}
\begin{split}
& \sAL_{\ul p}(\alpha, \beta) = \begin{pmatrix}- \alpha(g_{p_+,+} + g_{p_-,-}^-) 
+ f_{p_+ -1,+} - f_{p_- -1,-}^-\\  
\beta(g_{p_+,+}^- + g_{p_-,-}) + h_{p_+ -1,+}^- - h_{p_- -1,-}  \end{pmatrix}=0,  \\
& \hspace*{6.9cm}  \ul p=(p_-,p_+)\in\bbN_0^2. 
\end{split}
\end{align}

Explicitly (recalling $\gamma=1-\alpha\beta$ and taking $p_-=p_+$ for simplicity), 
\begin{align} \no
\sAL_{(0,0)} (\alpha, \beta) &=  \begin{pmatrix}  -c_{(0,0)} \alpha \\ 
c_{(0,0)}\beta\end{pmatrix} =0,\\ \no
\sAL_{(1,1)} (\alpha, \beta) &=  \begin{pmatrix} -\gamma (c_{0,-}\alpha^- + c_{0,+}\alpha^+) 
- c_{(1,1)} \alpha \\
 \gamma (c_{0,+}\beta^- + c_{0,-}\beta^+) +
c_{(1,1)} \beta\end{pmatrix}=0,\\ \no
\sAL_{(2,2)} (\alpha, \beta) &=  \begin{pmatrix}\begin{matrix}
-\gamma \big(c_{0,+}\alpha^{++} \gamma^+ + c_{0,-}\alpha^{--} \gamma^-
- \alpha (c_{0,+}\alpha^+\beta^- + c_{0,-}\alpha^-\beta^+)\\
- \beta (c_{0,-}(\alpha^-)^2 + c_{0,+}(\alpha^+)^2)\big)\end{matrix}\\[3mm] 
\begin{matrix}
 \gamma \big(c_{0,-}\beta^{++} \gamma^+ + c_{0,+}\beta^{--} \gamma^-
- \beta (c_{0,+}\alpha^+\beta^- + c_{0,-}\alpha^-\beta^+)\\
- \alpha (c_{0,+}(\beta^-)^2 + c_{0,-}(\beta^+)^2)\big)\end{matrix}\end{pmatrix}
 \\ & \quad+ \begin{pmatrix}
-\gamma (c_{1,-} \alpha^- + c_{1,+} \alpha^+) - c_{(2,2)} \alpha\\
 \gamma (c_{1,+} \beta^- + c_{1,-} \beta^+) + c_{(2,2)} \beta\end{pmatrix}
=0,  \, \text{ etc.,}
\end{align}
represent the first few equations of the stationary Ablowitz--Ladik hierarchy. 
Here we introduced  
\begin{equation}
c_{\ul p} = (c_{p,-} + c_{p,+})/2, \quad p_\pm\in\bbN_0.     \lb{ALdefcp}
\end{equation}
By definition, the set of solutions of \eqref{ALstat}, with $p_\pm$ ranging in $\bbN_0$ and $c_{\ell,\pm}\in\bbC$, $\ell\in\bbN_0$, represents the class of algebro-geometric Ablowitz--Ladik solutions. 

In the special case $\ul p=(1,1)$, $c_{0,\pm}=1$, and $c_{(1,1)}=-2$, 
one obtains the stationary version of the Ablowitz--Ladik system \eqref{AL1.12}
\begin{equation}
\begin{pmatrix} -\gamma (\alpha^- + \alpha^+) + 2 \alpha \\
 \gamma (\beta^- + \beta^+) - 2 \beta \end{pmatrix}=0. 
\end{equation}

Subsequently, it will also be useful to work with the corresponding homogeneous coefficients $\hat f_{\ell, \pm}$,
$\hat g_{\ell, \pm}$, and $\hat h_{\ell, \pm}$, defined by the vanishing of all summation constants $c_{k,\pm}$ for $k=1,\dots,\ell$, and choosing $c_{0,\pm}=1$,
\begin{align}
& \hat f_{0,+}=-\alpha^+, \quad \hat f_{0,-}=\alpha, \quad 
 \hat f_{\ell,\pm}=f_{\ell,\pm}|_{c_{0,\pm}=1, \, c_{j,\pm}=0, j=1,\dots,\ell},  \quad \ell\in\bbN, 
 \lb{AL2.04a} \\
& \hat g_{0,\pm}=\tfrac12, \quad 
\hat g_{\ell,\pm}=g_{\ell,\pm}|_{c_{0,\pm}=1, \, c_{j,\pm}=0, j=1,\dots,\ell}, 
\quad \ell\in\bbN,  \lb{AL2.04b} \\
& \hat h_{0,+}=\beta, \quad \hat h_{0,-}=-\beta^+,  \quad 
\hat h_{\ell,\pm}=h_{\ell,\pm}|_{c_{0,\pm}=1, \, c_{j,\pm}=0, j=1,\dots,\ell}, 
\quad \ell\in\bbN.  \lb{AL2.04c}
\end{align}
By induction one infers that
\begin{equation} \label{ALhat f}
f_{\ell, \pm} = \sum_{k=0}^\ell c_{\ell-k, \pm} \hat f_{k, \pm}, \quad
g_{\ell, \pm} = \sum_{k=0}^\ell c_{\ell-k, \pm} \hat g_{k, \pm}, \quad 
h_{\ell, \pm} = \sum_{k=0}^\ell c_{\ell-k, \pm} \hat h_{k, \pm}.  
\end{equation} 
In a slight abuse of notation we will occasionally stress the dependence of $f_{\ell,\pm}$, 
$g_{\ell,\pm}$, and $h_{\ell,\pm}$ on $\alpha, \beta$ by writing  
$f_{\ell,\pm}(\alpha,\beta)$, $g_{\ell,\pm}(\alpha,\beta)$, and 
$h_{\ell,\pm}(\alpha,\beta)$. 

\begin{remark}  \lb{rAL2.3} 
Using the nonlinear recursion relations \eqref{AL2.206aX}--\eqref{ALhmcheckX} recorded in 
Theorem~\ref{tALB.2A}, one infers inductively that all homogeneous elements 
$\hat f_{\ell,\pm}$, $\hat g_{\ell,\pm}$, and $\hat h_{\ell,\pm}$, $\ell\in\bbN_0$, are polynomials in $\alpha, \beta$, and some of their shifts. $($Alternatively, one can prove directly by induction that the nonlinear recursion relations  
\eqref{AL2.206aX}--\eqref{ALhmcheckX} are equivalent to that in 
\eqref{AL0+}--\eqref{ALh_l-} with all summation constants put equal to zero, 
$c_{\ell,\pm}=0$, $\ell\in\bbN$.$)$ 
\end{remark}

\begin{remark}\lb{rAL2.4}
As an efficient tool to later distinguish between nonhomogeneous and homogeneous 
quantities $f_{\ell,\pm}$, $g_{\ell,\pm}$, $h_{\ell,\pm}$, and $\hat f_{\ell,\pm}$, 
$\hat g_{\ell,\pm}$, $\hat h_{\ell,\pm}$, respectively, we now introduce the notion of degree as follows. Denote  
\begin{align}
f^{(r)}=S^{(r)}f, \quad f=\{f(n)\}_{n\in\bbZ}\in\bbC^{\bbZ}, \quad
   S^{(r)}&=\begin{cases}(S^+)^r, &\text{$r\ge 0$},\\
(S^-)^{-r}, &\text{$r< 0$},\end{cases} \quad
r\in \bbZ,    \lb{AL2.1AA}
\end{align}
and define
\begin{equation}
\deg \big(\alpha^{(r)}\big)=r, \quad \deg \big(\beta^{(r)}\big)=-r, \quad r\in\bbZ.  
\lb{3.2.15aa}
\end{equation}
This then results in 
\begin{align}
\begin{split}
\deg\big(\hat f_{\ell,+}^{(r)}\big)&= \ell+1+r, \quad \deg\big(\hat f_{\ell,-}^{(r)}\big)
= -\ell+r, \quad \deg\big(\hat g_{\ell, \pm}^{(r)}\big)= \pm\ell, \\
\deg\big(\hat h_{\ell,+}^{(r)}\big)&= \ell-r, \quad \deg\big(\hat h_{\ell,-}^{(r)}\big)
= -\ell-1-r, \quad \ell\in\bbN_0, \; r\in\bbZ, 
\end{split}      \lb{AL2.1AAa} 
\end{align}
using induction in the linear recursion relations \eqref{AL0+}--\eqref{ALh_l-}. 
\end{remark}

In accordance with our notation introduced in \eqref{AL2.04a}--\eqref{AL2.04c}, 
the corresponding homogeneous stationary Ablowitz--Ladik equations are defined by  
\begin{equation}
\shAL_{\ul p} (\alpha, \beta) 
= \sAL_{\ul p} (\alpha, \beta)\big|_{c_{0,\pm}=1, \, c_{\ell,\pm}=0, \, \ell=1,\dots,p_{\pm}}, 
\quad \ul p=(p_-,p_+)\in\bbN_0^2.   \lb{ALstathom}
\end{equation}

We also note the following useful result.

\begin{lemma} \lb{lAL2.5}
The coefficients $f_{\ell,\pm}$, $g_{\ell,\pm}$, and $h_{\ell,\pm}$ satisfy the relations
\begin{align} \lb{ALadd}
\begin{split}
g_{\ell,+} - g_{\ell,+}^- &= \alpha h_{\ell,+} + \beta f_{\ell,+}^-, \quad \ell\in\bbN_0, \\
g_{\ell,-} - g_{\ell,-}^- &= \alpha h_{\ell,-}^- + \beta f_{\ell,-}, \quad \ell\in\bbN_0.   
\end{split}
\end{align}
Moreover, we record the following symmetries, 
\begin{equation}\lb{ALsym}
\hat f_{\ell,\pm}(c_{0,\pm},\alpha,\beta)=\hat h_{\ell,\mp}(c_{0,\mp},\beta, \alpha), \quad 
\hat g_{\ell,\pm}(c_{0,\pm},\alpha,\beta)=\hat g_{\ell,\mp}(c_{0,\mp},\beta, \alpha), \quad \ell\in \bbN_0.
\end{equation}
\end{lemma}
\begin{proof}
The relations \eqref{ALadd}  are derived as follows:
\begin{align}
\alpha h_{\ell+1,+} + \beta f_{\ell+1,+}^- &= 
\alpha h_{\ell,+}^- + \alpha \beta (g_{\ell+1,+} + g_{\ell+1,+}^-) 
+ \beta f_{\ell,+} - \alpha \beta (g_{\ell+1,+} + g_{\ell+1,+}^-)  \no \\
&= \alpha h_{\ell,+}^- + \beta f_{\ell,+} = g_{\ell+1,+} - g_{\ell+1,+}^-,
\end{align}
and
\begin{align}
\alpha h_{\ell+1,-}^- + \beta f_{\ell+1,-} &= 
\alpha h_{\ell,-} - \alpha \beta (g_{\ell+1,-} + g_{\ell+1,-}^-)
+ \beta f_{\ell,-}^- + \alpha \beta (g_{\ell+1,-} + g_{\ell+1,-}^-)  \no \\
&= \alpha h_{\ell,-} + \beta f_{\ell,-}^- = g_{\ell+1,-} + g_{\ell+1,-}^-.
\end{align}
The statement \eqref{ALsym} follows by showing that $\hat h_{\ell,\mp}(\beta, \alpha)$ and $\hat g_{\ell,\mp}(\beta, \alpha)$ satisfy the same recursion relations as those of 
$\hat f_{\ell,\pm}(\alpha,\beta)$ and $\hat g_{\ell,\pm}(\alpha,\beta)$, respectively. That the recursion constants are the same, follows from the observation that the corresponding coefficients have the proper degree. 
\end{proof}

Next we turn to the Laurent polynomials $F_{\ul p}$, $G_{\ul p}$, $H_{\ul p}$, and $K_{\ul p}$ defined in
\eqref{ALF_p}--\eqref{ALH_p} and \eqref{ALK_pt}. Explicitly, one obtains
\begin{align}
\begin{split}
F_{(0,0)}&=0, \\
F_{(1,1)}&=c_{0,-}\alpha z^{-1} + c_{0,+}(-\alpha^+),   \\
F_{(2,2)}&=c_{0,-}\alpha z^{-2}+\big(c_{0,-}\big(\gamma \alpha^- - \alpha^2 \beta^+\big) 
+ c_{1,-} \alpha \big)z^{-1} \\
&\quad + c_{0,+}\big(- \gamma^+ \alpha^{++} + (\alpha^+)^2 \beta\big) 
+ c_{1,+} (-\alpha^+) + c_{0,+}(-\alpha^+) z ,  \\
G_{(0,0)}&=\tfrac{1}{2} c_{0,+}, \\
G_{(1,1)}&= \tfrac{1}{2}c_{0,-} z^{-1} + c_{0,+}(-\alpha^+ \beta) + \tfrac{1}{2}c_{1,+} 
+ \tfrac{1}{2} c_{0,+} z,  \\
G_{(2,2)}&= \tfrac{1}{2}c_{0,-} z^{-2} + \big(c_{0,-}(-\alpha \beta^+) + \tfrac{1}{2}c_{1,-}\big)z^{-1}  \\
&\quad + c_{0,+}\big((\alpha^+ \beta)^2- \gamma^+ \alpha^{++} \beta - \gamma \alpha^+ \beta^- \big) +  c_{1,+} (- \alpha^+ \beta) + \tfrac{1}{2}c_{2,+}  \\
&\quad + \big(c_{0,+}(-\alpha^+ \beta) + \tfrac{1}{2} c_{1,+}\big)z + \tfrac{1}{2} c_{0,+} z^2,   
\\
H_{(0,0)}&=0, \\
H_{(1,1)}&= c_{0,-}(-\beta^+) + c_{0,+}\beta z,  \\
H_{(2,2)}&= c_{0,-}(-\beta^+) z^{-1}  + c_{0,-}\big(- \gamma^+ \beta^{++} 
+ \alpha (\beta^+)^2 \big) + c_{1,-}(- \beta^+)  \\
&\quad + \big(c_{0,+}(\gamma \beta^- - \alpha^+ \beta^2) + c_{1,+}\beta \big)z
+ c_{0,+}\beta z^2,  \\
K_{(0,0)}&=\tfrac{1}{2} c_{0,-},  \\
K_{(1,1)}&= \tfrac{1}{2} c_{0,-} z^{-1} + c_{0,-}(-\alpha \beta^+) + \tfrac{1}{2}c_{1,-} 
+\tfrac{1}{2} c_{0,+} z ,   \\
K_{(2,2)}&= \tfrac{1}{2} c_{0,-} z^{-2} +\big(c_{0,-}(-\alpha \beta^+) + \tfrac{1}{2}c_{1,-}\big)z^{-1}  \\
&\quad + c_{0,-}\big((\alpha \beta^+)^2- \gamma^+ \alpha \beta^{++} - \gamma \alpha^- \beta^+\big) +  c_{1,-}(- \alpha \beta^+) + \tfrac{1}{2}c_{2,-}   \\
&\quad+ \big(c_{0,+}(-\alpha^+ \beta) + \tfrac{1}{2}c_{1,+}\big)z
+ \tfrac{1}{2} c_{0,+} z^2, \, \text{ etc.}
\end{split}
\end{align}


The corresponding homogeneous quantities are defined by ($\ell\in\bbN_0$)
\begin{align} 
\begin{split} 
\hatt F_{0,\mp} (z) & = 0, \quad 
\hatt F_{\ell,-}(z) = \sum_{k=1}^\ell \hat f_{\ell-k,-} z^{-k}, 
\quad  \hatt F_{\ell,+}(z) = \sum_{k=0}^{\ell-1} \hat f_{\ell-1-k,+}z^k,   \\ 
\hatt G_{0,-} (z) &  = 0,   \quad 
\hatt G_{\ell,-}(z) = \sum_{k=1}^\ell \hat g_{\ell-k,-}z^{-k}, \\ 
\hatt G_{0,+} (z) &= \f{1}{2},   \quad 
\hatt G_{\ell,+}(z) = \sum_{k=0}^\ell \hat g_{\ell-k,+}z^k,   \\ 
\hatt H_{0,\mp} (z) &  = 0, \quad 
\hatt H_{\ell,-}(z) = \sum_{k=0}^{\ell-1} \hat h_{\ell-1-k,-} z^{-k}, \quad  
\hatt H_{\ell,+}(z) = \sum_{k=1}^\ell \hat h_{\ell-k,+}z^k,   \label{ALhat_F_p} \\
\hatt K_{0,-} (z) & = \f{1}{2},  \quad    
\hatt K_{\ell,-}(z) = \sum_{k=0}^\ell \hat g_{\ell-k,-}z^{-k}  
= \hatt G_{\ell,-} (z)+\hat g_{\ell,-},  \\
\hatt K_{0,+} (z) &= 0, \quad \hatt K_{\ell,+}(z) = \sum_{k=1}^\ell \hat g_{\ell-k,+}z^k 
= \hatt G_{\ell,+} (z) - \hat g_{\ell,+}. 
\end{split} 
\end{align}

Similarly, with $F_{\ell_+,+}$, $G_{\ell_+,+}$, $H_{\ell_+,+}$, and $K_{\ell_+,+}$ denoting the polynomial parts of $F_{\ul \ell}$, $G_{\ul \ell}$, $H_{\ul \ell}$, and $K_{\ul \ell}$, respectively, and $F_{\ell_-,-}$, $G_{\ell_-,-}$, $H_{\ell_-,-}$, and $K_{\ell_-,-}$ denoting the Laurent parts of $F_{\ul \ell}$, $G_{\ul \ell}$, $H_{\ul \ell}$, and $K_{\ul \ell}$, 
$\ul \ell =(\ell_-,\ell_+)\in \bbN_0$, such that 
\begin{align}
\begin{split}
F_{\ul \ell}(z)&=F_{\ell_-,-}(z) + F_{\ell_+,+}(z), \quad \;\,
G_{\ul \ell}(z)=G_{\ell_-,-}(z) + G_{\ell_+,+}(z),   \lb{AL2.46} \\
H_{\ul \ell}(z)&=H_{\ell_-,-}(z) + H_{\ell_+,+}(z), \quad 
K_{\ul \ell}(z)=K_{\ell_-,-}(z) + K_{\ell_+,+}(z), \\
\end{split}
\end{align}
one finds that
\begin{align} \label{ALhatFpm}
\begin{split}
F_{\ell_\pm, \pm} &= \sum_{k=1}^{\ell_{\pm}} c_{\ell_{\pm} -k, \pm} \hatt F _{k, \pm}, \quad 
H_{\ell_\pm,\pm} = \sum_{k=1}^{\ell_\pm} c_{\ell_\pm -k,\pm} \hatt H_{k,\pm},  \\
G_{\ell_-,-} & = \sum_{k=1}^{\ell_-} c_{\ell_- -k,-} \hatt G_{k,-}, \quad 
G_{\ell_+,+} = \sum_{k=0}^{\ell_+} c_{\ell_+ -k,+} \hatt G_{k,+}, \\
K_{\ell_-,-} & = \sum_{k=0}^{\ell_-} c_{\ell_- -k,-} \hatt K_{k,-}, \quad 
K_{\ell_+,+} = \sum_{k=1}^{\ell_+} c_{\ell_+ -k,+} \hatt K_{k,+}.
\end{split}
\end{align} 
In addition, one immediately obtains the following relations from \eqref{ALsym}:

\begin{lemma} \label{lAL2.7} 
Let $\ell \in\bbN_0$. Then,
\begin{align}
\hatt F_{\ell,\pm}(\alpha,\beta,z,n) &= \hatt H_{\ell,\mp}(\beta,\alpha,z^{-1},n),\\
\hatt H_{\ell,\pm}(\alpha,\beta,z,n) &= \hatt F_{\ell,\mp}(\beta,\alpha,z^{-1},n),\\
\hatt G_{\ell,\pm}(\alpha,\beta,z,n) &= \hatt G_{\ell,\mp}(\beta,\alpha,z^{-1},n), \\
\hatt K_{\ell,\pm}(\alpha,\beta,z,n) &= \hatt K_{\ell,\mp}(\beta,\alpha,z^{-1},n).
\end{align}
\end{lemma}

Returning to the stationary Ablowitz--Ladik hierarchy, we will frequently assume in the following that $\alpha, \beta$ satisfy the $\ul p$th stationary Ablowitz--Ladik system  
$\sAL_{\ul p}(\alpha, \beta) = 0$, supposing a particular choice of summation constants $c_{\ell,\pm}\in\bbC$, 
$\ell=0,\dots,p_{\pm}$, $p_\pm\in\bbN_0$, has been made.

\begin{remark} \lb{rAL2.9}
$(i)$ The particular choice $c_{0,+}=c_{0,-}=1$ in \eqref{ALstat} yields the stationary Ablowitz--Ladik equation. Scaling $c_{0,\pm}$ with the same constant then amounts to scaling $V_{\ul p}$ with this constant which drops out in the stationary zero-curvature equation 
\eqref{ALstatzc}. \\ 
$(ii)$ Different ratios between $c_{0,+}$  and $c_{0,-}$ will lead to different stationary hierarchies. In particular, the choice $c_{0,+}=2$, $c_{0,-}=\dots=c_{p_- -1,-}=0$, 
$c_{p_-,-}\neq0$, yields the stationary
Baxter--Szeg\H o hierarchy considered in detail in \cite{GeronimoGesztesyHolden:2004}. However, in this case some parts from the recursion relation for the negative
coefficients still remain. In fact, \eqref{ALh_l-} reduces
to $g_{p_-,-}-g_{p_-,-}^- = \alpha h_{p_- -1,-}$, $h_{p_- -1,-}=0$ and thus
requires $g_{p_-,-}$ to be a constant in \eqref{ALstat} and \eqref{AL_p}. Moreover, 
$f_{p_- -1,-}=0$ in \eqref{ALstat} in this case. \\
$(iii)$ Finally, by Lemma \ref{lAL2.7}, the choice
$c_{0,+}=\dots=c_{p_+ -1,+}=0$, $c_{p_+,+}\neq0$, $c_{0,-}=2$ again yields the 
Baxter--Szeg\H o hierarchy, but with $\alpha$ and $\beta$ interchanged. 
\end{remark}

Next, taking into account \eqref{ALK_p}, one infers that 
the expression $R_{\ul p}$, defined as 
\begin{equation} \label{ALR}
R_{\ul p} = G_{\ul p}^2 - F_{\ul p} H_{\ul p}, 
\end{equation}
is a lattice constant, that is, $R_{\ul p} - R_{\ul p}^- = 0$, since taking determinants 
in the stationary zero-curvature equation \eqref{ALstatzc} immediately yields  
\begin{equation}
\gamma \big(-(G_{\ul p}^-)^2 + F_{\ul p}^- H_{\ul p}^- + G_{\ul p}^2 - F_{\ul p} H_{\ul p}\big) z = 0.
\end{equation}
Hence, $R_{\ul p}(z)$  only depends on $z$, and assuming in addition to 
\eqref{AL2.01} that 
\begin{equation}
c_{0,\pm} \in \bbC\setminus \{0\},  \quad  
\ul p=(p_-,p_+)\in\bbN_0^2\setminus\{(0,0)\},    \lb{ALc0}
\end{equation}
one may write $R_{\ul p}$ as\footnote{We use the convention that a product is to be interpreted equal to $1$ whenever the upper limit of the product is strictly less than its lower limit.}  
\begin{equation}  \label{ALE_m}
R_{\ul p}(z) = \bigg(\f{c_{0,+}}{2z^{p_-}}\bigg)^2
\prod_{m=0}^{2p+1}(z-E_m), \quad
\{E_m\}_{m=0}^{2p+1} \subset \bbC \setminus\{ 0\},   \;
p=p_- + p_+ -1 \in\bbN_0.
\end{equation}
Moreover, \eqref{ALR} also implies
\begin{equation}
\lim_{z \rightarrow 0} 4z^{2p_-} R_{\ul p}(z) = c_{0,+}^2 \prod_{m=0}^{2p+1}(-E_m)
= c_{0,-}^2,
\end{equation}
and hence,
\begin{equation} \label{ALprod E_m}
\prod_{m=0}^{2p+1}E_m = \frac{c_{0,-}^2}{c_{0,+}^2}.
\end{equation}

Relation \eqref{ALR} allows one to introduce a hyperelliptic curve 
$\calK_p$ of (arithmetic) genus $p=p_- + p_+ -1$ (possibly with a singular affine part), where
\begin{equation} \label{ALKp}
\calK_p \colon\calF_p (z,y) = y^2 - 4c_{0,+}^{-2}z^{2p_-}R_{\ul p}(z) 
= y^2 - \prod_{m=0}^{2p+1}(z-E_m) = 0, 
\quad p=p_- + p_+ -1. 
\end{equation}

\begin{remark}  \lb{lAL2.10}
In the special case $p_-=p_+$ and $c_{\ell,+}=c_{\ell,-}$, $\ell=0,\dots,p_-$, the 
symmetries of Lemma \ref{lAL2.7} also hold for
$F_{\ul p}$, $G_{\ul p}$, and $H_{\ul p}$ and thus $R_{\ul p}(1/z)=R_{\ul p}(z)$ and hence the numbers $E_m$, $m=0,\dots,2p+1$, come in pairs $(E_k,1/E_k)$, 
$k=1,\dots,p+1$.  
\end{remark}

Equations \eqref{AL1,1}--\eqref{AL2,1} and \eqref{ALR} permit one to derive nonlinear difference
equations for $F_{\ul p}$, $G_{\ul p}$, and $H_{\ul p}$ separately.  One obtains 
\begin{align} \no
&\big((\alpha^+ + z \alpha)^2 F_{\ul p} - z(\alpha^+)^2 \gam F_{\ul p}^-\big)^2
-2 z \alpha^2 \gam^+ 
\big((\alpha^+ + z \alpha)^2 F_{\ul p} + z(\alpha^+)^2 \gam F_{\ul p}^-\big) F_{\ul p}^+
\\ \label{ALnldeF}
& + z^2 \alpha^4 (\gam^+)^2 (F_{\ul p}^+)^2
=4(\alpha\alpha^+)^2(\alpha^+ +\alpha z)^2 R_{\ul p}, \\[1mm] 
& (\alpha^+ + z \alpha) (\beta+ z \beta^+) (z + \alpha^+ \beta)
(1 + z \alpha \beta^+) G_{\ul p}^2\no \\ \no
& + z (\alpha^+ \gam G_{\ul p}^- + z \alpha \gam^+ G_{\ul p}^+)
(z  \beta^+ \gam G_{\ul p}^- + \beta \gam^+ G_{\ul p}^+)\\ \no
& - z \gam \big((\alpha^+ \beta + z^2 \alpha \beta^+) (2-\gam^+) + 2z (1-\gam^+) (2-\gam)\big) G_{\ul p}^- G_{\ul p}  \\ \no
& - z \gam^+ \big(2 z (1-\gam) (2-\gam^+) + (\alpha^+ \beta + z^2 \alpha \beta^+) 
(2-\gam)\big) G_{\ul p}^+ G_{\ul p}   \\ \label{ALnldeG}
&= (\alpha^+ \beta - z^2 \alpha \beta^+)^2 R_{\ul p},  \\[1mm]
& z^2 \big((\beta^+)^2 \gam H_{\ul p}^- - \beta^2 \gam^+ H_{\ul p}^+\big)^2
- 2 z (\beta+ z \beta^+)^2 \big((\beta^+)^2 \gam H_{\ul p}^- 
+ \beta^2 \gam^+ H_{\ul p}^+\big) H_{\ul p}    \no \\ \label{ALnldeH}
& + (\beta+ z \beta^+)^4 H_{\ul p}^2
=4z^2(\beta\beta^+)^2 (\beta+\beta^+ z)^2 R_{\ul p}.
\end{align}

Equations analogous to \eqref{ALnldeF}--\eqref{ALnldeH} can be used to derive
nonlinear recursion relations for the homogeneous coefficients
$\hat f_{\ell,\pm}$, $\hat g_{\ell,\pm}$, and $\hat h_{\ell,\pm}$ (i.e., the ones satisfying \eqref{AL2.04a}--\eqref{AL2.04c} in the case of vanishing summation constants) as proved in Theorem \ref{tALB.2A} in Appendix \ref{ALApp.high}. This then yields a proof 
that $\hat f_{\ell,\pm}$, $\hat g_{\ell,\pm}$, and $\hat h_{\ell,\pm}$ are polynomials in 
$\alpha$, $\beta$, and some of their shifts (cf.\ Remark \ref{rAL2.3}). In addition, as proven in Theorem \ref{tALB.2}, \eqref{ALnldeF} leads to an explicit determination of the summation constants $c_{1,\pm},\dots,c_{p_{\pm},\pm}$ in 
\eqref{ALstat} in terms of the zeros $E_0,\dots,E_{2p+1}$ of the associated
Laurent polynomial $R_{\ul p}$ in \eqref{ALE_m}. In fact, one can prove (cf.\ 
\eqref{ALBc})
\begin{equation}
c_{\ell,\pm}= c_{0,\pm} c_\ell\big(\ul E^{\pm 1}\big), \quad \ell=0,\dots,p_\pm, \lb{ALcell}
\end{equation}
where   
\begin{align}
&c_{0}\big(\ul E^{\pm 1}\big)=1,\no \\
&c_{k}\big(\ul E^{\pm 1}\big)   \label{ALc_ell}  \\
&=-\!\!\!\!\!\!\!\sum_{\substack{j_0,\dots,j_{2p+1}=0\\
   j_0+\cdots+j_{2p+1}=k}}^{k}\!\!
\f{(2j_0)!\cdots(2j_{2p+1})!}
{2^{2k} (j_0!)^2\cdots (j_{2p+1}!)^2 (2j_0-1)\cdots(2j_{2p+1}-1)}
E_0^{\pm j_0}\cdots E_{2p+1}^{\pm j_{2p+1}},  \no  \\
& \hspace*{10.95cm} k\in\bbN,   \no
\end{align}
are symmetric functions of $\ul E^{\pm 1}=(E_0^{\pm 1},\dots,E_{2p+1}^{\pm 1})$ 
introduced in \eqref{ALB2.26g} and \eqref{ALB2.26h}.  

\begin{remark} \lb{rAL2.12}
If $\alpha, \beta$ satisfy one of the stationary Ablowitz--Ladik equations in
\eqref{ALstat} for a particular value of $\ul p$, $\sAL_{\ul p}(\alpha,\beta)=0$, then they
satisfy infinitely many such equations of order higher than $\ul p$ for
certain  choices of summation constants $c_{\ell,\pm}$. This can be shown as in 
\cite[Remark I.1.5]{GesztesyHolden:2003}.
\end{remark}

\bigskip
Finally we turn to the time-dependent Ablowitz--Ladik hierarchy. For that purpose the coefficients $\alpha$ and $\beta$ are now considered as functions of both the lattice point and time. For each system in the hierarchy, that is, for each $\ul p$, we introduce a deformation (time) parameter $t_{\ul p}\in\bbR$ in $\alpha, \beta$, replacing $\alpha(n), \beta(n)$ by $\alpha(n,t_{\ul p}), \beta(n,t_{\ul p})$. Moreover, the definitions 
\eqref{AL2.03},  \eqref{AL_v}, and \eqref{ALF_p}--\eqref{ALH_p}  of $U, V_{\ul p}$, and $F_{\ul p}, G_{\ul p}, H_{\ul p}, K_{\ul p}$, respectively, still apply; however, equation \eqref{ALK_p} now needs to be replaced by \eqref{ALK_pt} in the time-dependent context.

Imposing the zero-curvature relation
\begin{equation}
U_{t_{\ul p}} + U V_{\ul p} - V_{\ul p}^+ U =0, \quad \ul p\in\bbN_0^2,
\end{equation}
then results in the equations
\begin{align}  
0 &=  U_{t_{\ul p}} + U V_{\ul p} - V_{\ul p}^+ U  \no \\
&=i \begin{pmatrix} z (G_{\ul p}^- - G_{\ul p}) + z \beta F_{\ul p} + \alpha H_{\ul p}^- & 
- i\alpha_{t_{\ul p}} + F_{\ul p} - z F_{\ul p}^- - \alpha (G_{\ul p} + K_{\ul p}^-) \\[1mm]
 -iz\beta_{t_{\ul p}}+ z \beta (G_{\ul p}^- + K_{\ul p}) - z H_{\ul p} + H_{\ul p}^- &
  -z \beta F_{\ul p}^- - \alpha H_{\ul p} + K_{\ul p} - K_{\ul p}^-  \end{pmatrix}  \no \\
 &= i\begin{pmatrix}0&\begin{matrix} -i\alpha_{t_{\ul p}} 
 - \alpha(g_{p_+,+} + g_{p_-,-}^-) \\
 + f_{p_+ -1,+} - f_{p_- -1,-}^-\end{matrix}\\[2mm]
\begin{matrix}  z\big(-i\beta_{t_{\ul p}}+ \beta(g_{p_+,+}^- + g_{p_-,-})\\
 - h_{p_- -1,-} + h_{p_+ -1,+}^-\big)\end{matrix} &0      \end{pmatrix}, \label{ALzc p}
\end{align}
or equivalently,
\begin{align}  \label{ALalphat}
\alpha_{t_{\ul p}} &= i \big(z F_{\ul p}^- + \alpha (G_{\ul p} + K_{\ul p}^-) - F_{\ul p}\big),    \\ \label{ALbetat}
\beta_{t_{\ul p}} &= - i \big(\beta (G_{\ul p}^- + K_{\ul p}) - H_{\ul p} 
+ z^{-1} H_{\ul p}^-\big), \\ \label{AL1,1r}
0 &= z (G_{\ul p}^- - G_{\ul p}) + z\beta F_{\ul p} + \alpha H_{\ul p}^-,   \\ \label{AL2,2r}
0 &= z \beta F_{\ul p}^- + \alpha H_{\ul p} + K_{\ul p}^- - K_{\ul p}.
\end{align}
Varying $\ul p \in \bbN_0^2$, the collection of evolution equations   
\begin{align}   \label{AL_p}
\begin{split}
& \AL_{\ul p} (\alpha, \beta) = \begin{pmatrix}-i\alpha_{t_{\ul p}} 
- \alpha(g_{p_+,+} + g_{p_-,-}^-) + f_{p_+ -1,+} - f_{p_- -1,-}^-\\
  -i\beta_{t_{\ul p}}+ \beta(g_{p_+,+}^- + g_{p_-,-}) 
  - h_{p_- -1,-} + h_{p_+ -1,+}^- \end{pmatrix}=0,  \\
& \hspace*{6.44cm} t_{\ul p}\in\bbR, \; \ul p=(p_-,p_+)\in\bbN_0^2,   
\end{split}
\end{align}
then defines the time-dependent Ablowitz--Ladik hierarchy. Explicitly, taking $p_-=p_+$ for simplicity, 
\begin{align} \no
& \AL_{(0,0)} (\alpha, \beta) =  \begin{pmatrix} -i \alpha_{t_{(0,0)}}- c_{(0,0)}\alpha \\ 
-i\beta_{t_{(0,0)}}+c_{(0,0)}\beta \end{pmatrix} 
=0,\\ \no
& \AL_{(1,1)} (\alpha, \beta) =  \begin{pmatrix}  
-i \alpha_{t_{(1,1)}}- \gamma (c_{0,-}\alpha^- + c_{0,+}\alpha^+) 
- c_{(1,1)} \alpha \\
-i\beta_{t_{(1,1)}}+ \gamma (c_{0,+}\beta^- + c_{0,-}\beta^+) +
c_{(1,1)} \beta\end{pmatrix}=0,\\  
& \AL_{(2,2)} (\alpha, \beta)  \\
&\quad =  \begin{pmatrix}\begin{matrix}-i \alpha_{t_{(2,2)}}-
\gamma \big(c_{0,+}\alpha^{++} \gamma^+ + c_{0,-}\alpha^{--} \gamma^-
- \alpha (c_{0,+}\alpha^+\beta^- + c_{0,-}\alpha^-\beta^+)\\
- \beta (c_{0,-}(\alpha^-)^2 + c_{0,+}(\alpha^+)^2)\big)\end{matrix}\\[3mm] 
\begin{matrix}-i\beta_{t_{(2,2)}}+
 \gamma \big(c_{0,-}\beta^{++} \gamma^+ + c_{0,+}\beta^{--} \gamma^-
- \beta (c_{0,+}\alpha^+\beta^- + c_{0,-}\alpha^-\beta^+)\\
- \alpha (c_{0,+}(\beta^-)^2 + c_{0,-}(\beta^+)^2)\big)\end{matrix}\end{pmatrix}  \no \\ 
 & \qquad+ \begin{pmatrix}
-\gamma (c_{1,-} \alpha^- + c_{1,+} \alpha^+) - c_{(2,2)} \alpha\\
 \gamma (c_{1,+} \beta^- + c_{1,-} \beta^+) + c_{(2,2)} \beta\end{pmatrix}
=0, \, \text{ etc.,} \no
\end{align}
represent the first few equations of the time-dependent Ablowitz--Ladik hierarchy. 
Here we recall the definition of $c_{\ul p}$ in \eqref{ALdefcp}.

The special case $\ul p=(1,1)$, $c_{0,\pm}=1$, and $c_{(1,1)}=-2$, that is,
\begin{equation}
 \begin{pmatrix} -i \alpha_{t_{(1,1)}}- \gamma (\alpha^- + \alpha^+) + 2 \alpha  \\   
  -i\beta_{t_{(1,1)}}+ \gamma (\beta^- + \beta^+) - 2 \beta \end{pmatrix}=0,
\end{equation}
represents \textit{the} Ablowitz--Ladik system \eqref{AL1.12}.

The corresponding homogeneous equations are then defined by 
\begin{equation}  
\hAL_{\ul p} (\alpha, \beta) = 
\AL_{\ul p} (\alpha, \beta)\big|_{c_{0, \pm}=1, \, c_{\ell, \pm}=0, \, \ell=1,\dots,p_{\pm}}=0, 
\quad \ul p=(p_-,p_+) \in\bbN_0^2.  
\end{equation} 

By \eqref{AL_p}, \eqref{ALg_l+}, and \eqref{ALg_l-},
the time derivative of $\gamma=1-\alpha \beta$ is given by
\begin{equation} \lb{AL2.14}
\gamma_{t_{\ul p}} = i \gamma \big((g_{p_+,+} - g_{p_+,+}^-) 
- (g_{p_-,-} - g_{p_-,-}^-) \big).
\end{equation}
(Alternatively, this follows from computing the trace of 
$U_{t_{\ul p}}U^{-1}=V^+_{p}-UV_{\ul p}U^{-1}$.)  
For instance, if $\alpha$, $\beta$ satisfy $\AL_1(\alpha, \beta)=0$, then
\begin{equation}
\gamma_{t_1} = i \gamma \big(\alpha (c_{0,-}\beta^+ + c_{0,+}\beta^-) 
- \beta(c_{0,+}\alpha^+ + c_{0,-}\alpha^-)\big).
\end{equation}

\begin{remark}  \lb{rAL2.13}
From \eqref{AL1,1}--\eqref{AL2,1} and the explicit computations of the coefficients 
$f_{\ell,\pm}$, $g_{\ell,\pm}$, and $h_{\ell,\pm}$, one concludes that the zero-curvature equation \eqref{ALzc p} 
and hence the Ablowitz--Ladik hierarchy is invariant under the scaling transformation 
\begin{equation}
\alpha \rightarrow \alpha_c = \{c\, \alpha(n)\}_{n\in\bbZ}, \quad 
\beta \rightarrow \beta_c = \{ \beta(n)/c\}_{n\in\bbZ}, \quad c \in \bbC\setminus \{0\}.
\end{equation}
Moreover, $R_{\ul p}=G_{\ul p}^2 - H_{\ul p}F_{\ul p}$ and hence 
$\{E_m\}_{m=0}^{2p+1}$ are 
invariant under this transformation. Furthermore, choosing $c=e^{i c_{\ul p} t}$,
one verifies that it is no restriction to assume $c_{\ul p}=0$. This also shows 
that stationary solutions $\alpha, \beta$ can only be constructed up to a
multiplicative constant. 
\end{remark}

\begin{remark} \lb{rAL2.14}
$(i)$ The special choices $\beta=\pm\ol\alpha$, $c_{0,\pm}=1$ lead to the discrete nonlinear Schr\"odinger hierarchy. In particular, choosing $c_{(1,1)}=-2$ yields the discrete nonlinear Schr\"odinger equation in its usual form (see, e.g., 
\cite[Ch.\ 3]{AblowitzPrinariTrubatch:2004} and the references cited therein), 
\begin{equation}
-i\alpha_t - (1 \mp |\alpha|^2)(\alpha^- + \alpha^+) + 2\alpha = 0,   
\end{equation}
as its first nonlinear element. The choice $\beta = \ol \alpha$ is called the {\it defocusing} case, $\beta = - \ol \alpha$ represents the {\it focusing} case of the discrete nonlinear Schr\"odinger hierarchy. \\
$(ii)$ The alternative choice $\beta = \ol \alpha$, $c_{0,\pm} = \mp i$, leads to the hierarchy of Schur flows. In particular, choosing $c_{(1,1)} = 0$ yields  
\begin{equation}
\alpha_t - (1 - |\alpha|^2)(\alpha^+ - \alpha^-) = 0   
\end{equation}
as the first nonlinear element of this hierarchy (cf.\  \cite{AmmarGragg:1994}, 
\cite{FaybusovichGekhtman:1999}, \cite{FaybusovichGekhtman:2000},  
\cite{Golinskii:2006}, \cite{MukaihiraNakamura:2002}, \cite{Simon:2007}).
\end{remark}

\section{The Stationary Ablowitz--Ladik Formalism}  \lb{sAL3}

This section is devoted to a detailed study of the stationary
Ablowitz--Ladik hierarchy. Our principal tools are derived from combining the polynomial recursion formalism introduced in Section \ref{sAL2} and a fundamental meromorphic function $\phi$ on a hyperelliptic curve $\calK_p$. With the help of $\phi$ we
study the Baker--Akhiezer vector $\Psi$, and trace formulas for $\alpha$ and 
$\beta$. 

Unless explicitly stated otherwise, we suppose in this section that 
\begin{equation} \label{ALneq 0,1}
\alpha, \beta \in \bbC^\bbZ, \quad 
\alpha(n)\beta(n) \notin \{0,1\}, \; n \in \bbZ, 
\end{equation}
and assume  \eqref{AL2.03}, \eqref{AL_v}, \eqref{ALstatzc}, 
\eqref{ALF_p}--\eqref{ALK_p}, \eqref{AL0+}--\eqref{ALh_l-}, \eqref{ALminus},  
\eqref{ALstat}, \eqref{ALR}, \eqref{ALE_m}, keeping $p\in\bbN_0$ fixed.

We recall the hyperelliptic curve 
\begin{align} \label{ALcalK_p}
\begin{split}
& \calK_p \colon \calF_p (z,y) = y^2 - 4c_{0,+}^{-2}z^{2p_-}R_{\ul p}(z) 
= y^2 - \prod_{m=0}^{2p+1}(z-E_m) = 0, \\
& R_{\ul p}(z) = \bigg(\f{c_{0,+}}{2z^{p_-}}\bigg)^2 \prod_{m=0}^{2p+1}(z-E_m), 
\quad  \{E_m\}_{m=0}^{2p+1} \subset \bbC \setminus\{ 0\},  \; p=p_- +p_+ -1, 
\end{split}
\end{align}
as introduced in \eqref{ALKp}. Throughout this section we assume the affine part of 
$\calK_p$ to be nonsingular, that is, we suppose that 
\begin{equation}
E_m\neq E_{m'} \text{  for $m\neq m'$, \; $m,m'=0,1,\dots,2p+1$.}   \lb{ALEneqE}
\end{equation} 
$\calK_p$ is compactified by
joining two points $P_{\infty_\pm}$,
$P_{\infty_+}\neq P_{\infty_-}$, but for notational simplicity  the
compactification is also denoted by $\calK_p$. Points $P$ on
$\calK_p\setminus\{\Pinfp, \Pinfm\}$ are  represented as pairs $P=(z,y)$, where
$y(\dott)$ is the meromorphic function on $\calK_p$ satisfying
$\calF_p(z,y)=0$. The complex structure on $\calK_p$ is then defined in the usual manner. Hence, $\calK_p$ becomes a two-sheeted hyperelliptic Riemann surface of genus $p$ in a standard manner.

We also emphasize that by fixing the curve $\calK_p$ (i.e., by fixing
$E_0,\dots,E_{2p+1}$), the summation constants $c_{1,\pm},\dots,c_{p_\pm,\pm}$ in
$f_{p_\pm,\pm}$, $g_{p_\pm,\pm}$, and $h_{p_\pm,\pm}$ (and hence in the corresponding stationary $\sAL_{\ul p}$ equations) are uniquely determined as is clear from \eqref{ALcell}, \eqref{ALc_ell}, which establish the summation constants
$c_{\ell,\pm}$ as symmetric functions of $E_0^{\pm 1},\dots,E_{2p+1}^{\pm 1}$. 

For notational simplicity we will usually tacitly assume that $p\in\bbN$ and hence 
$\ul p\in\bbN_0^2\setminus\{(0,0),(0,1),(1,0)\}$.\ (The trivial case $\ul p=0$ is explicitly treated in Example \ref{eAL3.8}.)

We denote by $\{\mu_j(n)\}_{j=1,\dots,p}$ and $\{\nu_j(n)\}_{j=1,\dots,p}$ the zeros of $(\dott)^{p_-}F_{\ul p}(\dott,n)$ and $(\dott)^{p_- -1} H_{\ul p}(\dott,n)$, respectively. Thus, we may write 
\begin{align}   
F_{\ul p}(z)&= - c_{0,+}\alpha^+ z^{-p_-}\prod_{j=1}^{p}(z-\mu_j),   \label{ALmu(n)}  \\
H_{\ul p}(z)&= c_{0,+}\beta z^{-p_- +1}\prod_{j=1}^{p}(z-\nu_j), \label{ALnu(n)}
\end{align}
and we recall that (cf.\ \eqref{ALR})
\begin{equation}  
R_{\ul p} - G_{\ul p}^2 = - F_{\ul p} H_{\ul p}.   \lb{ALquad}
\end{equation}
The next step is crucial; it permits us to ``lift'' the zeros $\mu_j$ and $\nu_j$ from the complex plane $\bbC$ to the curve $\calK_p$.
From \eqref{ALquad} one infers that
\begin{equation}
R_{\ul p}(z) -G_{\ul p}(z)^2=0, \quad
z\in\{\mu_j,\nu_k\}_{j,k=1,\dots,p}. \lb{3.3.7A}
\end{equation}
We now introduce $\{ \hat \mu_j \}_{j=1,\dots,p}\subset \calK_p$ and
$\{ \hat \nu_j \}_{j=1,\dots,p}\subset \calK_p$ by
\begin{equation} \label{ALhmu}
\hat \mu_j(n)=(\mu_j(n), (2/c_{0,+})\mu_j(n)^{p_-} G_{\ul p}(\mu_j(n),n)), \quad j=1, \dots, p, 
\; n\in\bbZ,   
\end{equation}
and 
\begin{equation}  \label{ALhnu}
\hat \nu_j(n)=(\nu_j(n), - (2/c_{0,+})\nu_j(n)^{p_-} G_{\ul p}(\nu_j(n),n)), \quad j=1, \dots, p, 
\; n\in\bbZ.
\end{equation}

We also introduce the points $P_{0,\pm}$ by 
\begin{equation}
    \Pzpm=(0,\pm (c_{0,-}/c_{0,+}))\in\calK_p, \quad 
    \f{c_{0,-}^2}{c_{0,+}^2} = \prod_{m=0}^{2p+1} E_m.   \lb{AL3.10}
\end{equation}
We emphasize that $\Pzpm$ and $\Pinfpm$ are not necessarily on the same
sheet of $\calK_p$.  

Next, we briefly recall our conventions used in connection with divisors on 
$\calK_p$. A map, $\calD \colon \calK_p \to \bbZ$, is called a divisor on 
$\calK_p$ if $\calD(P)\neq0$ for only finitely many $P\in\calK_p$.  The set of divisors on $\calK_p$ is denoted by $\Div(\calK_p)$. We shall employ the
following (additive) notation for divisors,
\begin{align} \lb{A.17}
&\calD_{Q_0\ul Q}=\calD_{Q_0}+\calD_{\ul Q}, \quad \calD_{\ul
Q}=\calD_{Q_1}+\cdots +\calD_{Q_m}, \\
& {\ul Q}=\{Q_1, \dots ,Q_m\} \in \sym^m \calK_p,
\quad Q_0\in\calK_p, \; m\in\bbN, \no
\end{align}
where for any $Q\in\calK_p$,
\begin{equation} \lb{A.18}
\calD_Q \colon  \calK_p \to\bbN_0, \quad
P \mapsto  \calD_Q (P) = \begin{cases} 1 & \text{for $P=Q$},\\
0 & \text{for $P\in \calK_p \setminus \{Q\}$}, \end{cases}
\end{equation}
and $\sym^n \calK_p$ denotes the $n$th symmetric product of $\calK_p$. In particular, $\sym^m \calK_p$ can be identified with the set of nonnegative
divisors $0 \leq \calD \in \Div(\calK_p)$ of degree $m$. Moreover, for a nonzero, meromorphic function $f$ on $\calK_p$, the divisor of $f$ is denoted by $(f)$. Two
divisors $\calD$, $\calE\in \Div(\calK_p)$ are called equivalent, denoted by
$\calD \sim \calE$, if and only if $\calD -\calE=(f)$ for some
$f\in\calM (\calK_p) \setminus \{0\}$.  The divisor class
$[\calD]$ of $\calD$ is
then given by $[\calD]
=\{\calE \in \Div(\calK_p) \,|\, \calE \sim \calD\}$.  We
recall that
\begin{equation}
\deg ((f))=0, \; f\in\calM (\calK_p) \setminus \{0\},  \lb{a38}
\end{equation}
where the degree $\deg (\calD)$ of $\calD$ is given
by $\deg (\calD)
=\sum_{P\in \calK_p} \calD (P)$. 

Next we introduce the fundamental meromorphic function on $\calK_p$ by
\begin{align} 
\phi(P,n) &= \frac{(c_{0,+}/2)z^{-p_-} y + G_{\ul p}(z,n)}{F_{\ul p}(z,n)}  \label{ALphi} \\
&= \frac{-H_{\ul p}(z,n)}{(c_{0,+}/2)z^{-p_-} y - G_{\ul p}(z,n)},   \label{ALphi1}  \\
& \hspace*{.9cm}  P=(z,y)\in \calK_p, \; n\in \bbZ,   \no 
\end{align}
with divisor  $(\phi(\dott,n))$ of $\phi(\dott,n)$ given by
\begin{equation} \label{AL(phi)}
(\phi(\dott,n)) = \calD_{P_{0,-} \hunu(n)} - \calD_{\Pinfm \humu(n)},  
\end{equation}
using \eqref{ALmu(n)} and \eqref{ALnu(n)}. Here we abbreviated 
\begin{equation}
\humu = \{\hat \mu_1, \dots, \hat \mu_{p}\}, \, 
\hunu = \{\hat \nu_1, \dots, \hat \nu_{p}\} \in\symq.
\end{equation}
(The function $\phi$ is closely related to one of the variants of Weyl--Titchmarsh 
functions discussed in \cite{GesztesyZinchenko:2006}, 
\cite{GesztesyZinchenko:2006a}, \cite{Simon:2004a} in the special defocusing 
case $\beta=\bar{\alpha}$.)
Given $\phi(\dott,n)$, the meromorphic stationary Baker--Akhiezer vector 
$\Psi(\dott,n,n_0)$ on $\calK_p$ is then defined by
\begin{align} \no
\Psi(P,n,n_0) &= \binom{\psi_1(P,n,n_0)}{\psi_2(P,n,n_0)}, \\  \label{ALpsi1}
\psi_1(P,n,n_0) &= \begin{cases}      
\prod_{n'=n_0 + 1}^n \big(z + \alpha(n') \phi^-(P,n')\big), & n \geq n_0 +1, \\
1,                      &  n=n_0, \\
\prod_{n'=n + 1}^{n_0} \big(z + \alpha(n') \phi^-(P,n')\big)^{-1}, & n \leq n_0 -1,
\end{cases}   \\
\psi_2(P,n,n_0) &= \phi(P,n_0)
\begin{cases}      
\prod_{n'=n_0 + 1}^n \big(z \beta(n') \phi^-(P,n')^{-1} + 1\big), & n \geq n_0 +1, \\
1,                      &  n=n_0, \\
\prod_{n'=n + 1}^{n_0} \big(z \beta(n') \phi^-(P,n')^{-1} + 1\big)^{-1}, & n \leq n_0 -1.
\end{cases}         \label{ALpsi2}
\end{align}
Basic properties of $\phi$ and $\Psi$ are summarized in the following result.

\begin{lemma} \lb{lAL3.1}
Suppose that $\alpha, \beta$ satisfy \eqref{ALneq 0,1} and the $\ul p$th stationary 
Ablowitz--Ladik system \eqref{ALstat}. Moreover, assume \eqref{ALcalK_p} and \eqref{ALEneqE} and let
$P=(z,y) \in \calK_p\setminus \{\Pinfp, \Pinfm,\Pzp,\Pzm\}$, $(n, n_0) \in \bbZ^2$.
Then $\phi$ satisfies the Riccati-type equation
\begin{align} \label{ALriccati} 
& \alpha \phi(P)\phi^-(P) - \phi^-(P) + z \phi(P) = z \beta,   \\
\intertext{as well as}
  \label{ALphi 1}
& \phi(P) \phi(P^*) = \frac{H_{\ul p}(z)}{F_{\ul p}(z)},\\ \label{ALphi 2}
& \phi(P) + \phi(P^*) = 2\frac{G_{\ul p}(z)}{F_{\ul p}(z)},\\ \label{ALphi 3}
& \phi(P) - \phi(P^*) = c_{0,+}z^{-p_-} \frac{y(P)}{F_{\ul p}(z)}.
\end{align}
The vector $\Psi$ satisfies
\begin{align} 
& U(z) \Psi^-(P)=\Psi(P),  \label{ALpsi 2} \\ 
\label{ALpsi 3}
& V_{\ul p}(z)\Psi^-(P)= - (i/2)c_{0,+} z^{-p_-} y \Psi^-(P), \\ 
& \psi_2(P,n,n_0) = \phi(P,n) \psi_1(P,n,n_0),   \label{ALpsi 1} \\ \label{ALpsi 4}
& \psi_1(P,n,n_0) \psi_1(P^*,n,n_0) = z^{n-n_0} \frac{F_{\ul p}(z,n)}{F_{\ul p}(z,n_0)} 
\pgam(n,n_0),
\\ \label{ALpsi 5}
& \psi_2(P,n,n_0) \psi_2(P^*,n,n_0) = z^{n-n_0} \frac{H_{\ul p}(z,n)}{F_{\ul p}(z,n_0)} 
\pgam(n,n_0),\\
& \psi_1(P,n,n_0) \psi_2(P^*,n,n_0) +\psi_1(P^*,n,n_0) \psi_2(P,n,n_0) \label{ALpsi 6} \\
& \quad =2 z^{n-n_0} \frac{G_{\ul p}(z,n)}{F_{\ul p}(z,n_0)} 
\pgam(n,n_0),\no \\
& \psi_1(P,n,n_0) \psi_2(P^*,n,n_0) -\psi_1(P^*,n,n_0) \psi_2(P,n,n_0) \label{ALpsi 7} \\
& \quad =-c_{0,+} z^{n-n_0-p_-} \frac{y}{F_{\ul p}(z,n_0)}  \pgam(n,n_0), \no
\end{align}
where we used the abbreviation 
\begin{equation} \lb{ALpgam}
\pgam(n,n_0) = \begin{cases}      
\prod_{n'=n_0 + 1}^n \gamma(n'), & n \geq n_0 +1, \\
1,                      &  n=n_0, \\
\prod_{n'=n + 1}^{n_0} \gamma(n')^{-1},  & n \leq n_0 -1.
\end{cases}
\end{equation}
\end{lemma}
\begin{proof}
To prove \eqref{ALriccati} one uses the definition \eqref{ALphi} of $\phi$ and equations 
\eqref{AL1,1}, \eqref{AL1,2}, and \eqref{ALR} to obtain 
\begin{align}
& \alpha \phi(P)\phi^-(P) - \phi(P)^- + z \phi(P) - z \beta   \no \\
&\quad= 
\frac{1}{F_{\ul p} F_{\ul p}^-}\Big(\alpha G_{\ul p} G_{\ul p}^- + (c_{0,+}/2) z^{-p_-} y \alpha 
(G_{\ul p} + G_{\ul p}^-) + \alpha R_{\ul p}  \no \\
& \qquad  - (G_{\ul p}^- + (c_{0,+}/2) z^{-p_-} y) F_{\ul p} 
+ z (G_{\ul p} + (c_{0,+}/2) z^{-p_-} y) F_{\ul p}^- - z \beta F_{\ul p} F_{\ul p}^-\Big)  \no \\
& \quad = \frac{1}{F_{\ul p} F_{\ul p}^-}\Big(\alpha G_{\ul p} (G_{\ul p} + G_{\ul p}^-) + F_{\ul p}(-\alpha H_{\ul p} - G_{\ul p}^- - z \beta F_{\ul p}^-) + z F_{\ul p}^- G_{\ul p}\Big) = 0.
\end{align}
Equations \eqref{ALphi 1}--\eqref{ALphi 3} are clear from the definitions of $\phi$ and $y$.
By definition of $\psi$, \eqref{ALpsi 1} holds for $n=n_0$. By induction,
\begin{equation}
\frac{\psi_2(P,n,n_0)}{\psi_1(P,n,n_0)}= \frac{z \beta(n) \phi^-(P,n)^{-1} + 1}
{z + \alpha(n) \phi^-(P,n)} \frac{\psi_2^-(P,n,n_0)}{\psi_1^-(P,n,n_0)}
= \frac{z \beta(n) + \phi^-(P,n)}{z + \alpha(n) \phi^-(P,n)},
\end{equation}
and hence $\psi_2/\psi_1$ satisfies the Riccati-type equation \eqref{ALriccati}
\begin{equation}
\alpha(n) \phi^-(P,n) \frac{\psi_2(P,n,n_0)}{\psi_1(P,n,n_0)} - \phi^-(P,n) 
+ z \frac{\psi_2(P,n,n_0)}{\psi_1(P,n,n_0)} - z \beta(n) = 0.
\end{equation}
This proves \eqref{ALpsi 1}. 

The definition of $\psi$ implies
\begin{align}
\psi_1(P,n,n_0)&= (z + \alpha(n) \phi^-(P,n))\psi_1^-(P,n,n_0)  \no \\
&=z \psi_1^-(P,n,n_0) + \alpha(n) \psi_2^-(P,n,n_0),  \\
\psi_2(P,n,n_0) &= (z \beta(n) \phi^-(P,n)^{-1} + 1) \psi_2^-(P,n,n_0) \no \\
&= z \beta(n) \psi_1^-(P,n,n_0) + \psi_2^-(P,n,n_0),
\end{align}
which proves \eqref{ALpsi 2}. Property \eqref{ALpsi 3} follows from \eqref{ALpsi 1} and the definition of $\phi$. To prove \eqref{ALpsi 4} one can use \eqref{AL1,1} and \eqref{AL1,2}
\begin{align}
\psi_1(P) \psi_1(P^*) &= (z + \alpha \phi^-(P))(z + \alpha \phi^-(P^*)) 
\psi_1^-(P) \psi_1^-(P^*)   \no \\
&= \frac{1}{F_{\ul p}^-}(z^2 F_{\ul p}^- + 2z\alpha G_{\ul p}^- + \alpha^2 H_{\ul p}^-) \psi_1^-(P) \psi_1^-(P^*)   \no \\
&= \frac{1}{F_{\ul p}^-}(z^2 F_{\ul p}^- - z\alpha \beta F_{\ul p} + z \alpha(G_{\ul p} + G_{\ul p}^-)) \psi_1^-(P) \psi_1^-(P^*)  \no \\
&= z \gamma \frac{F_{\ul p}}{F_{\ul p}^-}\psi_1^-(P) \psi_1^-(P^*).
\end{align}
Equation \eqref{ALpsi 5} then follows from \eqref{ALphi 2} and \eqref{ALpsi 2}. Finally, equation \eqref{ALpsi 6} (resp.\ \eqref{ALpsi 7}) is proved by combining 
\eqref{ALphi 2} and \eqref{ALpsi 1} (resp.\ \eqref{ALphi 3} and \eqref{ALpsi 1}).
\end{proof}

Combining the Laurent polynomial recursion approach of Section \ref{sAL2} with 
\eqref{ALmu(n)} and \eqref{ALnu(n)} readily yields trace formulas for $f_{\ell,\pm}$ and $h_{\ell,\pm}$ in terms of symmetric functions of the zeros $\mu_j$ and 
$\nu_k$ of $(\dott)^{p_-}F_{\ul p}$ and $(\dott)^{p_- -1}H_{\ul p}$, respectively. For simplicity we just record the simplest cases.

\begin{lemma}  \lb{lAL3.2}
Suppose that $\alpha, \beta$ satisfy \eqref{ALneq 0,1} and the $\ul p$th stationary  
Ablowitz--Ladik system \eqref{ALstat}. Then, 
\begin{align} 
 \frac{\alpha}{\alpha^+}&= 
\prod_{j=1}^{p}\mu_j \bigg(\prod_{m=0}^{2p+1}E_m\bigg)^{-1/2},   \label{ALtr1} \\ 
\frac{\beta^+}{\beta}&= 
\prod_{j=1}^{p}\nu_j \bigg(\prod_{m=0}^{2p+1}E_m\bigg)^{-1/2},   \label{ALtr2} \\  
\sum_{j=1}^{p}\mu_j &= \alpha^+ \beta
- \gamma^+ \frac{\alpha^{++}}{\alpha^+} 
- \frac{c_{1,+}}{c_{0,+}},   \label{ALtr3} \\
\sum_{j=1}^{p}\nu_j &= \alpha^+ \beta
- \gamma \frac{\beta^-}{\beta} 
- \frac{c_{1,+}}{c_{0,+}}.  \label{ALtr4}
\end{align}
\end{lemma}
\begin{proof}
We compare coefficients in \eqref{ALF_p} and \eqref{ALmu(n)}
\begin{align}
z^{p_-} F_{\ul p}(z) &= f_{0,-} + \dots + z^{p_- + p_+ -2} f_{1,+} + z^{p_- + p_+ -1} f_{0,+}  \no \\
&= c_{0,+}\alpha^+ \bigg(\prod_{j=1}^{p}\mu_j + \dots + 
z^{p_- + p_+ -2} \sum_{j=1}^{p}\mu_j - z^{p_- + p_+ -1}\bigg)
\end{align}
and use $f_{0,-}=c_{0,-}\alpha$ and 
$f_{1,+}=c_{0,+}\big((\alpha^+)^2 \beta - \gamma^+ \alpha^{++}\big) - \alpha^+ c_{1,+}$
which yields \eqref{ALtr1} and \eqref{ALtr3}.
Similarly, one employs $h_{0,-} = - c_{0,-}\beta^+$ and 
$h_{1,+} = c_{0,+}\big(\gamma \beta^- - \alpha^+ \beta^2\big) + \beta c_{1,+}$ for the remaining formulas \eqref{ALtr2} and \eqref{ALtr4}.
\end{proof}

Next we turn to asymptotic properties of $\phi$ and $\Psi$ in a neighborhood of 
$\Pinfpm$ and $\Pzpm$.

\begin{lemma} \label{lAL3.3}
Suppose that $\alpha, \beta$ satisfy \eqref{ALneq 0,1} and the $\ul p$th stationary 
Ablowitz--Ladik system \eqref{ALstat}.  Moreover, let
$P=(z,y)\in\calK_p\setminus\{\Pinfp,\Pinfm,\Pzp,\Pzm\}$, $(n,n_0)\in\bbZ^2$. 
Then $\phi$ has the asymptotic behavior 
\begin{align}  
\phi(P) \underset{\zeta\to 0}{=}&  \begin{cases} 
\beta + \beta^-\gamma \zeta + \Oh(\zeta^2), & \quad  P \rightarrow P_{\infty_+}, \\
- (\alpha^+)^{-1} \zeta^{-1} + (\alpha^+)^{-2}\alpha^{++}\gamma^+
+ \Oh(\zeta), & \quad  P \rightarrow P_{\infty_-}, 
\end{cases}
\quad \zeta=1/z,  \label{ALphi infty} \\ 
\phi(P) \underset{\zeta\to 0}{=}&  \begin{cases} 
\alpha^{-1} - \alpha^{-2} \alpha^-\gamma \zeta  + \Oh(\zeta^2), 
& \quad P \rightarrow P_{0,+}, \\  
- \beta^+ \zeta - \beta^{++}\gamma^+ \zeta^2 + \Oh(\zeta^3), & \quad P \rightarrow P_{0,-},
\end{cases}
\quad \zeta=z.   \label{ALphi zero}
\end{align}
The components of the Baker--Akhiezer vector $\Psi$ have the asymptotic behavior 
\begin{align}
\psi_1(P,n,n_0) \underset{\zeta\to 0}{=}& \begin{cases} 
\zeta^{n_0-n}(1 + \Oh(\zeta)), & P \rightarrow P_{\infty_+}, \\ 
\frac{\alpha^+(n)}{\alpha^+(n_0)}
\pgam(n,n_0) + \Oh(\zeta),
&P \rightarrow P_{\infty_-},  
\end{cases}  \quad  \zeta=1/z,   \lb{ALpsi_1 infty}  \\ 
\psi_1(P,n,n_0) \underset{\zeta\to 0}{=}& \begin{cases} 
\frac{\alpha(n)}{\alpha(n_0)} + \Oh(\zeta), & P \rightarrow P_{0,+}, \\  
 \zeta^{n-n_0} \pgam(n,n_0)(1 + \Oh(\zeta)),
&P \rightarrow P_{0,-}, 
\end{cases} \quad   \zeta=z,  \label{ALpsi_1 zero}  \\ 
\psi_2(P,n,n_0) \underset{\zeta\to 0}{=}& \begin{cases} 
\beta(n) \zeta^{n_0-n}(1 + \Oh(\zeta)),
&P \rightarrow P_{\infty_+}, \\
- \frac{1}{\alpha^+(n_0)}
\pgam(n,n_0) \zeta^{-1} (1 + \Oh(\zeta)), 
& P \rightarrow P_{\infty_-}, 
\end{cases} \quad   \zeta=1/z,   \label{ALpsi_2 infty} \\ 
\psi_2(P,n,n_0) \underset{\zeta\to 0}{=}& \begin{cases} 
\frac{1}{\alpha(n_0)} + \Oh(\zeta),
& P \rightarrow P_{0,+},   \\
- \beta^+(n)
\pgam(n,n_0) \zeta^{n+1-n_0}(1 + \Oh(\zeta)), 
& P \rightarrow P_{0,-}, 
\end{cases} \quad   \zeta=z.   \lb{ALpsi_2 zero} 
\end{align}
The divisors $(\psi_j)$ of $\psi_j$, $j=1,2$, are given by
\begin{align} \label{ALpsi1aa}
(\psi_1(\dott,n,n_0)) &= \calD_{\humu(n)} - \calD_{\humu(n_0)} 
+ (n-n_0)(\calD_{P_{0,-}} - \calD_{\Pinfp}), \\ 
(\psi_2(\dott,n,n_0)) &= \calD_{\hunu(n)} - \calD_{\humu(n_0)} 
+ (n-n_0)(\calD_{P_{0,-}} - \calD_{\Pinfp}) 
+ \calD_{P_{0,-}} - \calD_{\Pinfm}.   \label{ALpsi2aa}
\end{align}
\end{lemma}
\begin{proof}
The existence of the asymptotic expansion of $\phi$ in terms of the local
coordinate $\zeta=1/z$ near $\Pinfpm$, respectively, $\zeta=z$ near $\Pzpm$ is
clear  from the explicit form of $\phi$ in \eqref{ALphi} and \eqref{ALphi1}. Insertion of
the Laurent polynomials $F_{\ul p}$ into \eqref{ALphi} and $H_{\ul p}$ into \eqref{ALphi1} then
yields the explicit expansion  coefficients in \eqref{ALphi infty} and \eqref{ALphi zero}.
Alternatively, and more efficiently, one can insert each of the following asymptotic expansions
\begin{align}
\begin{split}
\phi(P) & \underset{z\to\infty}{=} \phi_{-1}z + \phi_0 + \phi_1 z^{-1} + \Oh(z^{-2}), \\
\phi(P^*) & \underset{z\to\infty}{=} \phi_0 + \phi_1 z^{-1} + \Oh(z^{-2}), \\
\phi(P) & \underset{z\to 0}{=} \phi_0 + \phi_1 z + \Oh(z^2),  \\
\phi(P^*) & \underset{z\to 0}{=} \phi_1 z + \phi_2 z^2 + \Oh(z^3)   \lb{ALexp}
\end{split}
\end{align}
into the Riccati-type equation \eqref{ALriccati} and, upon comparing
coefficients of powers of $z$, which determines the expansion coefficients $\phi_k$ in \eqref{ALexp}, one concludes \eqref{ALphi infty} and \eqref{ALphi zero}.

Next we compute the divisor of $\psi_1$. By \eqref{ALpsi1} it suffices to compute the divisor of $z + \alpha \phi^-(P)$. First of all we note that 
\begin{equation}
z + \alpha \phi^-(P) = \left\{
\array{ll}
z + \Oh(1), & P \rightarrow P_{\infty_+}, \\
\frac{\alpha^+}{\alpha}\gamma + \Oh(z^{-1}), & P \rightarrow P_{\infty_-}, \\
\frac{\alpha}{\alpha^-} + \Oh(z), & P \rightarrow P_{0,+}, \\
\gamma z + \Oh(z^2), & P \rightarrow P_{0,-},
\endarray
\right.
\end{equation}
which establishes \eqref{ALpsi_1 infty} and \eqref{ALpsi_1 zero}. Moreover, the poles of 
the function $z + \alpha \phi^-(P)$ in 
$\calK_p\setminus\{P_{0,\pm}, P_{\infty_\pm}\}$
coincide with the ones of $\phi^-(P)$, and so it remains to compute the missing $p$ zeros in $\calK_p\setminus\{P_{0,\pm}, P_{\infty_\pm}\}$. Using \eqref{AL1,2}, 
\eqref{ALK_p}, \eqref{ALR},
and $y(\hat \mu_j)= (2/c_{0,+})\mu_j^{p_-} G_{\ul p}(\mu_j)$ (cf. \eqref{ALhmu}) one computes 
\begin{align}
& z + \alpha \phi^- (P) = z + \alpha \frac{(c_{0,+}/2) z^{-p_-} y + G_{\ul p}^-}{F_{\ul p}^-} \no \\
&\quad = \frac{F_{\ul p} + \alpha ((c_{0,+}/2) z^{-p_-} y - G_{\ul p})}{F_{\ul p}^-} \no \\
&\quad = \frac{F_{\ul p}}{F_{\ul p}^-} + \alpha \frac{(c_{0,+}/2)^2 z^{-2p_-} y^2 - G_{\ul p}^2}
{F_{\ul p}^-((c_{0,+}/2) z^{-p_-} y + G_{\ul p})}\no \\
&\quad = \frac{F_{\ul p}}{F_{\ul p}^-}\bigg(1 + \frac{\alpha H_{\ul p}}{(c_{0,+}/2) z^{-p_-} y + G_{\ul p}} \bigg)
\underset{P\to \hmu_j}{=} \frac{F_{\ul p}(P)}{F_{\ul p}^-(P)} \Oh(1).
\end{align}
Hence the sought after zeros are at $\hat \mu_j$, $j=1,\dots,p$ (with the possibility that a zero at $\hat \mu_j$ is cancelled by a pole at $\hat \mu_j^-$).

Finally, the behavior of $\psi_2$ follows immediately using $\psi_2=\phi \psi_1$.
\end{proof}

In addition to \eqref{ALphi infty}, \eqref{ALphi zero} one can use the Riccati-type equation \eqref{ALriccati} to derive a convergent expansion of $\phi$ around 
$\Pinfpm$ and $\Pzpm$ and recursively determine the coefficients as in Lemma \ref{lAL3.3}. Since this is not used later in this section, we omit further details at this point.

Since nonspecial divisors play a fundamental role in the derivation of theta function representations of algebro-geometric solutions of the AL hierarchy in \cite{GesztesyHoldenMichorTeschl:2007}, we now take a closer look at them.

\begin{lemma} \label{lAL3.4} 
Suppose that $\alpha, \beta$ satisfy \eqref{ALneq 0,1} and the
$\ul p$th stationary Ablowitz--Ladik system \eqref{ALstat}. Moreover, assume 
\eqref{ALcalK_p} and \eqref{ALEneqE} and let $n\in\bbZ$. Let $\calD_{\humu}$,
$\humu=\{\hmu_1,\dots,\hmu_{p}\}$, and $\calD_{\hunu}$,
$\hunu=\{\hunu_1,\dots,\hunu_{p}\}$, be the pole and zero divisors of degree
$p$, respectively, associated with $\alpha$, $\beta$, and $\phi$ defined
according to \eqref{ALhmu} and \eqref{ALhnu}, that is,
\begin{align}
\begin{split}
\hat\mu_j (n) &= (\mu_j (n), (2/c_{0,+}) \mu_j(n)^{p_-} G_{\ul p}(\mu_j(n),n)), 
\quad j=1,\dots,p,   \\
\hat\nu_j (n) &= (\nu_j (n),- (2/c_{0,+}) \nu_j(n)^{p_-} G_{\ul p}(\nu_j(n),n)), 
\quad j=1,\dots,p. 
\end{split}
\end{align}
Then $\calD_{\humu(n)}$ and $\calD_{\hunu(n)}$ are nonspecial for all
$n\in\bbZ$.
\end{lemma}
\begin{proof}
We provide a detailed proof in the case of $\calD_{\humu(n)}$.
By \cite[Thm.\ A.31]{GesztesyHolden:2005} 
(see also \cite[Thm.\ A.30]{GesztesyHolden:2003}), $\calD_{\humu(n)}$ 
is special if and only if
$\{\hmu_1(n),\dots,\hmu_{p}(n)\}$ contains at least one pair of the type
$\{\hat\mu(n),\hat\mu(n)^*\}$. Hence $\calD_{\humu(n)}$ is certainly
nonspecial as long as the projections $\mu_j(n)$ of $\hmu_j(n)$ are
mutually distinct, $\mu_j(n) \neq \mu_k(n)$ for $j\neq k$. On the other
hand, if two or more projections coincide for some $n_0\in\bbZ$, for
instance,
\begin{equation}
\mu_{j_1}(n_0)=\cdots=\mu_{j_N}(n_0)=\mu_0, \quad  N\in\{2,\dots,p\},
\lb{AL3.45AA}
\end{equation}
then $G_{\ul p}(\mu_0,n_0)\neq 0$ as long as $\mu_0\notin
\{E_0,\dots,E_{2p+1}\}$. This fact immediately follows from \eqref{ALR}
since $F_{\ul p}(\mu_0,n_0)=0$ but $R_{\ul p}(\mu_0)\neq 0$ by hypothesis. In
particular, $\hmu_{j_1}(n_0),\dots,\hmu_{j_N}(n_0)$ all meet on the same
sheet since
\begin{equation}
\hmu_{j_r}(n_0)=(\mu_0,(2/c_{0,+}) \mu_0^{p_-} G_{\ul p}(\mu_0,n_0)), 
\quad r=1,\dots,N,  \lb{AL3.45a}
\end{equation}
and hence no special divisor can arise in this manner. Remaining to be 
studied is the case where two or more projections collide at a branch point,
say at $(E_{m_0},0)$ for some $n_0\in\bbZ$. In this case one concludes
$F_{\ul p}(z,n_0)\underset{z\to E_{m_0}}{=}\Oh\big((z-E_{m_0})^2\big)$ and
\begin{equation}
G_{\ul p}(E_{m_0},n_0)=0 \lb{AL3.45c}
\end{equation}
using again \eqref{ALR} and $F_{\ul p}(E_{m_0},n_0)=R_{\ul p}(E_{m_0})=0$.
Since $G_{\ul p}(\dott,n_0)$ is a Laurent polynomial, \eqref{AL3.45c} implies
$G_{\ul p}(z,n_0)\underset{z\to E_{m_0}}{=}\Oh((z-E_{m_0}))$.
Thus, using \eqref{ALR} once more, one obtains the contradiction,
\begin{align}
\Oh\big((z-E_{m_0})^2\big)&\underset{z\to E_{m_0}}{=} R_{\ul p}(z)
\lb{AL3.45e}
\\ &\underset{z\to
E_{m_0}}{=} \bigg(\f{c_{0,+}}{2E_{m_0}^{p_-}}\bigg)^2 
(z-E_{m_0})\Bigg(\prod_{\substack{m=0\\m\neq m_0}}^{2p+1}
\big(E_{m_0}-E_m\big) +\Oh(z-E_{m_0})\Bigg). \no
\end{align}
Consequently, at most one $\hmu_j(n)$ can hit a branch point at a time
and again no special divisor arises. Finally, by our hypotheses on $\alpha, \beta$, 
$\hat\mu_j(n)$ stay finite for fixed $n\in\bbZ$ and hence never reach the points 
$\Pinfpm$. (Alternatively, by \eqref{ALphi infty},
$\hat\mu_j$ never reaches the point $\Pinfp$. Hence, if some
$\hmu_j$ tend to infinity, they all necessarily converge to $\Pinfm$.) 
Again no special divisor can arise in this manner.

The proof for $\calD_{\hunu(n)}$ is analogous
(replacing $F_{\ul p}$ by $H_{\ul p}$ and noticing that by \eqref{ALphi infty},
$\phi$ has no zeros near $\Pinfpm$), thereby completing the proof.
\end{proof}

The results of Sections \ref{sAL2} and \ref{sAL3} have been used extensively in 
\cite{GesztesyHoldenMichorTeschl:2007} to derive the class of stationary 
algebro-geometric solutions of the Ablowitz--Ladik hierarchy and the associated theta function representations of $\alpha$, $\beta$, $\phi$, and $\Psi$. These theta function representations also show that $\gamma(n)\notin \{0,1\}$ for all 
$n\in\bbZ$, and hence  condition \eqref{ALneq 0,1} is satisfied for the stationary algebro-geometric AL solutions discussed in this section, provided the associated divisors $\calD_{\humu(n)}$ and $\calD_{\hunu(n)}$ stay away from 
$\Pinfpm, \Pzpm$ for all $n\in\bbZ$. 

We conclude this section with the trivial case $\ul p=0$ excluded thus far.

\begin{example}  \lb{eAL3.8}
Assume $\ul p=0$ and $c_{0,+}=c_{0,-}=c_0\neq 0$ $($we recall that 
$g_{p_+,+}=g_{p_-,-}$$)$. Then,
\begin{align}
&F_{(0,0)}=\hatt F_{(0,0)}=H_{(0,0)}= \hatt H_{(0,0)}=0, 
\quad G_{(0,0)}=K_{(0,0)}=\frac12 c_{0},  \no  \\
& \hatt G_{(0,0)}=\hatt K_{(0,0)}=\frac12, \quad R_{(0,0)}=\frac14 c_{0}^2, \no \\
& \alpha = \beta = 0, \lb{AL_ex0} \\
& U= \begin{pmatrix}z& 0\\0& 1\end{pmatrix}, \quad
V_{(0,0)}=\frac{ic_{0}}2 \begin{pmatrix}1& 0\\0& -1\end{pmatrix}. \no
\end{align}
Introducing 
\begin{equation}
\Psi_+(z,n,n_0)= \begin{pmatrix}z^{n-n_0}\\0\end{pmatrix}, \quad  
\Psi_-(z,n,n_0)= \begin{pmatrix}0\\1\end{pmatrix}, \quad n, n_0 \in\bbZ,  \lb{AL_ex0b}
\end{equation}
one verifies the equations
\begin{equation}
U\Psi_{\pm}^-=\Psi_{\pm}, \quad 
V_{(0,0)}\Psi_{\pm}^-= \pm \frac{ic_0}2 \Psi_{\pm}^-.     \lb{AL_ex0a}
\end{equation}
\end{example}

\section{The Time-Dependent Ablowitz--Ladik Formalism} \label{sAL4}
 
In this section we extend the algebro-geometric analysis of Section \ref{sAL3}  
to the time-dependent Ablowitz--Ladik hierarchy.

For most of this section we assume the following hypothesis.

\begin{hypothesis} \label{hAL4.1}
$(i)$ Suppose that $\alpha, \beta$ satisfy
\begin{align}
\begin{split}
& \alpha(\dott,t), \beta(\dott,t)\in \bbC^\bbZ,\; t\in\bbR,
\quad 
\alpha(n,\dott), \, \beta(n,\dott) \in C^1(\bbR), \; n\in\bbZ,  \\
& \alpha(n,t)\beta(n,t)\notin\{0,1\}, \; (n,t)\in\bbZ\times\bbR.   \lb{AL4.1A}
\end{split}
\end{align}
$(ii)$  Assume that the hyperelliptic curve $\calK_p$ satisfies 
\eqref{ALcalK_p} and \eqref{ALEneqE}.
\end{hypothesis}

The basic problem in the analysis of algebro-geometric solutions of the 
Ablowitz--Ladik hierarchy consists of solving the time-dependent $\ul r$th 
Ablowitz--Ladik flow with initial data a stationary solution of the $\ul p$th system in the hierarchy. More precisely, given $\ul p \in \bbN_0^2\setminus\{(0,0)\}$ we consider a solution $\alpha^{(0)}$, $\beta^{(0)}$ of the $\ul p$th stationary Ablowitz--Ladik system  
$\sAL_{\ul p} (\alpha^{(0)}, \beta^{(0)}) = 0$, associated with the hyperelliptic curve 
$\calK_p$ and a corresponding set of summation constants 
$\{c_{\ell,\pm}\}_{\ell=1,\dots,p_\pm}\subset\bbC$. Next, let $\ul r=(r_-,r_+)\in\bbN_0^2$; we intend to construct a solution $\alpha, \beta$ of the $\ul r$th Ablowitz--Ladik flow  
$\AL_{\ul r}(\alpha,\beta)=0$ with $\alpha(t_{0,\ul r})=\alpha^{(0)}$, 
$\beta(t_{0,\ul r})=\beta^{(0)}$ for some $t_{0,\ul r}\in\bbR$.  To emphasize that the summation constants in the definitions of the stationary and the time-dependent Ablowitz--Ladik equations are independent of each other, we indicate
this by adding a tilde on all the time-dependent quantities. Hence we shall employ the notation $\ti V_{\ul r}$, $\ti F_{\ul r}$, $\ti G_{\ul r}$, $\ti H_{\ul r}$, $\ti K_{\ul r}$, $\tilde f_{s,\pm}$,
$\tilde g_{s,\pm}$, $\tilde h_{s,\pm}$, $\tilde c_{s,\pm}$, in order to distinguish them from $V_{\ul p}$, $F_{\ul p}$, $G_{\ul p}$, $H_{\ul p}$, $K_{\ul p}$, $f_{\ell,\pm}$, $g_{\ell,\pm}$, 
$h_{\ell,\pm}$, $c_{\ell,\pm}$, in the following. In addition, we will follow a more elaborate notation inspired by Hirota's $\tau$-function approach and indicate the individual $\ul r$th 
Ablowitz--Ladik flow by  a separate time variable $t_{\ul r} \in \bbR$. 

Summing up, we are interested in solutions $\alpha, \beta$ of the time-dependent
algebro-geometric initial value problem
\begin{align}
\begin{split}
& \ti \AL_{\ul r} (\alpha, \beta) = \begin{pmatrix}-i\alpha_{t_{\ul r}} 
- \alpha(\tilde g_{r_+,+} + \tilde g_{r_-,-}^-) + \tilde f_{r_+ -1,+} - \tilde f_{r_- -1,-}^-\\
-i\beta_{t_{\ul r}} + \beta(\tilde g_{r_+,+}^- + \tilde g_{r_-,-}) 
- \tilde h_{r_- -1,-} + \tilde h_{r_+ -1,+}^-     \end{pmatrix} = 0,  \lb{ALal_ivp}  \\
& (\alpha, \beta)\big|_{t=t_{0,\ul r}} = \big(\alpha^{(0)}, \beta^{(0)}\big) ,  
\end{split} \\
& \sAL_{\ul p} \big(\alpha^{(0)}, \beta^{(0)}\big) =  \begin{pmatrix} 
-\alpha^{(0)}(g_{p_+,+} + g_{p_-,-}^-) + f_{p_+ -1,+} - f_{p_- -1,-}^-\\
  \beta^{(0)}(g_{p_+,+}^- + g_{p_-,-}) - h_{p_- -1,-} + h_{p_+ -1,+}^-       
\end{pmatrix}=0   \lb{AL4.3A}
\end{align}
for some $t_{0,\ul r}\in\bbR$, where $\alpha=\alpha(n,t_{\ul r})$, $\beta=\beta(n,t_{\ul r})$ satisfy \eqref{AL4.1A} and a fixed curve $\calK_p$ is associated
with the stationary solutions $\alpha^{(0)}, \beta^{(0)}$ in \eqref{AL4.3A}. Here,
\begin{equation}
\ul p =(p_-,p_+) \in \bbN_0^2\setminus\{(0,0)\}, \quad 
\ul r =(r_-,r_+) \in \bbN_0^2, \quad p=p_- + p_+ -1. 
\end{equation}
In terms of the zero-curvature formulation this amounts to solving 
\begin{align} 
U_{t_{\ul r}}(z,t_{\ul r}) + U(z,t_{\ul r}) \ti V_{\ul r}(z,t_{\ul r}) - \ti V_{\ul r}^+(z,t_{\ul r}) U(z,t_{\ul r}) &= 0, 
\label{ALzc tilde}  \\
U(z,t_{0,\ul r}) V_{\ul p}(z,t_{0,\ul r}) - V_{\ul p}^+(z,t_{0,\ul r}) U(z,t_{0,\ul r}) &= 0. 
\lb{ALzcstat}
\end{align}
One can show (cf.\ \cite{GesztesyHoldenMichorTeschl:2007a}) that the stationary Ablowitz--Ladik system \eqref{ALzcstat} is actually satisfied for all times $t_{\ul r}\in\bbR$: 
Thus, we actually impose
\begin{align} 
U_{t_{\ul r}}(z,t_{\ul r}) + U(z,t_{\ul r}) \ti V_{\ul r}(z,t_{\ul r}) - \ti V_{\ul r}^+(z,t_{\ul r}) U(z,t_{\ul r}) & = 0, 
\label{ALzctilde}  \\ 
U(z,t_{\ul r}) V_{\ul p}(z,t_{\ul r}) - V_{\ul p}^+(z,t_{\ul r}) U(z,t_{\ul r}) & = 0,  \lb{ALzctstat}
\end{align}
instead of \eqref{ALzc tilde} and \eqref{ALzcstat}. 
For further reference, we recall the relevant quantities here 
(cf.\ \eqref{AL2.03}, \eqref{AL_v}, \eqref{ALF_p}--\eqref{ALK_pt}):
\begin{align}
\begin{split}
U(z) &= \begin{pmatrix}
z & \alpha     \\
z \beta & 1\\
\end{pmatrix},  \\
V_{\ul p}(z) &= i  \begin{pmatrix}
G_{\ul p}^-(z) & - F_{\ul p}^-(z)     \\[1.5mm]
H_{\ul p}^-(z)  & - G_{\ul p}^-(z)  \\
\end{pmatrix}, \quad 
\ti V_{\ul r}(z) = i  \begin{pmatrix}
\ti G_{\ul r}^-(z)  & - \ti F_{\ul r}^-(z)     \\[1.5mm]
\ti H_{\ul r}^-(z) & - \ti K_{\ul r}^-(z)  \\
\end{pmatrix}, \lb{ALv_osv}  
\end{split}
\end{align}
and 
\begin{align}
F_{\ul p}(z) &= \sum_{\ell=1}^{p_-} f_{p_- -\ell,-} z^{-\ell}  
+ \sum_{\ell=0}^{p_+ -1} f_{p_+ -1-\ell,+} z^\ell 
=- c_{0,+}\alpha^+ z^{-p_-}\prod_{j=1}^{p}(z-\mu_j),  \no \\ \no
G_{\ul p}(z) &= \sum_{\ell=1}^{p_-} g_{p_- -\ell,-} z^{-\ell} 
+ \sum_{\ell=0}^{p_+} g_{p_+ -\ell,+} z^\ell,  \\ \no
H_{\ul p}(z) &= \sum_{\ell=0}^{p_- -1} h_{p_- -1-\ell,-} z^{-\ell} 
+ \sum_{\ell=1}^{p_+} h_{p_+ -\ell,+} z^\ell 
= c_{0,+}\beta z^{-p_- +1}\prod_{j=1}^{p}(z-\nu_j), \\
\ti F_{\ul r}(z) &= \sum_{s=1}^{r_-} \tilde f_{r_- -s,-} z^{-s}  
+ \sum_{s=0}^{r_+ -1} \tilde f_{r_+ -1-s,+} z^s,   \lb{AL4.9}  \\ \no
\ti G_{\ul r}(z) &= \sum_{s=1}^{r_-} \tilde g_{r_- -s,-} z^{-s}  
+ \sum_{s=0}^{r_+} \tilde g_{r_+ -s,+} z^s,\\ \no
\ti H_{\ul r}(z) &= \sum_{s=0}^{r_- -1} \tilde h_{r_- -1-s,-} z^{-s}  
+ \sum_{s=1}^{r_+} \tilde h_{r_+ -s,+} z^s, \\
\ti K_{\ul r}(z) &= \sum_{s=0}^{r_-} \tilde g_{r_- -s,-} z^{-s}  
+  \sum_{s=1}^{r_+} \tilde g_{r_+ -s,+} z^s 
= \ti G_{\ul r}(z)+\tilde g_{r_-,-}-\tilde g_{r_+,+} 
\end{align}
for fixed $\ul p\in\bbN_0^2\setminus\{(0,0)\}$, $\ul r\in\bbN_0^2$. Here 
$f_{\ell,\pm}$, $\tilde f_{s,\pm}$,  
$g_{\ell,\pm}$, $\tilde g_{s,\pm}$, $h_{\ell,\pm}$, and 
$\tilde h_{s,\pm}$ are defined as in 
\eqref{AL0+}--\eqref{ALh_l-} with appropriate sets of summation constants 
$c_{\ell,\pm}$, $\ell\in\bbN_0$, and 
$\tilde c_{k,\pm}$, $k\in\bbN_0$. Explicitly, \eqref{ALzctilde} and 
\eqref{ALzctstat} are equivalent to (cf.\ \eqref{AL1,1}--\eqref{AL2,1},  
\eqref{ALalphat}--\eqref{AL2,2r}),
\begin{align}  \label{ALalpha_t}
\alpha_{t_{\ul r}} &= i \big(z \ti F_{\ul r}^- + \alpha (\ti G_{\ul r} + \ti{K}_{\ul r}^-)
- \ti F_{\ul r}\big),\\ \label{ALbeta_t}
\beta_{t_{\ul r}} &= - i \big(\beta (\ti G_{\ul r}^- + \ti{K}_{\ul r}) - \ti H_{\ul r} 
+ z^{-1} \ti H_{\ul r}^-\big), \\ \label{AL1,1 r}
0 &= z (\ti G_{\ul r}^- - \ti G_{\ul r}) + z\beta \ti F_{\ul r} + \alpha \ti H_{\ul r}^-,\\ \label{AL2,2 r}
0 &= z \beta \ti F_{\ul r}^- + \alpha \ti H_{\ul r} + \ti{K}_{\ul r}^- - \ti{K}_{\ul r}, \\
0 &= z (G_{\ul p}^- - G_{\ul p}) + z \beta F_{\ul p} + \alpha H_{\ul p}^-,  \lb{AL11} \\  
0 &=z \beta F_{\ul p}^- + \alpha H_{\ul p} - G_{\ul p} + G_{\ul p}^-,  \lb{AL12} \\
0 &= - F_{\ul p} + z F_{\ul p}^- + \alpha (G_{\ul p} + G_{\ul p}^-), \lb{AL21}  \\  
0 &= z \beta (G_{\ul p} + G_{\ul p}^-) - z H_{\ul p} + H_{\ul p}^-,  \lb{AL22}
\end{align}
respectively. In particular, \eqref{ALR} holds in the present $t_{\ul r}$-dependent setting, that is,  
\begin{equation} \label{ALR_t}
G_{\ul p}^2 - F_{\ul p} H_{\ul p} = R_{\ul p}.
\end{equation}

As in the stationary context \eqref{ALhmu}, \eqref{ALhnu} we introduce
\begin{align}
\begin{split}
\hat \mu_j(n,t_{\ul r})&=(\mu_j(n,t_{\ul r}), (2/c_{0,+}) \mu_j(n,t_{\ul r})^{p_-} 
G_{\ul p}(\mu_j(n,t_{\ul r}),n,t_{\ul r}))\in\calK_p, \\ 
& \hspace*{4.4cm} j=1, \dots, p, \; (n,t_{\ul r})\in\bbZ\times\bbR,   \lb{AL4.20}
\end{split}
\end{align}
and
\begin{align}
\begin{split}
\hat \nu_j(n,t_{\ul r})&=(\nu_j(n,t_{\ul r}), - (2/c_{0,+}) \nu_j(n,t_{\ul r})^{p_-} 
G_{\ul p}(\nu_j(n,t_{\ul r}),n,t_{\ul r}))\in\calK_p, \\
& \hspace*{4.6cm}  j=1, \dots, p, \; (n,t_{\ul r})\in\bbZ\times\bbR,   \lb{AL4.21}
\end{split}
\end{align}
and note that the regularity assumptions \eqref{AL4.1A} on $\alpha, \beta$ imply
continuity of $\mu_j$ and $\nu_k$ with respect to $t_{\ul r}\in\bbR$ (away from collisions of these zeros, $\mu_j$ and $\nu_k$ are of course $C^\infty$).  

In analogy to \eqref{ALphi}, \eqref{ALphi1}, one defines the following
meromorphic function $\phi (\dott,n,t_{\ul r})$ on $\calK_p$,
\begin{align}
\phi(P,n,t_{\ul r}) & = \frac{(c_{0,+}/2) z^{-p_-} y + G_{\ul p}(z,n,t_{\ul r})}{F_{\ul p}(z,n,t_{\ul r})}    \lb{AL4.22}  \\
& = \frac{-H_{\ul p}(z,n,t_{\ul r})}{(c_{0,+}/2) z^{-p_-} y - G_{\ul p}(z,n,t_{\ul r})},    \lb{AL4.23}  \\ 
& \hspace*{-.05cm} P=(z,y)\in\calK_p, \; (n,t_{\ul r})\in\bbZ\times\bbR,  \no
\end{align}
with divisor $(\phi(\dott,n,t_{\ul r}))$ of $\phi(\dott,n,t_{\ul r})$ given by 
\begin{equation} 
(\phi(\dott,n,t_{\ul r})) = \calD_{P_{0,-} \hunu(n,t_{\ul r})} - \calD_{\Pinfm \humu(n,t_{\ul r})}.  
\lb{AL4.24}
\end{equation}
The time-dependent Baker--Akhiezer vector is then defined in terms of $\phi$ by 
\begin{align}
& \Psi(P,n,n_0,t_{\ul r},t_{0,\ul r}) = \begin{pmatrix}\psi_1(P,n,n_0,t_{\ul r},t_{0,\ul r})\\
\psi_2(P,n,n_0,t_{\ul r},t_{0,\ul r})\end{pmatrix},   \lb{AL3.25}  \\ 
& \psi_1(P,n,n_0,t_{\ul r},t_{0,\ul r}) = \exp \bigg(i \int_{t_{0,\ul r}}^{t_{\ul r}} ds 
\big(\ti G_{\ul r} (z,n_0,s) - \ti F_{\ul r}(z,n_0,s) \phi(P,n_0,s)\big)\bigg) \no \\ 
&\quad \times 
\begin{cases}
\prod_{n'=n_0+1}^{n}\big(z + \alpha(n',t_{\ul r}) \phi^-(P,n',t_{\ul r})\big), & n \geq n_0 +1,\\
1, & n=n_0, \\
\prod_{n'=n+1}^{n_0}\big(z + \alpha(n',t_{\ul r}) \phi^-(P,n',t_{\ul r})\big)^{-1}, & n \leq n_0 -1, 
\end{cases}   \lb{AL4.26}
\\
& \psi_2(P,n,n_0,t_{\ul r},t_{0,\ul r}) = \exp \bigg(i \int_{t_{0,\ul r}}^{t_{\ul r}} ds 
\big(\ti G_{\ul r} (z,n_0,s) - \ti F_{\ul r}(z,n_0,s)  \phi(P,n_0,s)\big)\bigg)  \no \\ 
& \;\;  \times \phi(P,n_0,t_{\ul r})
\begin{cases}
\prod_{n'=n_0+1}^{n}\big(z \beta(n',t_{\ul r})\phi^-(P,n',t_{\ul r})^{-1} + 1\big), 
& n \geq n_0 +1,\\
1, & n=n_0, \\
\prod_{n'=n+1}^{n_0}\big(z \beta(n',t_{\ul r})\phi^-(P,n',t_{\ul r})^{-1} + 1\big)^{-1}, 
& n \leq n_0 -1, 
\end{cases}  \lb{AL4.27} \\ 
&\hspace*{2.3cm} P=(z,y)\in\calK_p\setminus\{\Pinfp,\Pinfm,\Pzp,\Pzm\}, 
\; (n,t_{\ul r})\in\bbZ\times\bbR.  \no
\end{align}
One observes that
\begin{align} 
\begin{split}
& \psi_1(P,n,n_0,t_{\ul r},\tilde t_{\ul r}) = \psi_1(P,n_0,n_0,t_{\ul r},\tilde t_{\ul r})\psi_1(P,n,n_0,t_{\ul r},t_{\ul r}), \label{ALpsi t n_0}  \\
&  P=(z,y)\in\calK_p\setminus\{\Pinfp,\Pinfm,\Pzp,\Pzm\}, 
\; (n,n_0,t_{\ul r},\tilde t_{\ul r})\in\bbZ^2\times\bbR^2.
\end{split}
\end{align}

The following lemma records basic properties of $\phi$ and $\Psi$ in analogy to the stationary case discussed in Lemma \ref{lAL3.1}. 

\begin{lemma} \lb{lAL4.2}
Assume Hypothesis \ref{hAL4.1} and suppose that 
\eqref{ALzctilde}, \eqref{ALzctstat} hold. In addition, let
$P=(z,y)\in\calK_p\setminus\{\Pinfp, \Pinfm\}$, 
$(n,n_0,t_{\ul r},t_{0,\ul r})\in\bbZ^2\times\bbR^2$. Then $\phi$ satisfies 
\begin{align} \label{ALriccati_time} 
& \alpha \phi(P)\phi^-(P) - \phi^-(P) + z \phi(P) = z \beta, \\
& \phi_{t_{\ul r}}(P) =i \ti F_{\ul r} \phi^2(P) - 
i \big(\ti G_{\ul r}(z) + \ti K_{\ul r}(z)\big)\phi(P) +  i \ti H_{\ul r}(z),    \label{ALphi_t}  \\
  \label{ALphi 1t}
& \phi(P) \phi(P^*) = \frac{H_{\ul p}(z)}{F_{\ul p}(z)},\\ \label{ALphi 2t}
& \phi(P) + \phi(P^*) = 2\frac{G_{\ul p}(z)}{F_{\ul p}(z)},\\ \label{ALphi 3t}
& \phi(P) - \phi(P^*) = c_{0,+} z^{-p_-} \frac{y(P)}{F_{\ul p}(z)}.
\end{align}
Moreover, assuming 
$P=(z,y)\in\calK_p\setminus\{\Pinfp, \Pinfm,\Pzp,\Pzm\}$, then 
$\Psi$ satisfies
\begin{align}  \label{ALpsi t 1}
& \psi_2(P,n,n_0,t_{\ul r},t_{0,\ul r})= \phi(P,n,t_{\ul r}) \psi_1(P,n,n_0,t_{\ul r},t_{0,\ul r}),\\ 
\label{ALpsi 2t}
& U(z) \Psi^-(P)=\Psi(P),\\ \label{ALpsi 3t}
& V_{\ul p}(z)\Psi^-(P)= - (i/2) c_{0,+} z^{-p_-} y \Psi^-(P), \\
& \Psi_{t_{\ul r}}(P) = \ti V_{\ul r}^+(z) \Psi(P),   
\lb{ALtime_AL} \\
& \psi_1(P,n,n_0,t_{\ul r},t_{0,\ul r}) \psi_1(P^*,n,n_0,t_{\ul r},t_{0,\ul r}) = z^{n-n_0} \frac{F_{\ul p}(z,n,t_{\ul r})}{F_{\ul p}(z,n_0,t_{0,\ul r})}  \pgam(n,n_0,t_{\ul r}), \label{ALpsi 4t}\\ 
& \psi_2(P,n,n_0,t_{\ul r},t_{0,\ul r}) \psi_2(P^*,n,n_0,t_{\ul r},t_{0,\ul r}) 
= z^{n-n_0} \frac{H_{\ul p}(z,n,t_{\ul r})}{F_{\ul p}(z,n_0,t_{0,\ul r})}  \pgam(n,n_0,t_{\ul r}),
\label{ALpsi 5t} \\
& \psi_1(P,n,n_0,t_{\ul r},t_{0,\ul r}) \psi_2(P^*,n,n_0,t_{\ul r},t_{0,\ul r}) 
+\psi_1(P^*,n,n_0,t_{\ul r},t_{0,\ul r}) \psi_2(P,n,n_0,t_{\ul r},t_{0,\ul r}) \no \\
&\quad =2 z^{n-n_0} \frac{G_{\ul p}(z,n,t_{\ul r})}{F_{\ul p}(z,n_0,t_{0,\ul r})}
\pgam(n,n_0,t_{\ul r}), \label{ALpsi 6t} \\
& \psi_1(P,n,n_0,t_{\ul r},t_{0,\ul r}) \psi_2(P^*,n,n_0,t_{\ul r},t_{0,\ul r}) 
-\psi_1(P^*,n,n_0,t_{\ul r},t_{0,\ul r}) \psi_2(P,n,n_0,t_{\ul r},t_{0,\ul r}) \no\\
& \quad =-c_{0,+} z^{n-n_0-p_-} \frac{y}{F_{\ul p}(z,n_0,t_{0,\ul r})} \pgam(n,n_0,t_{\ul r}), 
\label{ALpsi 7t}
\end{align}
where 
\begin{equation} \lb{ALpgamt}
\pgam(n,n_0,t_{\ul r}) = \begin{cases}      
\prod_{n'=n_0 + 1}^n \gamma(n',t_{\ul r}), & n \geq n_0 +1, \\
1,                      &  n=n_0, \\
\prod_{n'=n + 1}^{n_0} \gamma(n',t_{\ul r})^{-1},  & n \leq n_0 -1.
\end{cases}
\end{equation}
In addition, as long as the zeros $\mu_j(n_0,s)$ of $(\dott)^{p_-}F_{\ul p} (\dott,n_0,s)$ are all simple and distinct from zero for $s \in\calI_{\mu}$, 
$\calI_{\mu}\subseteq\bbR$ an open interval, 
$\Psi(\dott,n,n_0,t_{\ul r},t_{0,\ul r})$ is meromorphic on $\calK_p\setminus \{\Pinfp,\Pinfm,$ $\Pzp,\Pzm\}$ for $(n,t_{\ul r},t_{0,\ul r})\in\bbZ\times\calI_{\mu}^2$.
\end{lemma}
\begin{proof}
Equations \eqref{ALriccati_time}, \eqref{ALphi 1t}--\eqref{ALpsi 3t}, and 
\eqref{ALpsi 4t}--\eqref{ALpsi 7t} are proved as in the stationary case, see 
Lemma \ref{lAL3.1}. Thus, we turn to the proof of \eqref{ALphi_t} and 
\eqref{ALtime_AL}: Differentiating the Riccati-type equation 
\eqref{ALriccati_time} yields
\begin{align}
0 &= \big( \alpha \phi \phi^- - \phi^- + z \phi - z \beta \big)_{t_{\ul r}}  \no \\
&= \alpha_{t_{\ul r}} \phi \phi^- + (\alpha \phi^- + z) \phi_{t_{\ul r}} + (\alpha \phi -1) \phi_{t_{\ul r}}^-
- z \beta_{t_{\ul r}}  \no \\
&= \big((\alpha \phi^- + z) + (\alpha \phi -1)S^-\big)\phi_{t_{\ul r}}
+ i \phi \phi^- \big( \alpha (\ti G_{\ul r} + \ti K_{\ul r}^-) 
+ z \ti F_{\ul r}^- - \ti F_{\ul r} \big)  \no \\
& \quad + i z \beta (\ti G_{\ul r}^- + \ti K_{\ul r}) + i (z \ti H_{\ul r} - \ti H_{\ul r}^-), 
\end{align}
using \eqref{ALalpha_t} and \eqref{ALbeta_t}.
Next, one employs \eqref{ALriccati} to rewrite
\begin{equation}
(\alpha \phi^- + z) + (\alpha \phi -1)S^- = \frac{1}{\phi}(z \beta + \phi^-) 
+ \frac{z}{\phi^-}(\beta - \phi)S^-.
\end{equation}
This allows one to calculate the right-hand side of \eqref{ALphi_t} using 
\eqref{AL1,1 r} and \eqref{AL2,2 r}
\begin{align}
&\big((\alpha \phi^- + z) + (\alpha \phi -1)S^-\big) \big(\ti H_{\ul r} + \ti F_{\ul r} \phi^2 - 
(\ti G_{\ul r} + \ti K_{\ul r})\phi\big) \no  \\
&= (\alpha \phi^- + z)\ti H_{\ul r} + (\alpha \phi -1)\ti H_{\ul r}^- 
+ \phi (z \beta + \phi^-) \ti F_{\ul r} + z \phi^- (\beta - \phi) \ti F_{\ul r}^-  \no \\
& \quad - (z \beta + \phi^-)(\ti G_{\ul r} + \ti K_{\ul r}) 
- z(\beta - \phi)(\ti G_{\ul r}^- + \ti K_{\ul r}^-)  \no \\
&= \phi \phi^- (\ti F_{\ul r} - z \ti F_{\ul r}^-) + z\ti H_{\ul r} - \ti H_{\ul r}^-
+ \phi^- (\alpha \ti H_{\ul r} + z \beta \ti F_{\ul r}^-) 
+ \phi (\alpha \ti H_{\ul r}^- + z \beta \ti F_{\ul r})  \no \\
& \quad - z \beta (\ti G_{\ul r} + \ti K_{\ul r} + \ti G_{\ul r}^- + \ti K_{\ul r}^-)
- z \phi (\ti G_{\ul r}^- + \ti K_{\ul r}^-) - \phi^- (\ti G_{\ul r} + \ti K_{\ul r})  \no \\
&= \phi \phi^- (\ti F_{\ul r} - z \ti F_{\ul r}^-) + z\ti H_{\ul r} - \ti H_{\ul r}^-
- z \beta(\ti G_{\ul r}^- + \ti K_{\ul r}) + (z \phi - \phi^- - z \beta)(\ti G_{\ul r} + \ti K_{\ul r}^-)  \no \\
&= \phi \phi^- (\ti F_{\ul r} - z \ti F_{\ul r}^-) + z\ti H_{\ul r} - \ti H_{\ul r}^- 
- z \beta(\ti G_{\ul r}^- + \ti K_{\ul r}) 
- \alpha \phi \phi^- (\ti G_{\ul r} + \ti K_{\ul r}^-).
\end{align}
Hence,
\begin{equation} \lb{AL4.40}
\Big(\frac{1}{\phi}(z \beta + \phi^-) + \frac{z}{\phi^-}(\beta - \phi)S^-\Big) 
\big(\phi_{t_{\ul r}} - i \ti H_{\ul r} - i \ti F_{\ul r} \phi^2 + i (\ti G_{\ul r} + \ti K_{\ul r})\phi\big)=0.
\end{equation}
Solving the first-order difference equation \eqref{AL4.40} then yields
\begin{align}
&\phi_{t_{\ul r}}(P,n,t_{\ul r})-i\ti F_{\ul r}(z,n,t_{\ul r})\phi(P,n,t_{\ul r})^2 \no \\
&\qquad +i(\ti G_{\ul r}(z,n,t_{\ul r})+\ti K_{\ul r}(z,n,t_{\ul r}))\phi(P,n,t_{\ul r}) 
-i\ti H_{\ul r}(z,n,t_{\ul r}) \no \\
&\quad=C(P,t_{\ul r})\begin{cases}
\prod_{n'=1}^n B(P,n',t_{\ul r})/A(P,n',t_{\ul r}), & \text{$n\ge 1$}, \\
1,& \text{$n=0$}, \\
\prod_{n'=n+1}^0 A(P,n',t_{\ul r})/B(P,n',t_{\ul r}), & \text{$n\le -1$} 
\end{cases} \lb{AL4.42}
\end{align}
for some $n$-independent function $C(\dott,t_{\ul r})$ meromorphic on
$\calK_p$, where
\begin{equation}
A=\phi^{-1}(z\beta+\phi^{-}), \quad B=-z(\phi^-)^{-1}(\beta-\phi).  
\end{equation}
The asymptotic behavior of $\phi(P,n,t_{\ul r})$ in \eqref{ALphi infty} then
yields (for $t_{\ul r}\in\bbR$ fixed)
\begin{equation}  \lb{AL4.44}
\frac{B(P)}{A(P)}\underset{P\to\Pinfp}{=}
-(1-\alpha\beta)(\beta^-)^{-1}z^{-1}+\Oh(z^{-2}).
\end{equation}
Since the left-hand side of \eqref{AL4.42} is of order $\Oh(z^{r_+})$ as
$P\to\Pinfp$, and $C$ is meromorphic, insertion of \eqref{AL4.44}
into \eqref{AL4.42}, taking $n\ge 1$ sufficiently large, then yields
a contradiction unless $C=0$.  This proves \eqref{ALphi_t}.

Proving \eqref{ALtime_AL} is equivalent to showing 
\begin{align} \label{ALpsi_1 t}
\psi_{1,t_{\ul r}} &= i (\ti G_{\ul r} - \phi \ti F_{\ul r}) \psi_1, \\ \label{ALpsi_2 t}
\psi_1  \phi_{t_{\ul r}} + \phi \psi_{1,t_{\ul r}} & = i (\ti H_{\ul r} - \phi \ti K_{\ul r}) \psi_1,
\end{align}
using \eqref{ALpsi t 1}.
Equation \eqref{ALpsi_2 t} follows directly from \eqref{ALpsi_1 t} and from 
\eqref{ALphi_t},
\begin{align}
\psi_1  \phi_{t_{\ul r}} + \phi \psi_{1,t_{\ul r}} &= 
\psi_1 \big(i \ti H_{\ul r} + i \ti F_{\ul r} \phi^2 - 
i (\ti G_{\ul r} + \ti K_{\ul r})\phi + i (\ti G_{\ul r} - \phi \ti F_{\ul r})\phi\big)  \no \\
&=i (\ti H_{\ul r} - \phi \ti K_{\ul r}) \psi_1.
\end{align}
To prove \eqref{ALpsi_1 t} we start from 
\begin{align}
& (z + \alpha \phi^-)_{t_{\ul r}} = \alpha_{t_{\ul r}} \phi^- + \alpha\phi_{t_{\ul r}}^-  \no \\
& \quad = \phi^- i \big(z \ti F_{\ul r}^- + \alpha (\ti G_{\ul r} + \ti K_{\ul r}^-)
- \ti F_{\ul r}\big) + \alpha i \big(\ti H_{\ul r}^- + \ti F_{\ul r}^- (\phi^-)^2 - 
(\ti G_{\ul r}^- + \ti K_{\ul r}^-)\phi^-\big)  \no \\
& \quad = i \alpha \phi^- (\ti G_{\ul r} - \ti G_{\ul r}^-) 
+ i (z + \alpha \phi^-) \phi^- \ti F_{\ul r}^-
- i \phi^- \ti F_{\ul r} + i \alpha \ti H_{\ul r}^-  \no \\
& \quad = i (z + \alpha \phi^-) \big(\ti G_{\ul r} - \phi \ti F_{\ul r} 
- (\ti G_{\ul r}^- - \phi^- \ti F_{\ul r}^-) \big), 
\end{align}
where we used \eqref{AL1,1 r} and \eqref{ALriccati} to rewrite
\begin{equation}
i \alpha \ti H_{\ul r}^- - i \phi^- \ti F_{\ul r} = i z (\ti G_{\ul r} - \ti G_{\ul r}^-)
- \alpha \phi \phi^- \ti F_{\ul r} - z \phi \ti F_{\ul r}.
\end{equation}
Abbreviating 
\begin{equation}
\sigma(P, n_0,t_{\ul r}) = i \int_0^{t_{\ul r}} ds \big(\ti G_{\ul r} (z,n_0,s) -  
\ti F_{\ul r}(z,n_0,s) \phi(P,n_0,s)\big),
\end{equation}
one computes for $n \geq n_0 +1$, 
\begin{align}
 \psi_{1,t_{\ul r}} &= \bigg(\exp(\sigma) 
\prod_{n'=n_0+1}^n (z+\alpha \phi^-)(n') \bigg)_{t_{\ul r}}   \no \\
&= \sigma_{t_{\ul r}} \psi_{1} + \exp(\sigma) \sum_{n'=n_0+1}^n (z+\alpha \phi^-)_{t_{\ul r}}(n')
\prod_{\substack{n''=1 \\ n'' \neq n'}}^n (z+\alpha \phi^-)(n'')  \no \\
&= \psi_1 \bigg(\sigma_{t_{\ul r}} + i \sum_{n'=n_0+1}^n \big((\ti G_{\ul r} -\ti F_{\ul r} \phi )(n') 
- (\ti G_{\ul r} - \ti F_{\ul r}\phi )(n'-1)\big) \bigg)  \no \\
&= i (\ti G_{\ul r} - \ti F_{\ul r} \phi) \psi_1.
\end{align}
The case $n\leq n_0$ is handled analogously establishing \eqref{ALpsi_1 t}.
 
That $\Psi(\dott,n,n_0,t_{\ul r},t_{0,\ul r})$ is meromorphic on 
$\calK_p\setminus\{\Pinfpm,\Pzpm\}$ 
if $F_{\ul p} (\dott,n_0,t_{\ul r})$ has only simple zeros distinct from zero is a consequence of \eqref{AL4.26},  \eqref{AL4.27}, and of 
\begin{equation}
-i \widetilde F_{\ul r}(z,n_{0},s)\phi(P,n_{0},s)
\underset{P\to\hat\mu_{j}(n_{0},s)}{=}
\partial_s\ln\big(F_{\ul p}(z,n_{0},s)\big)+\Oh(1),   \lb{AL4.56}
\end{equation}
using \eqref{AL4.20}, \eqref{AL4.24}, and \eqref{ALF_t}. 
(Equation \eqref{ALF_t} in Lemma \ref{lAL4.3} follows from \eqref{ALphi_t}, 
\eqref{ALphi 2t}, and \eqref{ALphi 3t} which have already been proven.) 
\end{proof}

Next we consider the $t_{\ul r}$-dependence of $F_{\ul p}$, $G_{\ul p}$, and $H_{\ul p}$.

\begin{lemma}  \lb{lAL4.3}
Assume Hypothesis \ref{hAL4.1} and suppose that 
\eqref{ALzctilde}, \eqref{ALzctstat} hold. In addition, let
$(z,n,t_{\ul r})\in\bbC\times\bbZ\times\bbR$. Then, 
\begin{align} \label{ALF_t}
F_{\ul p,t_{\ul r}} &= - 2 i G_{\ul p} \ti F_{\ul r} 
+ i \big(\ti G_{\ul r} + \ti K_{\ul r}\big)F_{\ul p},   \\ \label{ALG_t}
G_{\ul p,t_{\ul r}} &= i F_{\ul p} \ti H_{\ul r} - i H_{\ul p} \ti F_{\ul r}, \\  \label{ALH_t}
H_{\ul p,t_{\ul r}} &= 2 i G_{\ul p} \ti H_{\ul r} - i \big(\ti G_{\ul r} + \ti K_{\ul r}\big)H_{\ul p}.
\end{align}
In particular, \eqref{ALF_t}--\eqref{ALH_t} are equivalent to 
\begin{equation} \label{ALWp t}
V_{\ul p,t_{\ul r}} = \big[\ti V_{\ul r}, V_{\ul p}\big].
\end{equation}
\end{lemma}
\begin{proof}
To prove \eqref{ALF_t} one first differentiates equation \eqref{ALphi 3t}
\begin{equation}
\phi_{t_{\ul r}}(P) - \phi_{t_{\ul r}}(P^*) = - c_{0,+} z^{-p_-} y F_{\ul p}^{-2} F_{\ul p,t_{\ul r}}. 
\end{equation}
The time derivative of $\phi$ given in \eqref{ALphi_t} and \eqref{ALphi 2t} yield  
\begin{align}
\phi_{t_{\ul r}}(P) - \phi_{t_{\ul r}}(P^*) &= i \big(\ti H_{\ul r} + \ti F_{\ul r} \phi(P)^2 - 
\big(\ti G_{\ul r} + \ti K_{\ul r}\big)\phi(P) \big)   \no \\
& \quad - i \big(\ti H_{\ul r} + \ti F_{\ul r} \phi(P^*)^2 - 
\big(\ti G_{\ul r} + \ti K_{\ul r}\big)\phi(P^*) \big)   \no \\
&= i \ti F_{\ul r} (\phi(P) + \phi(P^*))(\phi(P) - \phi(P^*))   \no \\
& \quad - i \big(\ti G_{\ul r} + \ti K_{\ul r}\big)(\phi(P) - \phi(P^*))   \no \\
&= 2 i c_{0,+} z^{-p_-} \ti F_{\ul r} y G_{\ul p} F_{\ul p}^{-2} 
-  i c_{0,+} z^{-p_-} \big(\ti G_{\ul r} + \ti K_{\ul r}\big) y F_{\ul p}^{-1},
\end{align}
and hence 
\begin{equation}
F_{\ul p,t_{\ul r}} = - 2 i G_{\ul p} \ti F_{\ul r} + i \big(\ti G_{\ul r} + \ti K_{\ul r}\big)F_{\ul p}.
\end{equation}
Similarly, starting from \eqref{ALphi 2t}
\begin{equation}
\phi_{t_{\ul r}}(P) + \phi_{t_{\ul r}}(P^*) = 2 F_{\ul p}^{-2}(F_{\ul p}  G_{\ul p,t_{\ul r}} - F_{\ul p,t_{\ul r}} G_{\ul p})
\end{equation}
yields \eqref{ALG_t} and
\begin{equation}
0 = R_{\ul p, t_{\ul r}} = 2 G_{\ul p} G_{\ul p,t_{\ul r}} - F_{\ul p,t_{\ul r}} H_{\ul p}   - F_{\ul p}  H_{\ul p,t_{\ul r}}
\end{equation}
proves \eqref{ALH_t}.
\end{proof}

Next we turn to the Dubrovin equations for the time variation of the zeros $\mu_j$ of 
$(\dott)^{p_-}F_{\ul p}$ and $\nu_j$ of $(\dott)^{p_- -1}H_{\ul p}$ governed by the $\widetilde{\AL}_{\ul r}$ flow. 

\begin{lemma}  \lb{lAL4.5}
Assume Hypothesis \ref{hAL4.1} and suppose that \eqref{ALzctilde}, 
\eqref{ALzctstat} hold on $\bbZ\times \calI_{\mu}$ with 
$\calI_{\mu} \subseteq\bbR$ an open interval. In addition, assume that the
zeros $\mu_j$, $j=1,\dots,p$, of $(\dott)^{p_-}F_{\ul p}(\dott)$ remain distinct and nonzero on 
$\bbZ\times \calI_{\mu}$. Then $\{\hmu_j\}_{j=1,\dots,p}$, defined in 
\eqref{AL4.20}, satisfies the following first-order system of differential equations on $\bbZ\times \calI_{\mu}$, 
\begin{equation}
\mu_{j,t_{\ul r}} = - i \ti F_{\ul r}(\mu_j)  y(\hmu_j) (\alpha^+)^{-1} 
\prod_{\substack{k=1\\k\neq j}}^{p} (\mu_j-\mu_k)^{-1},  \quad 
j=1,\dots,p,   \lb{AL4.69a}
\end{equation}
with 
\begin{equation}
\hmu_j(n,\cdot)\in C^\infty(\calI_\mu,\calK_p), \quad j=1,\dots,p, 
\; n\in\bbZ.   \lb{AL4.69b}
\end{equation} 
For the zeros $\nu_j$, $j=1,\dots,p$, of $(\dott)^{p_- -1}H_{\ul p}(\dott)$, identical statements hold with $\mu_j$ and $\calI_{\mu}$ replaced by $\nu_j$ and $\calI_{\nu}$, etc. $($with 
$\calI_{\nu} \subseteq\bbR$ an open interval\,$)$. In particular, 
$\{\hat \nu_j\}_{j=1,\dots,p}$, defined in \eqref{AL4.21}, satisfies the first-order 
system on $\bbZ\times I_{\nu}$, 
\begin{equation}
\nu_{j,t_{\ul r}} = i \ti H_{\ul r}(\nu_j) y(\hnu_j) ( \beta \nu_j)^{-1} 
\prod_{\substack{k=1\\k\neq j}}^{p} (\nu_j-\nu_k)^{-1},  \quad 
j=1,\dots,p,   \lb{AL4.71a}
\end{equation}
with 
\begin{equation}
\hnu_j(n,\cdot)\in C^\infty(\calI_\nu,\calK_p), \quad j=1,\dots,p, 
\; n\in\bbZ.   \lb{AL4.71b}
\end{equation} 
\end{lemma}
\begin{proof}
It suffices to consider \eqref{AL4.69a} for $\mu_{j,t_{\ul r}}$. Using the product representation for $F_{\ul p}$ in \eqref{AL4.9} and employing \eqref{AL4.20} and 
\eqref{ALF_t}, one computes
\begin{align}
\begin{split}
F_{\ul p,t_{\ul r}}(\mu_j) & =\Bigg(c_{0,+} \alpha^+ \mu_j^{-p_-} 
\prod_{\substack{k=1 \\ k\neq j}}^{p}(\mu_j - \mu_k)\Bigg)\mu_{j,t_{\ul r}} 
= -2i G_{\ul p}(\mu_j) \ti F_{\ul r}(\mu_j)  \\
& = -i c_{0,+} \mu_j^{-p_-} y(\hmu_j) \ti F_{\ul r}(\mu_j),  \quad j=1,\dots,p,
\end{split}
\end{align}
proving \eqref{AL4.69a}. The case of \eqref{AL4.71a} for $\nu_{j,t_{\ul r}}$ is of course analogous using the product representation for $H_{\ul p}$ in \eqref{AL4.9} and employing  \eqref{AL4.21} and \eqref{ALH_t}. 
\end{proof}

When attempting to solve the Dubrovin systems \eqref{AL4.69a} and \eqref{AL4.71a}, they must be augmented with appropriate divisors 
$\calD_{\humu(n_0,t_{0,\ul r})}\in\sym^{p} \calK_p$, $t_{0,\ul r}\in \calI_\mu$, and 
$\calD_{\hunu(n_0,t_{0,\ul r})}\in\sym^{p} \calK_p$, $t_{0,\ul r}\in \calI_\nu$,  
as initial conditions.

Since the stationary trace formulas for $f_{\ell,\pm}$ and $h_{\ell,\pm}$ in terms of symmetric functions of the zeros $\mu_j$ and $\nu_k$ of $(\dott)^{p_-}F_{\ul p}$ and 
$(\dott)^{p_- -1}H_{\ul p}$ in 
Lemma \ref{lAL3.2} extend line by line to the corresponding time-dependent setting, we next record their $t_{\ul r}$-dependent analogs without proof. For simplicity we again confine ourselves to the simplest cases only. 

\begin{lemma}  \lb{lAL4.5a}
Assume Hypothesis \ref{hAL4.1} and suppose that \eqref{ALzctilde}, \eqref{ALzctstat} hold. Then, 
\begin{align} 
 \frac{\alpha}{\alpha^+}&= 
\prod_{j=1}^{p}\mu_j \bigg(\prod_{m=0}^{2p+1}E_m\bigg)^{-1/2},   \label{ALtr1t} \\ 
\frac{\beta^+}{\beta}&= 
\prod_{j=1}^{p}\nu_j \bigg(\prod_{m=0}^{2p+1}E_m\bigg)^{-1/2},   \label{ALtr2t} \\  
\sum_{j=1}^{p}\mu_j &= \alpha^+ \beta
- \gamma^+ \frac{\alpha^{++}}{\alpha^+} 
- \frac{c_{1,+}}{c_{0,+}},   \label{ALtr3t} \\
\sum_{j=1}^{p}\nu_j &= \alpha^+ \beta
- \gamma \frac{\beta^-}{\beta} 
- \frac{c_{1,+}}{c_{0,+}}.  \label{ALtr4t}
\end{align}
\end{lemma}

Next, we turn to the asymptotic expansions of $\phi$ and $\Psi$ in a neighborhood of 
$\Pinfpm$ and $\Pzpm$. 

\begin{lemma}  \lb{lAL4.6}
Assume Hypothesis \ref{hAL4.1} and suppose that \eqref{ALzctilde}, \eqref{ALzctstat} hold. Moreover, let $P=(z,y)\in\calK_p\setminus\{\Pinfp,\Pinfm,\Pzp,\Pzm\}$, 
$(n,n_0,t_{\ul r},t_{0,\ul r})\in\bbZ^2\times\bbR^2$. Then $\phi$ has the asymptotic behavior 
\begin{align}  
\phi(P) \underset{\zeta\to 0}{=}&  \begin{cases} 
\beta + \beta^-\gamma \zeta + \Oh(\zeta^2), & \quad  P \rightarrow P_{\infty_+}, \\
- (\alpha^+)^{-1} \zeta^{-1} + (\alpha^+)^{-2}\alpha^{++}\gamma^+
+ \Oh(\zeta), & \quad  P \rightarrow P_{\infty_-}, 
\end{cases}
\quad \zeta=1/z,  \label{ALphi inftyt} \\ 
\phi(P) \underset{\zeta\to 0}{=}&  \begin{cases} 
\alpha^{-1} - \alpha^{-2} \alpha^-\gamma \zeta  + \Oh(\zeta^2), 
& \quad P \rightarrow P_{0,+}, \\  
- \beta^+ \zeta - \beta^{++}\gamma^+ \zeta^2 + \Oh(\zeta^3), & \quad P \rightarrow P_{0,-},
\end{cases}
\quad \zeta=z.   \label{ALphi zerot}
\end{align}
The component $\psi_1$ of the Baker--Akhiezer vector $\Psi$ has the asymptotic behavior 
\begin{align} 
& \psi_1(P,n,n_0,t_{\ul r},t_{0,\ul r}) \underset{\zeta\to 0}{=} 
\exp\bigg(\pm \frac{i}{2}(t_{\ul r}-t_{0,\ul r})\sum_{s=0}^{r_+} \tilde c_{r_+ -s,+} 
\zeta^{-s} \bigg)
(1+\Oh(\zeta))  \no \\
&\quad \times\begin{cases}
\zeta^{n_0-n}, &  P\to\Pinfp, \\[1mm]
\begin{matrix}\pgam(n,n_0,t_{\ul r})\frac{\alpha^+(n,t_{\ul r})}{\alpha^+(n_0,t_{0,\ul r})}\\
\times\exp\Big(i \int_{t_{0,\ul r}}^{t_{\ul r}} ds 
\big(\tilde g_{r_+,+}(n_0,s) - \tilde g_{r_-,-}(n_0,s)\big)\Big),
\end{matrix} &P\to\Pinfm, 
\end{cases} 
\quad \zeta=1/z,  \label{ALpsi infty t}\\
& \psi_1(P,n,n_0,t_{\ul r},t_{0,\ul r}) \underset{\zeta\to 0}{=}
\exp\bigg(\pm \frac{i}{2}(t_{\ul r}-t_{0,\ul r})\sum_{s=0}^{r_-} \tilde c_{r_- -s,-}\zeta^{-s} \bigg)
(1 + \Oh(\zeta)) \no \\
&\quad \times\begin{cases}
\frac{\alpha(n,t_{\ul r})}{\alpha(n_0,t_{0,\ul r})}, & P\to P_{0,+}, \\[1mm]
\begin{matrix} \pgam(n,n_0,t_{\ul r})\zeta^{n-n_0} \\
\times \exp\Big(i \int_{t_{0,\ul r}}^{t_{\ul r}} ds 
\big(\tilde g_{r_+,+}(n_0,s) - \tilde g_{r_-,-}(n_0,s)\big)\Big),
\end{matrix} &P\to P_{0,-}, 
\end{cases} 
\quad \zeta=z.  \label{ALpsi 0 t}
\end{align} 
\end{lemma}
\begin{proof} 
Since by the definition of $\phi$ in \eqref{AL4.22} the time parameter $t_{\ul r}$ can be viewed as an additional but fixed parameter, the asymptotic behavior of $\phi$ remains the same as in Lemma \ref{lAL3.3}. Similarly, also the asymptotic behavior of 
$\psi_1(P,n,n_0,t_{\ul r},t_{\ul r})$ is derived in an identical fashion to that in Lemma \ref{lAL3.3}.
This proves \eqref{ALpsi infty t} and \eqref{ALpsi 0 t} for $t_{0,\ul r}=t_{\ul r}$, that is, 
\begin{align}
\psi_1(P,n,n_0,t_{\ul r},t_{\ul r}) 
&\underset{\zeta\to 0}{=} \begin{cases}
\zeta^{n_0-n}(1+\Oh(\zeta)), & P\to\Pinfp, \\
\pgam(n,n_0,t_{\ul r})\frac{\alpha^+(n,t_{\ul r})}{\alpha^+(n_0,t_{\ul r})}+\Oh(\zeta), 
&  P\to\Pinfm, \end{cases}  
\qquad \zeta=1/z,    \\
\psi_1(P,n,n_0,t_{\ul r},t_{\ul r})
&\underset{\zeta\to 0}{=} \begin{cases}
\frac{\alpha(n,t_{\ul r})}{\alpha(n_0,t_{\ul r})} +\Oh(\zeta), &  P\to\Pzp, \\
\pgam(n,n_0,t_{\ul r})\zeta^{n-n_0}(1+\Oh(\zeta)), 
& P\to\Pzm, \end{cases}  
\qquad \zeta=z.   
\end{align}
It remains to investigate 
\begin{equation}
\psi_1(P,n_0,n_0,t_{\ul r},t_{0,\ul r}) = \exp \bigg(i \int_{t_{0,\ul r}}^{t_{\ul r}} dt 
\big(\ti G_{\ul r} (z,n_0,t) - \ti F_{\ul r}(z,n_0,t) \phi(P,n_0,t)\big) \bigg).   \lb{AL4.83A}
\end{equation}
The asymptotic expansion of the integrand is derived using Theorem \ref{tALB.2}. 
Focusing on the homogeneous coefficients first, one computes as 
$P \to P_{\infty_{\pm}}$,
\begin{align}
&\hatt G_{s,+} - \hatt F_{s,+} \phi = \hatt G_{s,+} - \hatt F_{s,+} 
\frac{G_{\ul p} + (c_{0,+}/2) z^{-p_-}y}{F_{\ul p}}   \no \\
& \quad =
\hatt G_{s,+} - \hatt F_{s,+} \bigg(\f{2z^{p_-}}{c_{0,+}} \frac{G_{\ul p}}{y} + 1 \bigg)
\bigg(\f{2z^{p_-}}{c_{0,+}} \frac{F_{\ul p}}{y}\bigg)^{-1}   \no \\
& \;\; \underset{\zeta\to 0}{=}
\pm \frac{1}{2} \zeta^{-s} + \frac{\hat g_{0,+} \mp \tfrac{1}{2}}{\hat f_{0,+}}
\hat f_{s,+} + \Oh(\zeta), \quad P \to P_{\infty_{\pm}}, \; \zeta=1/z.
\end{align}
Since
\begin{equation}
\ti F_{\ul r} \underset{\zeta\to 0}{=} \sum_{s=0}^{r_+} \tilde c_{r_+ -s,+}\hatt F_{s,+} 
+ \Oh(\zeta),  \quad
\ti G_{\ul r} \underset{\zeta\to 0}{=} \sum_{s=0}^{r_+} \tilde c_{r_+ -s,+}\hatt G_{s,+} 
+ \Oh(\zeta),
\end{equation}
one infers from \eqref{ALphi inftyt}
\begin{equation}
\ti G_{\ul r} - \ti F_{\ul r} \phi \underset{\zeta\to 0}{=}
\f{1}{2}\sum_{s=0}^{r_+} \tilde c_{r_+ -s,+} \zeta^{-s} + \Oh(\zeta),
\quad  P \to P_{\infty_+}, \; \zeta=1/z.    \lb{AL4.85a}
\end{equation}
Insertion of \eqref{AL4.85a} into \eqref{AL4.83A} then proves \eqref{ALpsi infty t} as
$P\to\Pinfp$.

As $P \to P_{\infty_-}$, we need one additional term in the asymptotic expansion of
$\ti F_{\ul r}$, that is, we will use
\begin{equation}
\ti F_{\ul r} \underset{\zeta\to 0}{=} \sum_{s=0}^{r_+} \tilde c_{r_+ -s,+}\hatt F_{s,+}
 + \sum_{s=0}^{r_-} \tilde c_{r_- -s,-} \hat f_{s-1,-} \zeta + \Oh(\zeta^2).
\end{equation}
This then yields
\begin{equation}
\ti G_{\ul r} - \ti F_{\ul r} \phi \underset{\zeta\to 0}{=}
-\f{1}{2}\sum_{s=0}^{r_+} \tilde c_{r_+ -s,+} \zeta^{-s} 
- (\alpha^+)^{-1}(\tilde f_{r_+,+} - \tilde f_{r_- -1,-})
+ \Oh(\zeta).  \lb{AL4.85b}
\end{equation}
Invoking \eqref{ALf_l+} and  \eqref{ALal_ivp} one concludes that 
\begin{equation}
\tilde f_{r_- -1,-} - \tilde f_{r_+,+}  = -i \alpha_{t_{\ul r}}^+  
+ \alpha^+ (\tilde g_{r_+,+} - \tilde g_{r_-,-}) \lb{AL4.85c}
\end{equation}
and hence 
\begin{equation}
\ti G_{\ul r} - \ti F_{\ul r} \phi \underset{\zeta\to 0}{=}
-\f{1}{2}\sum_{s=0}^{r_+} \tilde c_{r_+ -s,+} \zeta^{-s} 
 - \f{i \alpha_{t_{\ul r}}^+}{\alpha^+}  + \tilde g_{r_+,+} - \tilde g_{r_-,-} + \Oh(\zeta).  
 \lb{AL4.85d}
\end{equation}
Insertion of \eqref{AL4.85d} into \eqref{AL4.83A} then proves 
\eqref{ALpsi infty t} as $P\to\Pinfm$.

Using Theorem \ref{tALB.2} again, one obtains in the same manner as
$P \to P_{0,\pm}$,
\begin{equation} \label{AL4.87a}
\hatt G_{s,-} - \hatt F_{s,-} \phi \underset{\zeta\to 0}{=}  \pm \frac{1}{2} \zeta^{-s} - \hat g_{s,-} +
\frac{\hat g_{0,-} \pm \tfrac{1}{2}}{\hat f_{0,-}}
\hat f_{s,-} + \Oh(\zeta).
\end{equation}
Since
\begin{align}
\ti F_{\ul r} & \underset{\zeta\to 0}{=} \sum_{s=0}^{r_-} \tilde c_{r_- -s,-}
\hatt F_{s,-}  + \tilde f_{r_+ -1,+} + \Oh(\zeta),
\quad P \to \Pzpm, \; \zeta=z,   \lb{AL4.88A } \\
\ti G_{\ul r} & \underset{\zeta\to 0}{=} \sum_{s=0}^{r_-} \tilde c_{r_- -s,-}
\hatt G_{s,-} + \tilde g_{r_+,+} + \Oh(\zeta),
\quad P \to \Pzpm, \; \zeta=z,   \lb{AL4.88B}
\end{align}
\eqref{AL4.87a}--\eqref{AL4.88B} yield
\begin{equation}
\ti G_{\ul r} - \ti F_{\ul r}  \phi \underset{\zeta\to 0}{=}
\pm \frac{1}{2} \sum_{s=0}^{r_-} \tilde c_{r_- -s,-} \zeta^{-s} + \tilde g_{r_+,+} 
- \tilde g_{r_-,-}
- \frac{\hat g_{0,-} \pm \tfrac{1}{2}}{\hat f_{0,-}}
(\tilde f_{r_+ -1,+} - \tilde f_{r_-,-}) + \Oh(\zeta),   
\end{equation}
where we again used \eqref{ALphi zerot}, \eqref{ALhat f}, and \eqref{ALal_ivp}. 
As $P\to\Pzm$, one thus obtains
\begin{align}
\ti G_{\ul r} - \ti F_{\ul r}  \phi \underset{\zeta\to 0}{=}
- \frac{1}{2} \sum_{s=0}^{r_-} \tilde c_{r_- -s,-} \zeta^{-s} + \tilde g_{r_+,+} 
- \tilde g_{r_-,-},  \quad P \to \Pzm, \; \zeta=z.  \lb{AL4.95}
\end{align}
Insertion of \eqref{AL4.95} into \eqref{AL4.83A} then proves \eqref{ALpsi 0 t} 
as $P\to\Pzm$.

As $P\to\Pzp$, one obtains  
\begin{align}
\ti G_{\ul r} - \ti F_{\ul r}  \phi  & \underset{\zeta\to 0}{=}
\frac{1}{2} \sum_{s=0}^{r_-} \tilde c_{r_- -s,-} \zeta^{-s} + \tilde g_{r_+,+} 
- \tilde g_{r_-,-}
- \frac{1}{\alpha} (\tilde f_{r_+ -1,+} - \tilde f_{r_-,-}) + \Oh(\zeta),   \no \\
& \underset{\zeta\to 0}{=}
\frac{1}{2} \sum_{s=0}^{r_-} \tilde c_{r_- -s,-} \zeta^{-s} 
- \frac{i \alpha_{t_{\ul r}}}{\alpha} + \Oh(\zeta),   \quad P \to \Pzp, \; \zeta=z,  \lb{AL4.96}
\end{align}
using $\tilde f_{r_-,-}=\tilde f_{r_- -1,-}^- + \alpha(\tilde g_{r_-,-} - \tilde g_{r_-,-}^-)$ 
(cf.\ \eqref{ALf_l-}) and \eqref{ALal_ivp}. 
Insertion of \eqref{AL4.96} into \eqref{AL4.83A} then proves \eqref{ALpsi 0 t} 
as $P\to\Pzp$. 
\end{proof}

Next, we note that Lemma \ref{lAL3.4} on nonspecial divisors in the stationary context extends to the present time-dependent situation without a change. Indeed, since $t_{\ul r} \in \bbR$ just plays the role of a parameter, the proof of Lemma 
\ref{lAL3.4} extends line by line and is hence omitted.

\begin{lemma} \label{lAL4.6a} 
Assume Hypothesis \ref{hAL4.1} and suppose that \eqref{ALzctilde}, \eqref{ALzctstat} hold. Moreover, let $(n,t_{\ul r})\in\bbZ\times\bbR$. Denote by $\calD_{\humu}$,
$\humu=\{\hmu_1,\dots,\hmu_{p}\}$ and $\calD_{\hunu}$,
$\hunu=\{\hnu_1,\dots,\hnu_{p}\}$, the pole and zero divisors of degree
$p$, respectively, associated with $\alpha$, $\beta$, and $\phi$ defined
according to \eqref{AL4.20} and \eqref{AL4.21}, that is,
\begin{align}
\hat\mu_j (n,t_{\ul r}) &= (\mu_j (n,t_{\ul r}),(2/c_{0,+}) \mu_j(n,t_{\ul r})^{p_-} 
G_{\ul p}(\mu_j(n,t_{\ul r}),n,t_{\ul r})), 
\quad  j=1,\dots,p, \\
\hat\nu_j (n,t_{\ul r}) &= (\nu_j (n,t_{\ul r}),- (2/c_{0,+}) \nu_j(n,t_{\ul r})^{p_-} 
G_{\ul p}(\nu_j(n,t_{\ul r}),n,t_{\ul r})), 
\quad  j=1,\dots,p. 
\end{align}
Then $\calD_{\humu(n,t_{\ul r})}$ and $\calD_{\hunu(n,t_{\ul r})}$ are nonspecial for all
$(n,t_{\ul r})\in\bbZ\times\bbR$.
\end{lemma}

\medskip

Finally, we note that 
\begin{align}
\begin{split}
& \pgam(n,n_0,t_{\ul r})= \pgam(n,n_0,t_{0,\ul r})\\
& \quad  \times\exp\bigg(i \int_{t_{0,\ul r}}^{t_{\ul r}} ds  
\big(\tilde g_{r_+,+}(n,s)-\tilde g_{r_+,+}(n_0,s) -
\tilde g_{r_-,-}(n,s)+\tilde g_{r_-,-}(n_0,s)\big)\bigg),  \lb{AL4.128}
\end{split}
\end{align}
which follows from \eqref{AL2.14}, \eqref{ALpgam}, and from 
\begin{align}
& \pgam(n,n_0,t_{\ul r})_{t_{\ul r}}=\sum_{j=n_0+1}^n \gam(j,t_{\ul r})_{t_{\ul r}} 
\prod_{\substack{k=n_0+1\\k\neq j}}^n\gam(j,t_{\ul r})  \\
& \quad = i \big(\tilde g_{r_+,+}(n,t_{\ul r})-\tilde g_{r_+,+}(n_0,t_{\ul r}) 
-\tilde g_{r_-,-}(n,t_{\ul r})+\tilde g_{r_-,-}(n_0,t_{\ul r})\big)\pgam(n,n_0,t_{\ul r})  \no 
\end{align}
after integration with respect to $t_{\ul r}$.

The results of Sections \ref{sAL2}--\ref{sAL4} have been used extensively in 
\cite{GesztesyHoldenMichorTeschl:2007} to derive the class of time-dependent algebro-geometric solutions of the Ablowitz--Ladik hierarchy and the associated theta function representations of $\alpha$, $\beta$, $\phi$, and $\Psi$. 
These theta function representations also show that 
$\gamma(n,t_{\ul r})\notin \{0,1\}$ for all $(n,t_{\ul r})\in\bbZ\times\bbR$, and hence  condition \eqref{AL4.1A} is satisfied for the time-dependent algebro-geometric AL solutions discussed in this section, provided the associated divisors 
$\calD_{\humu(n,t_{\ul r})}$ and $\calD_{\hunu(n,t_{\ul r})}$ stay away from 
$\Pinfpm, \Pzpm$ for all $(n,t_{\ul r})\in\bbZ\times\bbR$.

\appendix
\section{Asymptotic Spectral Parameter Expansions  and Nonlinear
Recursion Relations} \lb{ALApp.high}
\renewcommand{\theequation}{A.\arabic{equation}}
\renewcommand{\thetheorem}{A.\arabic{theorem}}
\setcounter{theorem}{0}
\setcounter{equation}{0}

In this appendix we consider asymptotic spectral parameter expansions of
$F_{\ul p}/y$, $G_{\ul p}/y$, and $H_{\ul p}/y$ in a neighborhood of $\Pinfpm$ and $\Pzpm$, 
the resulting recursion relations for
the homogeneous coefficients  $\hat f_\ell$, $\hat g_\ell$, and $\hat
h_\ell$, their connection with the nonhomogeneous coefficients $f_\ell$,
$g_\ell$, and $h_\ell$, and the connection between $c_{\ell,\pm}$ and
$c_{\ell}(\ul E^{\pm 1})$. We will employ the notation
\begin{equation}
 \ul E^{\pm 1}=\big(E_0^{\pm 1},\dots,E_{2p+1}^{\pm 1}\big).   \lb{ALEpm}
\end{equation}

We start with the following elementary results (consequences of the  binomial
expansion) assuming $\eta\in\bbC$ such that
$|\eta|<\min\{|E_0|^{-1},\dots, |E_{2p+1}|^{-1}\}$:
\begin{equation}
\left(\prod_{m=0}^{2p+1} \big(1-{E_m}{\eta}\big)
\right)^{-1/2}=\sum_{k=0}^{\infty}\hat c_k(\ul
E)\eta^{k}, \lb{ALB2.26d}
\end{equation}
where
\begin{align}
\hat c_0(\ul E)&=1,\no \\
\hat c_k(\ul E)&=\!\!\!\!\!\!\!\sum_{\substack{j_0,\dots,j_{2p+1}=0\\
       j_0+\cdots+j_{2p+1}=k}}^{k}\!\!
\f{(2j_0)!\cdots(2j_{2p+1})!}
{2^{2k} (j_0!)^2\cdots (j_{2p+1}!)^2}E_0^{j_0}\cdots
E_{2p+1}^{j_{2p+1}}, \quad  k\in\bbN. \label{ALB2.26e}
\end{align}
The first few coefficients explicitly read
\begin{equation}
\hat c_0(\ul E)=1, \;
\hat c_1(\ul E)=\f12\sum_{m=0}^{2p+1} E_m,
\; \hat c_2(\ul E)=\f14\sum_{\substack{m_1,m_2=0\\ m_1< m_2}}^{2p+1}
E_{m_1} E_{m_2}+\f38 \sum_{m=0}^{2p+1} E_m^2,
\quad \text{etc.} \lb{ALB2.26f}
\end{equation}
Similarly,
\begin{equation}
\left(\prod_{m=0}^{2p+1} \big(1-{E_m}{\eta}\big)
\right)^{1/2}=\sum_{k=0}^{\infty}c_k(\ul
E)\eta^{k}, \lb{ALB2.26g}
\end{equation}
where
\begin{align}
c_0(\ul E)&=1,\no \\
c_k(\ul E)&=\!\!\!\!\!\!\!\!\sum_{\substack{j_0,\dots,j_{2p+1}=0\\
       j_0+\cdots+j_{2p+1}=k}}^{k}\!\!
\f{(2j_0)!\cdots(2j_{2p+1})!\, E_0^{j_0}\cdots E_{2p+1}^{j_{2p+1}}}
{2^{2k} (j_0!)^2\cdots (j_{2p+1}!)^2 (2j_0-1)\cdots(2j_{2p+1}-1)},
\quad k\in\bbN.  \label{ALB2.26h}
\end{align}
The first few coefficients explicitly are given by
\begin{align}
c_0(\ul E)=1, \;
c_1(\ul E)=-\f12\sum_{m=0}^{2p+1} E_m, \;
c_2(\ul E)=\f14\sum_{\substack{m_1,m_2=0\\ m_1< m_2}}^{2p+1}
E_{m_1} E_{m_2}-\f18 \sum_{m=0}^{2p+1} E_m^2,
\quad \text{etc.} \lb{ALB2.26i}
\end{align}
Multiplying \eqref{ALB2.26d} and \eqref{ALB2.26g} and comparing coefficients of 
$\eta^k$ one finds
\begin{equation}
\sum_{\ell=0}^{k} \hat c_{k-\ell}(\ul E) c_\ell(\ul E)=\delta_{k,0}, \quad k\in\bbN_0. \lb{ALB2.26j}
\end{equation}

Next, we turn to asymptotic expansions of various quantities in the case of the 
Ablowitz--Ladik hierarchy assuming $\alpha, \beta \in \bbC^{\bbZ}$, 
$\alpha(n)\beta(n) \notin \{0,1\}$, 
$n\in\bbZ$. Consider a fundamental system of solutions 
$\Psi_{\pm}(z,\dott)=(\psi_{1,\pm}(z,\dott), \psi_{2,\pm}(z,\dott))^\top$ of 
$U(z)\Psi^-_{\pm}(z) = \Psi_{\pm}(z)$ for $z\in\bbC$ (or in some subdomain of $\bbC$), with $U$ given by \eqref{AL2.03}, such that
\begin{equation}
\det(\Psi_-(z),\Psi_+(z)) \neq 0.     \lb{ALpsipm}
\end{equation}
Introducing
\begin{equation}
\phi_{\pm} (z,n) = \f{\psi_{2,\pm}(z,n)}{\psi_{1,\pm}(z,n)}, \quad z\in\bbC, \; n\in\bbN, 
\lb{ALphipm}
\end{equation}
then $\phi_{\pm}$ satisfy the Riccati-type equation
\begin{equation}
\alpha \phi_{\pm}\phi^-_{\pm} - \phi^-_{\pm} +z\phi_{\pm} = z \beta,  \lb{ALRicc}
\end{equation}
and one introduces in addition,
\begin{align}
\frf&= \f{2}{\phi_+ - \phi_-},   \lb{ALgothf}  \\
\frg&= \f{\phi_+ + \phi_-}{\phi_+ - \phi_-},    \lb{ALgothg}  \\
\frh&= \f{2\phi_+  \phi_-}{\phi_+ - \phi_-}.    \lb{ALgothh}
\end{align}
Using the Riccati-type equation \eqref{ALRicc} and its consequences,
\begin{align}
\alpha(\phi_+\phi^-_+ - \phi_- \phi^-_-) - (\phi^-_+ - \phi_-^-) + 
z (\phi_+ - \phi_-) &= 0,   \lb{ALRicc1} \\
\alpha(\phi_+\phi^-_+ + \phi_- \phi^-_-) - (\phi^-_+ + \phi_-^-) + 
z (\phi_+ + \phi_-) &=2z \beta,   \lb{ALRicc2}
\end{align}
one derives the identities
\begin{align} 
 z(\frg^- - \frg) + z \beta \frf +\alpha \frh^- & =0,    \lb{ALlin1} \\
 z \beta \frf^- +\alpha \frh - \frg + \frg^- & =0,    \lb{ALlin2} \\
- \frf + z \frf^- + \alpha (\frg + \frg^-) & =0,     \lb{ALlin3} \\
z \beta (\frg^- + \frg) -z \frh + \frh^- & =0,      \lb{ALlin4} \\
\frg^2 -\frf \frh & = 1.  \lb{ALquadr}
\end{align}
Moreover, \eqref{ALlin1}--\eqref{ALlin4} and \eqref{ALquadr} also permit one to derive nonlinear difference equations for $\frf$, $\frg$, and $\frh$ separately, and one obtains
\begin{align} \no
&\big((\alpha^+ + z \alpha)^2 \frf - z(\alpha^+)^2 \gam \frf^-\big)^2
-2 z \alpha^2 \gam^+ 
\big((\alpha^+ + z \alpha)^2 \frf + z(\alpha^+)^2 \gam \frf^-\big) \frf^+
\\ \label{ALquadrf}
& \qquad + z^2 \alpha^4 (\gam^+)^2 (\frf^+)^2
=4(\alpha\alpha^+)^2(\alpha^+ +\alpha z)^2, \\[1mm] 
& (\alpha^+ + z \alpha) (\beta+ z \beta^+) (z + \alpha^+ \beta)
(1 + z \alpha \beta^+) \frg^2\no \\ \no
& \qquad + z (\alpha^+ \gam \frg^- + z \alpha \gam^+ \frg^+)
(z  \beta^+ \gam \frg^- + \beta \gam^+ \frg^+)\\ \no
& \qquad - z \gam \big((\alpha^+ \beta + z^2 \alpha \beta^+) (2-\gam^+) + 2z (1-\gam^+) (2-\gam)\big) \frg^- \frg  \\ \no
& \qquad - z \gam^+ \big(2 z (1-\gam) (2-\gam^+) + (\alpha^+ \beta + z^2 \alpha 
\beta^+) (2-\gam)\big) \frg^+ \frg   \\ \label{ALquadrg}
&\quad = (\alpha^+ \beta - z^2 \alpha \beta^+)^2,  \\[1mm]
& z^2 \big((\beta^+)^2 \gam \frh^- - \beta^2 \gam^+ \frh^+\big)^2
- 2 z (\beta+ z \beta^+)^2 \big((\beta^+)^2 \gam \frh^- 
+ \beta^2 \gam^+ \frh^+\big) \frh    \no \\ \label{ALquadrh}
& \qquad + (\beta+ z \beta^+)^4 \frh^2
=4z^2(\beta\beta^+)^2 (\beta+\beta^+ z)^2.
\end{align}

For the precise connection between $\frf$, $\frg$, $\frh$ and the Green's function of the Lax difference expression underlying the AL hierarchy, we refer to 
\cite[App.\ C]{GesztesyHolden:2005}, \cite{GesztesyHoldenMichorTeschl:2007b}.

Next, we assume the existence of the following asymptotic expansions of $\frf$, $\frg$, and $\frh$ near $1/z=0$ and $z=0$. More precisely, near $1/z=0$ we assume that  
\begin{align}
\begin{split}
& \frf (z) \underset{\substack{|z|\to\infty\\ \, z\in C_R}}{=} 
- \sum_{\ell=0}^\infty\hat \frf_{\ell,+} z^{-\ell-1},  \quad 
\frg (z) \underset{\substack{|z|\to\infty\\ \, z\in C_R}}{=} 
- \sum_{\ell=0}^\infty\hat \frg_{\ell,+} z^{-\ell}, \\
& \frh (z) \underset{\substack{|z|\to\infty\\ \, z\in C_R}}{=} 
- \sum_{\ell=0}^\infty\hat \frh_{\ell,+} z^{-\ell},   \lb{ALasympfghinf}
\end{split}
\end{align}
for $z$ in some cone $C_R$ with apex at $z=0$ and some opening angle  in $(0,2\pi]$, 
exterior to a disk centered at $z=0$ of sufficiently large radius $R>0$, for some set of coefficients $\hat \frf_{\ell,+}$, $\hat \frg_{\ell,+}$, and $\hat\frh_{\ell,+}$, $\ell\in\bbN_0$. Similarly, near $z=0$ we assume that
\begin{align}
\begin{split}
& \frf (z) \underset{\substack{|z|\to 0 \\ \, z\in C_r}}{=} 
- \sum_{\ell=0}^\infty\hat \frf_{\ell,-} z^{\ell},  \quad 
\frg (z) \underset{\substack{|z|\to 0 \\ \, z\in C_r}}{=} 
- \sum_{\ell=0}^\infty\hat \frg_{\ell,-} z^{\ell}, \\
& \frh (z) \underset{\substack{|z|\to 0 \\ \, z\in C_r}}{=} 
- \sum_{\ell=0}^\infty\hat \frh_{\ell,-} z^{\ell+1},   \lb{ALasympfghzero}
\end{split}
\end{align}
for $z$ in some cone $C_r$ with apex at $z=0$ and some opening angle  in $(0,2\pi]$, interior to a disk centered at $z=0$ of sufficiently small radius $r>0$, for some set of coefficients $\hat \frf_{\ell,-}$, $\hat \frg_{\ell,-}$, and $\hat\frh_{\ell,-}$, $\ell\in\bbN_0$. 
Then one can prove the following result. 

\begin{theorem} \lb{tALB.2A}
Assume $\alpha, \beta \in \bbC^{\bbZ}$, $\alpha(n)\beta(n)\notin \{0,1\}$, $n\in\bbZ$, and the existence of the asymptotic expansions \eqref{ALasympfghinf} and 
\eqref{ALasympfghzero}. Then $\frf$, $\frg$, and $\frh$ have the following asymptotic  expansions as $|z|\to\infty$, $z\in C_R$, respectively, $|z|\to 0$, $z\in C_r$, 
\begin{align}
\begin{split}
& \frf (z) \underset{\substack{|z|\to\infty\\ \, z\in C_R}}{=} 
- \sum_{\ell=0}^\infty\hat f_{\ell,+} z^{-\ell-1},  \quad 
\frg (z) \underset{\substack{|z|\to\infty\\ \, z\in C_R}}{=} 
- \sum_{\ell=0}^\infty\hat g_{\ell,+} z^{-\ell}, \\
& \frh (z) \underset{\substack{|z|\to\infty\\ \, z\in C_R}}{=} 
- \sum_{\ell=0}^\infty\hat h_{\ell,+} z^{-\ell},   \lb{ALasymptotfghinf}
\end{split}
\end{align}
and 
\begin{align}
\begin{split}
& \frf (z) \underset{\substack{|z|\to 0 \\ \, z\in C_r}}{=} 
- \sum_{\ell=0}^\infty\hat f_{\ell,-} z^{\ell},  \quad 
\frg (z) \underset{\substack{|z|\to 0 \\ \, z\in C_r}}{=} 
- \sum_{\ell=0}^\infty\hat g_{\ell,-} z^{\ell}, \\
& \frh (z) \underset{\substack{|z|\to 0 \\ \, z\in C_r}}{=} 
- \sum_{\ell=0}^\infty\hat h_{\ell,-} z^{\ell+1},   \lb{ALasymptotfghzero}
\end{split}
\end{align}
where $\hat f_{\ell,\pm}$, $\hat g_{\ell,\pm}$, and $\hat h_{\ell,\pm}$ are the homogeneous versions of the coefficients $f_{\ell,\pm}$, $g_{\ell,\pm}$, and 
$h_{\ell,\pm}$ defined in \eqref{AL2.04a}--\eqref{AL2.04c}. In particular, 
$\hat f_{\ell,\pm}$, $\hat g_{\ell,\pm}$, and $\hat h_{\ell,\pm}$
can be computed from the following nonlinear recursion relations\footnote{We recall, a sum is interpreted as zero whenever the upper limit in the sum is strictly less than its lower limit.} 
\begin{align}
&\hat f_{0,+}=-\alpha^+,\quad \hat f_{1,+}
= (\alpha^+)^2\beta-\gamma^+\alpha^{++},  \no \\
&\hat f_{2,+}=- (\alpha^{+})^3 \beta^2+\gamma (\alpha^{+})^2\beta^- 
+\gamma^+\big((\alpha^{++})^2\beta^+ - \gamma^{++}\alpha^{+++} + 2\alpha^{+}\alpha^{++}\beta  \big), \no \\
&\alpha^4\alpha^+\hat f_{\ell,+}=\f12\bigg((\alpha^+)^4\sum_{m = 0}^{\ell-4}{\hat f}_{m,+} {\hat f}_{\ell-m-4,+} +\alpha^4\sum_{m = 1}^{\ell-1} {\hat f}_{m,+}  {\hat f}_{\ell-m,+} \no \\
& \quad- 2 (\alpha^+)^2\sum_{m = 0}^{\ell-3}{\hat f}_{m,+}
\big(-2\alpha \alpha^+ {\hat f}_{\ell-m-3,+} + (\alpha^+)^2\gamma  {\hat f}^-_{\ell-m-3,+}
+\alpha^2  \gamma^+ {\hat f}^+_{\ell-m-3,+} \big)\no \\
& \quad+\sum_{m = 0}^{\ell-2}\big(\alpha^4 (\gamma^+)^2
 {\hat f}^+_{m,+} {\hat f}^+_{\ell-m-2,+} + (\alpha^+)^2\gamma {\hat f}^-_{m,+}((\alpha^+)^2\gamma 
{\hat f}^-_{\ell-m-2,+}  \no \\
&  \hspace*{1.2cm}  - 2\alpha^2 \gamma^+ {\hat f}^+_{\ell-m-2,+})\no \\
 &\quad  \; \; - 2\alpha \alpha^+{\hat f}_{m,+}(-3\alpha \alpha^+  {\hat f}_{\ell-m-2,+} 
 + 2(\alpha^+)^2\gamma {\hat f}^-_{\ell-m-2,+} 
 + 2\alpha^2 \gamma^+ {\hat f}^+_{\ell-m-2,+})\big)\no \\
& \quad - 2\alpha^2\sum_{m = 0}^{\ell-1}{\hat f}_{m,+}\big(-2\alpha \alpha^+ {\hat f}_{\ell-m-1,+} 
 + (\alpha^+)^2\gamma {\hat f}^-_{\ell-m-1,+} + \alpha^2 \gamma^+{\hat f}^+_{\ell-m-1,+}\big)\bigg),  \no \\
& \hspace*{10cm} \ell \geq 3,   \lb{AL2.206aX} \\
&\hat f_{0,-}= \alpha,\quad \hat f_{1,-}=\gamma\alpha^{-}- \alpha^2\beta^+ , \no \\
&\hat f_{2,-} = \alpha^3 (\beta^+)^2-\gamma^+ \alpha^2\beta^{++} 
-\gamma \big((\alpha^{-})^2\beta - \gamma^{-}\alpha^{--} + 2\alpha^{-}\alpha\beta^+  \big), \no \\
&\alpha(\alpha^+)^4\hat f_{\ell,-}=-\f12\bigg(
\alpha^4\sum_{m=0}^{\ell-4}{\hat f}_{m,-}{\hat f}_{\ell-m-4,-}
+ (\alpha^+)^4\sum_{m=1}^{\ell-1} {\hat f}_{m,-} {\hat f}_{\ell-m,-}\no \\
& \quad -  2 \alpha^2 \sum_{m=0}^{\ell-3}{\hat f}_{m,-} \big(-2
 \alpha \alpha^+ {\hat f}_{\ell-m-3,-} + (\alpha^+)^2 \gamma {\hat f}^-_{\ell-m-3,-} 
 + \alpha^2 \gamma^+   {\hat f}^+_{\ell-m-3,-} \big)\no \\
& \quad + \sum_{m=0}^{\ell-2}\big(\alpha^4 (\gamma^+)^2 {\hat f}^+_{m,-} {\hat f}^+_{\ell-m-2,-}  \no \\
& \hspace*{1.55cm}
   + (\alpha^+)^2 \gamma {\hat f}^-_{m,-}((\alpha^+)^2 \gamma {\hat f}^-_{\ell-m-2,-}
       - 2 \alpha^2 \gamma^+ {\hat f}^+_{\ell-m-2,-}) \no \\
&\quad  \; \;    - 2 \alpha \alpha^+{\hat f}_{m,-}(-3 \alpha \alpha^+ {\hat f}_{\ell-m-2,-} 
       + 2(\alpha^+)^2 \gamma {\hat f}^-_{\ell-m-2,-} 
       + 2 \alpha^2\gamma^+{\hat f}^+_{\ell-m-2,-} )\big)\no \\
  & \quad    - 2 (\alpha^+)^2\sum_{m=0}^{\ell-1}{\hat f}_{m,-}\big(-2\alpha\alpha^+ 
          {\hat f}_{\ell-m-1,-} + (\alpha^+)^2 \gamma {\hat f}^-_{\ell-m-1,-} + \alpha^2 \gamma^+ 
{\hat f}^+_{\ell-m-1,-}\big) \bigg), \no \\
& \hspace*{10cm} \ell \geq 3,  \lb{ALfmcheckX} \\
& \hat g_{0,+} = \f{1}{2}, \quad  \hat g_{1,+}= -\alpha^+\beta,  \no \\
&\hat g_{2,+} =(\alpha^{+}\beta)^2-\gamma^+\alpha^{++}\beta - \gamma\alpha^+\beta^+,\no \\
&(\alpha \beta^+)^2{\hat g}_{\ell,+}=-\bigg(
(\alpha^+)^2\beta^2\sum_{m = 0}^{\ell-4}{\hat g}_{m,+} {\hat g}_{\ell-m-4,+} 
+ \alpha^2(\beta^+)^2\sum_{m = 1}^{\ell-1} {\hat g}_{m,+} {\hat g}_{\ell-m,+} \no \\
& \quad +  \alpha^+\beta\sum_{m = 0}^{\ell-3}\big(\gamma\gamma^+ 
          {\hat g}^-_{m,+} {\hat g}^+_{\ell-m-3,+} + {\hat g}_{m,+}
          ((1 + \alpha \beta)(1 + \alpha^+\beta^+) {\hat g}_{\ell-m-3,+} \no \\
          &\hspace*{2.3cm}  -(\gamma + \alpha^+ \beta^+\gamma) 
          {\hat g}^-_{\ell-m-3,+} +(-2 + \gamma) \gamma^+ {\hat g}^+_{\ell-m-3,+})\big) \no \\
& \quad + \sum_{m = 0}^{\ell-2} \big(\alpha^+\beta^+ \gamma^2{\hat g}^-_{m,+} 
          {\hat g}^-_{\ell-m-2,+} + \alpha \beta (\gamma^+)^2 {\hat g}^+_{m,+} {\hat g}^+_{\ell-m-2,+} \no \\
&\hspace*{1.55cm}  
+ {\hat g}_{m,+}((\alpha^+ \beta^+ + \alpha^2 \alpha^+ \beta^2 \beta^+ 
        + \alpha \beta (1 + \alpha^+ \beta^+)^2) {\hat g}_{\ell-m-2,+} \no \\
&\hspace*{1.55cm}  - 2(\alpha^+(1 + \alpha\beta) \beta^+ \gamma {\hat g}^-_{\ell-m-2,+} 
                    + \alpha \beta (1 + \alpha^+\beta^+)\gamma^+{\hat g}^+_{\ell-m-2,+})\big) \no \\
& \quad  + \alpha\beta^+ \sum_{m =0}^{\ell-1}\big(\gamma \gamma^+ {\hat g}^-_{m,+} 
      {\hat g}^+_{\ell-m-1,+} + {\hat g}_{m,+}((1 + \alpha\beta)(1 + \alpha^+\beta^+)
{\hat g}_{\ell-m-1,+} \no \\
&\hspace*{1.8cm} - (\gamma + \alpha^+ \beta^+ \gamma){\hat g}^-_{\ell-m-1,+} 
+(-2 + \gamma) \gamma^+ {\hat g}^+_{\ell-m-1,+})\big) \bigg),  \quad 
\ell \geq 3,  \lb{AL2.206bX} \\
& \hat g_{0,-} = \f{1}{2}, \quad  \hat g_{1,-}= -\alpha\beta^+,   \no \\
& \hat g_{2,-} =(\alpha\beta^{+})^2-\gamma^+\alpha\beta^{++} - \gamma\alpha^-\beta^+,\no \\
&(\alpha^+)^2\beta^2 {\hat g}_{\ell,-}= -\bigg(\alpha^2 
    (\beta^+)^2 \sum_{m=0}^{\ell-4}{\hat g}_{m,-} {\hat g}_{\ell-m-4,-} 
    + (\alpha^+)^2 \beta^2 \sum_{m=1}^{\ell-1} {\hat g}_{m,-} {\hat g}_{\ell-m,-} \no \\
& \quad + \alpha \beta^+ \sum_{m=0}^{\ell-3}\big(\gamma\gamma^+{\hat g}^-_{m,-} 
            {\hat g}^+_{\ell-m-3,-} + {\hat g}_{m,-}((1 + \alpha\beta)(1 + \alpha^+ 
                  \beta^+) {\hat g}_{\ell-m-3,-} \no \\
    &\hspace*{2.3cm} - (\gamma + \alpha^+ 
          \beta^+\gamma) {\hat g}^-_{\ell-m-3,-} + (-2+ \gamma)\gamma^+{\hat g}^+_{\ell-m-3,-})\big) \no \\
& \quad +\sum_{m=0}^{\ell-2} \big(\alpha^+ \beta^+ \gamma^2 {\hat g}^-_{m,-} 
              {\hat g}^-_{\ell-m-2,-} + \alpha \beta (\gamma^+)^2 {\hat g}^+_{m,-} {\hat g}^+_{\ell-m-2,-} \no \\
 & \hspace*{1.55cm} + {\hat g}_{m,-}((\alpha^+ \beta^+ + \alpha^2 \alpha^+ \beta^2 \beta^+ 
           +   \alpha \beta (1 + \alpha^+\beta^+)^2) {\hat g}_{\ell-m-2,-} \no \\
 & \hspace*{1.55cm}- 2(\alpha^+(1 + \alpha \beta) \beta^+ \gamma {\hat g}^-_{\ell-m-2,-} 
              + \alpha \beta (1 + \alpha^+\beta^+) \gamma^+ {\hat g}^+_{\ell-m-2,-}))\big)\no \\
& \quad + \alpha^+ \beta \sum_{m=0}^{\ell-1} \big(\gamma \gamma^+ 
          {\hat g}^-_{m,-} {\hat g}^+_{\ell-m-1,-} + {\hat g}_{m,-}((1 + \alpha \beta)(
          1 + \alpha^+ \beta^+) {\hat g}_{\ell-m-1,-} \no \\
&\hspace*{1.7cm} -( \gamma + \alpha^+ \beta^+ \gamma) {\hat g}^-_{\ell-m-1,-}+ (-2 + 
            \gamma) \gamma^+ {\hat g}^+_{\ell-m-1,-})\big)
\bigg), \quad \ell \geq 3,  \lb{ALgmcheckX} \\ 
&\hat h_{0,+}=\beta,  \quad \hat h_{1,+}= \gamma\beta^- -\alpha^+\beta^2,  \no \\
&\hat h_{2,+}=(\alpha^{+})^2 \beta^3 - \gamma^+ \alpha^{++}\beta^2 
-\gamma\big(\alpha(\beta^-)^2 - \gamma^{-}\beta^{--} + 2\alpha^{+}\beta^{-}\beta  \big),  \no \\
&\beta(\beta^+)^4{\hat h}_{\ell,+}=-\f12\bigg(
\beta^4 \sum_{m=0}^{\ell-4}{\hat h}_{m,+}{\hat h}_{\ell-m-4,+}
+ (\beta^+)^4\sum_{m=1}^{\ell-1}  {\hat h}_{m,+} {\hat h}_{\ell-m,+} \no \\
& \quad -  2 \beta^2 \sum_{m=0}^{\ell-3}{\hat h}_{m,+} \big(-2
\beta \beta^+ {\hat h}_{\ell-m-3,+} + (\beta^+)^2 \gamma {\hat h}^-_{\ell-m-3,+} 
+ \beta^2 \gamma^+ {\hat h}^+_{\ell-m-3,+}\big) \no \\
& \quad
+ \sum_{m=0}^{\ell-2} \big(\beta^4 (\gamma^+)^2 {\hat h}^+_{m,+} {\hat h}^+_{\ell-m-2,+}  \no \\
& \hspace*{1.55cm} 
+ (\beta^+)^2 \gamma {\hat h}^-_{m,+}((\beta^+)^2 \gamma {\hat h}^-_{\ell-m-2,+}
 - 2\beta^2 \gamma^+ {\hat h}^+_{\ell-m-2,+}) \no \\
&\qquad - 2\beta \beta^+ {\hat h}_{m,+}(-3\beta\beta^+ {\hat h}_{\ell-m-2,+} 
+ 2 (\beta^+)^2 \gamma {\hat h}^-_{\ell-m-2,+} + 2 \beta^2 \gamma^+ {\hat h}^+_{\ell-m-2,+})\big) \no \\
& \quad - 2(\beta^+)^2\sum_{m=0}^{\ell-1}{\hat h}_{m,+} \big(-2 \beta \beta^+ 
          {\hat h}_{\ell-m-1,+} + (\beta^+)^2\gamma {\hat h}^-_{\ell-m-1,+} + 
\beta^2\gamma^+ {\hat h}^+_{\ell-m-1,+}\big)\bigg),  \no \\
& \hspace*{10cm} \ell \geq 3,  \lb{AL2.206cX} \\
&\hat h_{0,-}=-\beta^+,  \quad \hat h_{1,-}= -\gamma^+\beta^{++} +\alpha(\beta^{+})^2, \no  \\
&\hat h_{2,-}=-\alpha^2 (\beta^+)^3 + \gamma \alpha^{-}(\beta^+)^2 
+\gamma\big(\alpha^+(\beta^{++})^2 - \gamma^{++}\beta^{+++} + 2\alpha\beta^{+}\beta^{++} \big), \no \\
&\beta^+\beta^4{\hat h}_{\ell,-}=\f12\bigg((\beta^+)^4 \sum_{m=0}^{\ell-4}{\hat h}_{m,-}
 {\hat h}_{\ell-m-4,-}+ \beta^4\sum_{m=1}^{\ell-1} {\hat h}_{m,-} {\hat h}_{\ell-m,-} \no \\
& \quad 
- 2 (\beta^+)^2 \sum_{m=0}^{\ell-3}{\hat h}_{m,-} \big(-2 \beta \beta^+ {\hat h}_{\ell-m-3,-}
 + (\beta^+)^2 \gamma {\hat h}^-_{\ell-m-3,-} + \beta^2\gamma^+\ {\hat h}^+_{\ell-m-3,-}\big)  \no \\
& \quad
+ \sum_{m=0}^{\ell-2}\big(\beta^4 (\gamma^+)^2 {\hat h}^+_{m,-}{\hat h}^+_{\ell-m-2,-}  \no \\
& \qquad 
+ (\beta^+)^2 \gamma {\hat h}^-_{m,-} ((\beta^+)^2 \gamma 
                        {\hat h}^-_{\ell-m-2,-} - 2 \beta^2 \gamma^+ {\hat h}^+_{\ell-m-2,-})\no \\
& \qquad       - 2 \beta \beta^+ {\hat h}_{m,-}(-3\beta \beta^+\ {\hat h}_{\ell-m-2,-} +
 2 (\beta^+)^2 \gamma {\hat h}^-_{\ell-m-2,-} 
                    + 2\beta^2 \gamma^+ {\hat h}^+_{\ell-m-2,-})\big)\no \\
   & \quad - 2 \beta^2 \sum_{m=0}^{\ell-1}{\hat h}_{m,-} \big(-2\beta \beta^+ {\hat h}_{\ell-m-1,-}  + (\beta^+)^2 \gamma {\hat h}^-_{\ell-m-1,-} + \beta^2 \gamma^+  {\hat h}^+_{\ell-m-1,-}\big) \bigg),  \no \\
& \hspace*{10cm} \ell \geq 3.  \lb{ALhmcheckX}
\end{align}
\end{theorem}
\begin{proof}
We first consider the expansions \eqref{ALasymptotfghinf} near $1/z=0$ and the nonlinear recursion relations \eqref{AL2.206aX}, \eqref{AL2.206bX}, and 
\eqref{AL2.206cX} in detail.  
Inserting expansion \eqref{ALasympfghinf} for $\frf$ into \eqref{ALquadrf}, the expansion 
\eqref{ALasympfghinf} for $\frg$ into \eqref{ALquadrg}, and the expansion 
\eqref{ALasympfghinf} for $\frh$ into \eqref{ALquadrh}, then yields the nonlinear recursion relations \eqref{AL2.206aX}, \eqref{AL2.206bX}, and \eqref{AL2.206cX}, but with 
$\hat f_{\ell,+}$, $\hat g_{\ell,+}$, and $\hat h_{\ell,+}$ replaced by $\hat \frf_{\ell,+}$, 
$\hat \frg_{\ell,+}$, and $\hat \frh_{\ell,+}$, respectively. From the leading asymptotic behavior one finds that 
$\hat \frf_{0,+}=-\alpha^+$, ${\hat \frg}_{0,+}=\frac12$, and ${\hat \frh}_{0,+}=\beta$.

Next, inserting the expansions \eqref{ALasympfghinf} for $\frf$, $\frg$, and $\frh$ into 
\eqref{ALlin1}--\eqref{ALlin4}, and coparing powers of $z^{-\ell}$ as $|z|\to\infty$, 
$z\in C_R$, one infers that $\frf_{\ell,+}$, $\frg_{\ell,+}$, and $\frh_{\ell,+}$ satisfy the linear recursion relations \eqref{AL0+}--\eqref{ALh_l+}. Here we have used 
\eqref{ALK_p}. The coefficients $\hat \frf_{0,+}$, $\hat \frg_{0,+}$, and $\hat \frh_{0,+}$ are  consistent with \eqref{AL0+} for  $c_{0,+}=1$. Hence one concludes that
\begin{equation}
\hat \frf_{\ell,+} = f_{\ell,+}, \quad \hat \frg_{\ell,+} =g_{\ell,+}, \quad 
\hat \frh_{\ell,+} = h_{\ell,+}, \quad \ell\in\bbN_0,
\end{equation}
for certain values of the summation constants $c_{\ell,+}$. To conclude that actually, 
${\hat \frf}_{\ell,+}=\hat f_{\ell,+}$, ${\hat \frg}_{\ell,+}=\hat g_{\ell,+}$,  
${\hat \frh}_{\ell,+}=\hat h_{\ell,+}$, $\ell\in\bbN_0$, and hence all $c_{\ell,+}$, 
$\ell\in\bbN$, vanish, we now rely on the notion of degree as introduced in Remark 
\ref{rAL2.4}. To this end we recall that   
\begin{equation}
\deg\big(\hat f_{\ell,+}\big)=\ell+1, \quad \deg\big(\hat g_{\ell,+}\big)=\ell, \quad 
\deg\big(\hat h_{\ell,+}\big)=\ell, \quad \ell\in\bbN_0,
\end{equation}
(cf.\ \eqref{AL2.1AAa}). Similarly, the nonlinear recursion relations \eqref{AL2.206aX}, \eqref{AL2.206bX}, and \eqref{AL2.206cX} yield inductively that 
\begin{equation}
\deg\big(\hat \frf_{\ell,+}\big)=\ell+1, \quad \deg\big(\hat \frg_{\ell,+}\big)=\ell, \quad 
\deg\big(\hat \frh_{\ell,+}\big)=\ell, \quad \ell\in\bbN_0.
\end{equation}
Hence one concludes 
\begin{equation}
\hat \frf_{\ell,+} = \hat f_{\ell,+}, \quad \hat \frg_{\ell,+} =\hat g_{\ell,+}, \quad 
\hat \frh_{\ell,+} =\hat h_{\ell,+}, \quad \ell\in\bbN_0.
\end{equation}

The proof of the corresponding asymptotic expansion \eqref{ALasymptotfghzero} and the nonlinear recursion relations \eqref{ALfmcheckX}, \eqref{ALgmcheckX}, and 
\eqref{ALhmcheckX}  follows precisely the same strategy and is hence omitted. 
\end{proof}

Given this general result on asymptotic expansions, we now specialize to the algebro-geometric case at hand. We recall our conventions 
$y(P)= \mp (\zeta^{-p-1}+\Oh(\zeta^{-p}))$ for $P$ near 
$\Pinfpm$ (where $\zeta=1/z$) and $y(P) = \pm ((c_{0,-}/c_{0,+})+\Oh(\zeta))$ for $P$ 
near $\Pzpm$ (where $\zeta=z$). 

\begin{theorem} \lb{tALB.2}
Assume \eqref{ALneq 0,1}, $\sAL_{\ul p}(\alpha,\beta)=0$, and suppose
$P=(z,y)\in\calK_p\setminus\{\Pinfp,\Pinfm\}$. Then $z^{p_-} F_{\ul p}/y$,
$z^{p_-} G_{\ul p}/y$, and $z^{p_-} H_{\ul p}/y$ have the following convergent expansions as $P\to \Pinfpm$, respectively, $P\to\Pzpm$, 
\begin{align} 
\frac{z^{p_-}}{c_{0,+}} \frac{F_{\ul p}(z)}{y} &= \begin{cases} 
\mp \sum_{\ell=0}^\infty \hat f_{\ell,+} \zeta^{\ell+1},  &
P\to \Pinfpm, \qquad \zeta=1/z, \\
\pm \sum_{\ell=0}^\infty \hat f_{\ell,-} \zeta^\ell,  &
P\to \Pzpm, \qquad \zeta=z,
\end{cases}  \label{ALF/y 0} \\
\frac{z^{p_-}}{c_{0,+}} \frac{G_{\ul p}(z)}{y} &= \begin{cases}
\mp \sum_{\ell=0}^\infty \hat g_{\ell,+} \zeta^\ell,  &
P\to \Pinfpm, \qquad \zeta=1/z, \\
\pm \sum_{\ell=0}^\infty \hat g_{\ell,-} \zeta^\ell,  &
P\to \Pzpm, \qquad \zeta=z, 
\end{cases}    \label{ALG/y 0} \\
\frac{z^{p_-}}{c_{0,+}} \frac{H_{\ul p}(z)}{y} &= \begin{cases}
\mp \sum_{\ell=0}^\infty \hat h_{\ell,+} \zeta^\ell,  &
P\to \Pinfpm, \qquad \zeta=1/z, \\
\pm \sum_{\ell=0}^\infty \hat h_{\ell,-} \zeta^{\ell+1},  &
P\to \Pzpm, \qquad \zeta=z,
\end{cases}    \label{ALH/y 0} 
\end{align}
where $\zeta=1/z$ $($resp., $\zeta=z)$ is the local coordinate near $\Pinfpm$ 
$($resp., $\Pzpm)$ and $\hat f_{\ell,\pm}$, $\hat g_{\ell,\pm}$, and $\hat h_{\ell,\pm}$ are the homogeneous versions\footnote{Strictly speaking, the coefficients 
$\hat f_{\ell,\pm}$, $\hat g_{\ell,\pm}$, and $\hat h_{\ell,\pm}$ in 
\eqref{ALF/y 0}--\eqref{ALH/y 0} no longer have a well-defined degree and hence represent a slight abuse of notation since we assumed that $\sAL_{\ul p}(\alpha,\beta)=0$. At any rate, they are explicitly given by \eqref{ALstorf}--\eqref{ALstorh}.} of the coefficients 
$f_{\ell,\pm}$, $g_{\ell,\pm}$, and $h_{\ell,\pm}$ as introduced in 
\eqref{AL2.04a}--\eqref{AL2.04c}. 
Moreover, one infers for the $E_m$-dependent summation constants
$c_{\ell,\pm}$, $\ell=0,\dots, p_\pm$, in $F_{\ul p}$, $G_{\ul p}$, and $H_{\ul p}$ that 
\begin{equation}
c_{\ell,\pm}= c_{0,\pm} c_\ell\big(\ul E^{\pm 1}\big), \quad \ell=0,\dots,p_\pm.   \lb{ALBc}
\end{equation}
In addition, one has the following relations between the homogeneous and  
nonhomogeneous recursion coefficients:
\begin{align}
 f_{\ell,\pm}&=c_{0,\pm}\sum_{k=0}^{\ell} c_{\ell-k}\big(\ul E^{\pm1}\big)\hat f_{k,\pm}, 
 \quad \ell=0,\dots,p_\pm,  \lb{ALB2.26p1} \\
 g_{\ell,\pm}&=c_{0,\pm}\sum_{k=0}^{\ell} c_{\ell-k}\big(\ul E^{\pm 1}\big) \hat g_{k,\pm}, 
 \quad \ell=0,\dots,p_\pm, \lb{ALB2.26p2} \\
 h_{\ell,\pm}&=c_{0,\pm} \sum_{k=0}^{\ell}  c_{\ell-k}\big(\ul E^{\pm 1}\big) h_{k,\pm}, 
 \quad \ell=0,\dots,p_\pm. \lb{ALB2.26p3}
\end{align}
Furthermore, one has
\begin{align}
c_{0,\pm}\hat f_{\ell,\pm}&=\sum_{k=0}^{\ell} \hat c_{\ell-k}\big(\ul  
E^{\pm1}\big) f_{k,\pm},
\quad \ell=0,\dots,p_\pm-1, \lb{ALB2.26q1} \\
c_{0,\pm}\hat f_{p_{\pm},\pm}&=\sum_{k=0}^{p_{\pm}-1} \hat c_{p_{\pm}-k}
\big(\ul E^{\pm1}\big) f_{k,\pm}+\hat c_{0}(\ul E^{\pm1}) f_{p_{\mp}-1,\mp},  \no \\[1mm]
c_{0,\pm}\hat g_{\ell,\pm}&=\sum_{k=0}^{\ell}  \hat c_{\ell-k}\big 
(\ul E^{\pm 1}\big) g_{k,\pm},
\quad \ell=0,\dots,p_\pm-1,\lb{ALB2.26q2} \\
c_{0,\pm}\hat g_{p_{\pm},\pm}&=\sum_{k=0}^{p_{\pm}-1}  \hat c_{p_{\pm}-k}
\big(\ul E^{\pm 1}\big) g_{k,\pm}+\hat c_{0}(\ul E^{\pm 1}) g_{p_{\mp},\mp},    \no \\[1mm]
c_{0,\pm}\hat h_{\ell,\pm}&=\sum_{k=0}^{\ell}  \hat c_{\ell-k}\big 
(\ul E^{\pm 1}\big) h_{k,\pm},
\quad \ell=0,\dots,p_\pm-1, \lb{ALB2.26q3} \\
c_{0,\pm}\hat h_{p_{\pm},\pm}&=\sum_{k=0}^{p_{\pm}-1}  \hat c_{p_{\pm}-k}
\big(\ul E^{\pm 1}\big) h_{k,\pm} + \hat c_{0}(\ul E^{\pm 1}) h_{p_{\mp}-1,\mp}.  \no
\end{align}
For general $\ell$ $($not restricted to $\ell\le p_\pm$$)$ one 
has\footnote{$m\vee n = \max\{m,n\}$.}
\begin{align}
&c_{0,\pm}\hat f_{\ell,\pm} = \begin{cases}
\sum_{k=0}^{\ell} \hat c_{\ell-k}\big(\ul E^{\pm1}\big) f_{k,\pm} , & \ell=0,\dots,p_\pm-1,\\
\begin{array}{ll}\sum_{k=0}^{p_{\pm}-1} \hat c_{\ell-k}\big(\ul E^{\pm1}\big) f_{k,\pm} \\
+\sum_{k=(p-\ell)\vee 0}^{p_{\mp}-1} \hat c_{\ell+k-p}\big(\ul E^{\pm1}\big) 
f_{k,\mp},
\end{array} & \ell \geq p_\pm,   
\end{cases}   \lb{ALstorf}   \\[2mm]
&c_{0,\pm}\hat g_{\ell,\pm} = \begin{cases}
\sum_{k=0}^{\ell} \hat c_{\ell-k}\big(\ul E^{\pm1}\big) g_{k,\pm} , 
& \ell=0,\dots,p_\pm-\dpm,\\
\begin{array}{ll}\sum_{k=0}^{p_\pm -\dpm} \hat c_{\ell-k}\big(\ul E^{\pm1}\big) 
g_{k,\pm} \\
+\sum_{k=(p-\ell)\vee 0}^{p_{\mp}-\dpm} \hat c_{\ell+k-p}\big(\ul E^{\pm1}\big) 
g_{k,\mp},
\end{array} & \ell \geq p_\pm-\dpm+1,
\end{cases} \lb{ALstorg} \\[2mm]  
&c_{0,\pm}\hat h_{\ell,\pm}  = \begin{cases}
\sum_{k=0}^{\ell} \hat c_{\ell-k}\big(\ul E^{\pm1}\big) h_{k,\pm} ,  
& \ell=0,\dots,p_\pm -1,\\
\begin{array}{ll}\sum_{k=0}^{p_{\pm}-1} \hat c_{\ell-k}\big(\ul E^{\pm1}\big) 
h_{k,\pm} \\
+\sum_{k=(p-\ell)\vee 0}^{p_{\mp}-1} \hat c_{\ell+k-p}\big(\ul E^{\pm1}\big) 
h_{k,\mp},
\end{array} & \ell \geq p_\pm.  
\end{cases}  \lb{ALstorh} 
\end{align}
Here we used the convention 
\begin{equation} \lb{ALdelta}
\dpm= \begin{cases} 0, & +, \\ 1, & -. 
\end{cases}
\end{equation}
\end{theorem}
\begin{proof}
Identifying 
\begin{equation}
\Psi_+(z,\dott) \, \text{ with } \, \Psi(P,\dott,0) \, \text{ and } \, 
\Psi_-(z,\dott) \, \text{ with } \, \Psi(P^*,\dott,0),   \lb{ALpsiid}
\end{equation}
recalling that $W(\Psi(P,\dott,0),\Psi(P^*,\dott,0))=-c_{0,+} z^{n-n_0-p_-}y 
F_{\ul p}(z,0)^{-1}
\Gamma(n,n_0)$ (cf.\ \eqref{ALpsi 7}), and similarly, identifying 
\begin{equation}
\phi_+(z,\dott) \, \text{ with } \, \phi(P,\dott) \, \text{ and } \, 
\phi_-(z,\dott) \, \text{ with } \, \phi(P^*,\dott),   \lb{ALphiid}
\end{equation} 
a comparison of \eqref{ALphipm}--\eqref{ALgothh} and the results of Lemmas 
\ref{lAL3.1} and \ref{lAL3.3} shows that we may also identify 
\begin{equation}
\frf \, \text{ with } \, \mp \f{2 F_{\ul p}}{c_{0,+} z^{-p_-}y}, \, \quad \frg \, \text{ with } \, 
\mp \f{2 G_{\ul p}}{c_{0,+} z^{-p_-}y}, \text{ and } \, \frh \, \text{ with } \, 
\mp \f{2 H_{\ul p}}{c_{0,+} z^{-p_-}y},   \lb{ALFGid}
\end{equation}
the sign depending on whether $P$ tends to $\Pinfpm$ or to $\Pzpm$. 
In particular, \eqref{ALlin1}--\eqref{ALquadrh} then correspond to 
\eqref{AL1,1}--\eqref{AL2,1}, \eqref{ALR}, \eqref{ALnldeF}--\eqref{ALnldeH}, respectively.  
Since $z^{p_-} F_{\ul p}/y$, $z^{p_-}G_{\ul p}/y$, and $z^{p_-}H_{\ul p}/y$  clearly have asymptotic (in fact, even convergent) expansions as $|z|\to\infty$ and as $|z|\to 0$, the results of Theorem \ref{tALB.2A} apply. Thus, as $P\to\Pinfpm$, one obtains the following expansions using \eqref{ALB2.26d} and \eqref{ALF_p}--\eqref{ALH_p}: 
\begin{align}
& \frac{z^{p_-}}{c_{0,+}} \frac{F_{\ul p}(z)}{y} 
\underset{\zeta\to 0}{=} \mp\frac{1}{c_{0,+}}\bigg(\sum_{k=0}^\infty \hat c_k(\ul E) \zeta^k\bigg)
\bigg(\sum_{\ell=1}^{p_-} f_{p_- -\ell,-} \zeta^{p_+ +\ell} 
+ \sum_{\ell=0}^{p_+ -1} f_{p_+ -1-\ell,+}\zeta^{p_+ -\ell} \bigg),  \no \\
&\quad \underset{\zeta\to 0}{=} \mp\sum_{\ell=0}^\infty{\hat f}_{\ell,+} \zeta^{\ell+1},  
\lb{ALF_0}  \\
& \frac{z^{p_-}}{c_{0,+}} \frac{G_{\ul p}(z)}{y} 
\underset{\zeta\to 0}{=} \mp\frac{1}{c_{0,+}}\bigg(\sum_{k=0}^\infty 
\hat c_k(\ul E) \zeta^{k}\bigg)
\bigg(\sum_{\ell=1}^{p_-} g_{p_- -\ell,-}\zeta^{p_+ +\ell}  
+ \sum_{\ell=0}^{p_+} g_{p_+ -\ell,+} \zeta^{p_+ -\ell} \bigg)\notag\\
&\quad \underset{\zeta\to 0}{=} \mp\sum_{\ell=0}^\infty{\hat g}_{\ell,+} \zeta^{\ell}, 
\lb{ALG_0}\\
& \frac{z^{p_-}}{c_{0,+}} \frac{H_{\ul p}(z)}{y}  
\underset{\zeta\to 0}{=} \mp\frac{1}{c_{0,+}}\bigg(\sum_{k=0}^\infty 
\hat c_k(\ul E) \zeta^k\bigg)
\bigg(\sum_{\ell=0}^{p_- -1} h_{p_- -1-\ell,-}\zeta^{p_+ +\ell}  
+ \sum_{\ell=1}^{p_+} h_{p_+-\ell,+}\zeta^{p_+ -\ell} \bigg)\notag\\
&\quad \underset{\zeta\to 0}{=} \mp\sum_{\ell=0}^\infty{\hat h}_{\ell,+} \zeta^{\ell}.  
\lb{ALH_0}
\end{align}
This implies \eqref{ALF/y 0}--\eqref{ALH/y 0} as $P\to\Pinfpm$.

Similarly, as $P\to\Pzpm$,  \eqref{ALB2.26d} and \eqref{ALF_p}--\eqref{ALH_p}, and \eqref{ALprod E_m} imply 
\begin{align}
&\frac{z^{p_-}}{c_{0,+}} \frac{F_{\ul p}(z)}{y}  
\underset{\zeta\to 0}{=} \pm\frac{1}{c_{0,-}}\bigg(\sum_{k=0}^\infty \hat c_k(\ul E^{-1}) \zeta^k\bigg)  \no \\
& \hspace*{2.4cm}
\times \bigg(\sum_{\ell=1}^{p_-} f_{p_- -\ell,-} \zeta^{p_+ -\ell} 
+ \sum_{\ell=0}^{p_+ -1} f_{p_+ -1-\ell,+} \zeta^{p_+ +\ell} \bigg)\notag\\
&\quad \underset{\zeta\to 0}{=} \pm\sum_{\ell=0}^\infty{\hat f}_{\ell,-} \zeta^{\ell}, 
\lb{ALF_0m}  \\
&\frac{z^{p_-}}{c_{0,+}} \frac{G_{\ul p}(z)}{y}  
\underset{\zeta\to 0}{=} \pm\frac{1}{c_{0,-}}\bigg(\sum_{k=0}^\infty \hat c_k(\ul E^{-1}) \zeta^k\bigg)   
\bigg(\sum_{\ell=1}^{p_-} g_{p_- -\ell,-}\zeta^{p_+ -\ell}  
+ \sum_{\ell=0}^{p_+} g_{p_+ -\ell,+} \zeta^{p_+ +\ell} \bigg)   \notag\\
&\quad 
\underset{\zeta\to 0}{=} \pm\sum_{\ell=0}^\infty{\hat g}_{\ell,-} \zeta^{\ell}, 
\lb{ALG_0m}  \\
&\frac{z^{p_-}}{c_{0,+}} 
\frac{H_{\ul p}(z)}{y} 
\underset{\zeta\to 0}{=} \pm\frac{1}{c_{0,-}}\bigg(\sum_{k=0}^\infty \hat c_k(\ul E^{-1}) \zeta^k\bigg)   \no \\
& \hspace*{2.4cm} 
\times \bigg(\sum_{\ell=0}^{p_- -1} h_{p_- -1-\ell,-}\zeta^{p_+ -\ell}  
+ \sum_{\ell=1}^{p_+} h_{p_+ -\ell,+} \zeta^{p_+ +\ell} \bigg)\notag\\
&\quad \underset{\zeta\to 0}{=} \pm\sum_{\ell=0}^\infty{\hat h}_{\ell,-} \zeta^{\ell+1}. 
\lb{ALH_0m}
\end{align} 
Thus, \eqref{ALF/y 0}--\eqref{ALH/y 0} hold as $P\to\Pzpm$.

Next, comparing powers of $\zeta$ in the second and third term of \eqref{ALF_0}, formula \eqref{ALB2.26q1} follows (and hence \eqref{ALstorf} as well). Formulas 
\eqref{ALB2.26q2} and \eqref{ALB2.26q3} follow by using \eqref{ALG_0} and 
\eqref{ALH_0}, respectively.

To prove \eqref{ALB2.26p1} one uses \eqref{ALB2.26j} and finds 
\begin{align}
c_{0,\pm}\sum_{m=0}^\ell c_{\ell-m}\big(\ul E^{\pm1}\big)\hat f_{m,\pm}
=\sum_{m=0}^\ell c_{\ell-m}(\ul E)\sum_{k=0}^{m} \hat c_{m-k}\big(\ul E^{\pm1}\big) 
f_{k,\pm} = f_{\ell,\pm}.
\end{align}
The proofs of \eqref{ALB2.26p2} and \eqref{ALB2.26p3} and those of 
\eqref{ALstorg} and \eqref{ALstorh} are analogous.
\end{proof}

Finally, we also mention the following system of recursion relations for the homogeneous coefficients $\hat f_{\ell,\pm}$, $\hat g_{\ell,\pm}$, 
and $\hat h_{\ell,\pm}$.  

\begin{lemma} \label{lALB.1}
The homogeneous coefficiens $\hat f_{\ell,\pm}$, $\hat g_{\ell,\pm}$, and 
$\hat h_{\ell,\pm}$ are uniquely defined by the following recursion relations:
\begin{align}
\begin{split} \lb{ALB1.9}
\hat g_{0,+} &= \frac{1}{2}, \quad \hat f_{0,+} = -\alpha^+, \quad \hat h_{0,+} = \beta, \\
\hat g_{l+1,+} &= \sum_{k=0}^l \hat f_{l-k,+} \hat h_{k,+}
- \sum_{k=1}^l \hat g_{l+1-k,+} \hat g_{k,+}, \\
\hat f_{l+1,+}^- &= \hat f_{l,+} - \alpha (\hat g_{l+1,+} + \hat g_{l+1,+}^-),\\
\hat h_{l+1,+} &= \hat h_{l,+}^- + \beta (\hat g_{l+1,+} + \hat g_{l+1,+}^-), 
\end{split}
\end{align}
and
\begin{align}
\begin{split}
\hat g_{0,-} &= \frac{1}{2}, \quad \hat f_{0,-} = \alpha, \quad \hat h_{0,-} = -\beta^+, \\
\hat g_{l+1,-} &= \sum_{k=0}^l \hat f_{l-k,-} \hat h_{k,-}
- \sum_{k=1}^l \hat g_{l+1-k,-} \hat g_{k,-}, \\
\hat f_{l+1,-} &= \hat f_{l,-}^- + \alpha (\hat g_{l+1,-} + \hat g_{l+1,-}^-),\\
\hat h_{l+1,-}^- &= \hat h_{l,-} - \beta (\hat g_{l+1,-} + \hat g_{l+1,-}^-). 
\end{split}
\end{align}
\end{lemma}
\begin{proof}
One verifies that the coefficients defined via these recursion relations satisfy 
\eqref{AL0+}--\eqref{ALh_l+} (respectively, \eqref{AL0-}--\eqref{ALh_l-}). Since they are homogeneous of the required degree this completes the proof. 
\end{proof}

{\bf Acknowledgments.} 
F.G., J.M., and G.T. gratefully acknowledge the extraordinary hospitality of the Department of Mathematical Sciences of the Norwegian University of Science and
Technology, Trondheim, during extended stays in the summers of 2004--2006,
where parts of this paper were written. F.G. and G.T. would like to thank all organizers of the international conference on Operator Theory and Mathematical Physics (OTAMP), Lund, June 2006, and especially, Pavel Kurasov, for their kind invitation and  the stimulating atmosphere during the meeting. We are indebted to the anonymous referee for constructive remarks. 

\bigskip\bigskip


\end{document}